\documentclass[fleqn,usenatbib]{mnras}
\usepackage{newtxtext,newtxmath}
\usepackage{soul}
\usepackage[T1]{fontenc}
\usepackage{gensymb}
\usepackage{afterpage}
\DeclareRobustCommand{\VAN}[3]{#2}
\let\VANthebibliography\thebibliography
\def\thebibliography{\DeclareRobustCommand{\VAN}[3]{##3}\VANthebibliography}
\usepackage{graphicx}	
\usepackage{amsmath}	
\usepackage{siunitx}

\title[MeerKLASS UHF DR1]{The MeerKLASS UHF On-the-Fly Continuum Survey -- Data Release I}

\author[S. Paul et al.]{Sourabh Paul$^1$\thanks{sourabh.paul@manchester.ac.uk},
Keith Grainge$^1$,
Mario G. Santos$^{2,3}$,
Suman Chatterjee$^2$,
Sarvesh Mangla$^4$,
Laura Wolz$^1$,
\newauthor
Joseph J. Mohr$^4$,
Oleg Smirnov$^{5,3,6}$,
Cyril Tasse$^{7,8}$,
Kristof Rozgonyi$^4$,
Matthias Hoeft$^9$,
Yvette Perrott$^{10}$
\\
\\
$^1$Jodrell Bank Centre for Astrophysics, Department of Physics \& Astronomy, The University of Manchester, Manchester M13 9PL, UK\\
$^2$Department of Physics and Astronomy, University of the Western Cape, Robert Sobukwe Road, Cape Town 7535, South Africa\\
$^3$South African Radio Astronomy Observatory (SARAO), Cape Town, 7925, South Africa\\
$^4$University Observatory, LMU Faculty of Physics, Scheinerstr. 1, 81679, Munich, Germany\\
$^5$Centre for Radio Astronomy Techniques and Technologies (RATT), Department of Physics and Electronics, Rhodes University, Makhanda, 6140, South Africa\\
$^6$Institute for Radioastronomy, National Institute of Astrophysics (INAF IRA), Via Gobetti 101, 40129 Bologna, Italy\\
$^7$GEPI \& ORN, Observatoire de Paris, Université PSL, CNRS, 5 Place Jules Janssen, 92190 Meudon, France\\
$^8$Department of Physics \& Electronics, Rhodes University, PO Box 94, Grahamstown, 6140, South Africa\\
$^9$Thüringer Landessternwarte, Sternwarte 5, 07778 Tautenburg, Germany\\
$^{10}$School of Chemical and Physical Sciences, Victoria University of Wellington, Wellington 6012, New Zealand\\
}

\date{Accepted XXX. Received YYY; in original form ZZZ}

\pubyear{\the\year{}}

\begin{document}
\label{firstpage}
\pagerange{\pageref{firstpage}--\pageref{lastpage}}
\maketitle

\begin{abstract}
We present the first public data release (DR1) from the interferometric component of the MeerKAT Large Area Synoptic Survey (MeerKLASS) UHF survey, a legacy program demonstrating a novel on-the-fly (OTF) mapping technique. This release is based on 12 hours of early science observations covering approximately 800\,deg$^2$ of the southern sky. We describe the data processing pipeline developed to calibrate and image these fast-scanning observations, producing high-fidelity continuum images at a central frequency of 816\,MHz. The resulting mosaic reaches an RMS sensitivity of $\sim35\,\mu$Jy\,beam$^{-1}$ in its deepest regions, with a typical angular resolution of $\sim32'' \times 17''$. In these images, we identify $95483$ radio sources. We validate the catalogue through cross-matching with external surveys, confirming sub-arcsecond astrometric accuracy and a robust flux density scale. We compute the differential source counts, finding excellent agreement with existing measurements and validating our end-to-end processing. The success of this pilot study serves as a crucial proof of concept for the OTF observing strategy, and the public release of the images and source catalogue provides a valuable resource for a wide range of astrophysical studies. This work paves the way for the full MeerKLASS OTF survey and future large-area survey projects with the SKA.
\end{abstract}


\section{Introduction}
The MeerKAT Large Area Synoptic Survey (MeerKLASS; \citealt{meerklass}) is a flagship radio–continuum and neutral hydrogen (H\,I) intensity–mapping survey conducted with the MeerKAT telescope \citep{jonas2009}. Executed in the UHF band (544--1088\,MHz), MeerKLASS combines wide–area coverage, moderate depth, and high–fidelity imaging to address key science goals aligned with the Square Kilometer Array (SKA) precursor and pathfinder programmes. It has two central objectives: (i) studying the growth of cosmic structure via H\,I intensity mapping, and (ii) assembling a statistical census of the radio–source population across cosmic time. The survey is configured to leverage MeerKAT’s commensal observing: while the single–dish H\,I mode drives a constant–elevation, fast–scanning strategy that necessitates OTF observing, the array simultaneously records interferometric visibilities, enabling a wide–area, high–sensitivity continuum survey. This dual–mode strategy delivers contiguous, high–fidelity coverage across a large fraction of the southern sky (outside the Galactic plane), unifying cosmological and astrophysical science in a single observing programme. By capitalising on MeerKAT’s dense core configuration, low system temperature, and advanced calibration capabilities, MeerKLASS achieves sensitivity to both diffuse emission and compact sources over thousands of square degrees.

Operationally, the constant–elevation OTF strategy enables efficient mapping of large areas, mitigates direction-dependent calibration artefacts that often arise from static mosaics, and produces images with uniform sensitivity and a stable point spread function (PSF) across wide fields. OTF mapping \citep{Mangum2007} has been implemented successfully for single-dish surveys and has also been deployed interferometrically in a handful of studies \citep{Mooley2016, lacy2020}, with many practical considerations shared with drift–scan interferometric modes (e.g. \citealt{Perrott2013}). Our implementation adapts these ideas to MeerKAT’s UHF band and commensal observing on a substantially larger \emph{operational scale}. The present data release is based on OTF observations covering $\sim 800$ deg$^2$ within the Dark Energy Spectroscopic Instrument \citep[DESI;][]{DESI} survey footprint, with a total integration time of approximately $\sim12$ hours.

In its single-dish mode, MeerKLASS targets large-scale fluctuations in the redshifted 21~cm emission from unresolved galaxies, employing the HI intensity mapping technique \citep[e.g.][]{Chang_2010,Masui_2013,Anderson_2018,Wolz_2022,CHIME_2022,Cunnington_2023,Paul_2023}. By measuring the clustering of neutral hydrogen at redshifts $ z \lesssim 1.6$, the survey aims to probe the underlying matter distribution and detect the imprint of baryon acoustic oscillations (BAO)—a robust cosmological standard ruler that can be employed to study the expansion history of the Universe. These measurements will provide independent constraints on dark energy and modified gravity models, while also laying the groundwork for future SKA-era cosmological surveys.

In parallel, a single one-hour pass of the UHF-band OTF interferometric data typically images $\sim300$\,deg$^{2}$ at $\sim15\arcsec$ resolution and achieves an rms of $\sim100~\mu$Jy\,beam$^{-1}$; multiple passes reduce the noise approximately as $t^{-1/2}$. The wide area and frequency range permit statistically robust investigations of the evolution and spectral properties of star-forming galaxies (SFGs), radio-loud active galactic nuclei (AGN), and diffuse radio sources such as cluster halos, cluster relics, and remnant AGN lobes. The survey is particularly well suited to identify useful samples of giant radio galaxies, dying AGN, or high-redshift radio galaxies—that are under-represented in smaller-area, deep surveys.

MeerKLASS builds upon and complements a rich landscape of current and legacy radio–continuum surveys. At low frequencies (120–168\,MHz), the LOFAR Two-metre Sky Survey \citep[LoTSS;][]{shimwell2022} provides high-resolution ($\sim 6\arcsec$) imaging of the northern sky with typical rms $\sim$70–100\,$\mu$Jy\,beam$^{-1}$, while the TIFR GMRT Sky Survey \citep[TGSS;][]{intema2017} covers $\approx 90\%$ of the sky (Dec $> -53\degree$) at similar frequencies with $\sim 25\arcsec$ resolution and a typical sensitivity of $3.5$~mJy~beam$^{-1}$. The Galactic and Extra-Galactic All-Sky MWA Survey \citep[GLEAM;][72–231\,MHz]{wayth2015} offers wide-band southern-sky coverage with $\sim$2–3\arcmin resolution (e.g. $\sim 2.5\arcmin$ at 154\,MHz) and typical wide-band rms of $\sim$10\,mJy\,beam$^{-1}$. In the southern hemisphere, the Sydney University Molonglo Sky Survey \citep[SUMSS;][]{mauch2003} provides 843~MHz imaging at $45\arcsec$ resolution with an RMS sensitivity of $\sim$1~mJy~beam$^{-1}$, complementing the NRAO VLA Sky Survey \citep[NVSS;][]{condon1998}, which covers the sky north of $-40\degree$ declination at 1.4~GHz and $45\arcsec$ resolution with a typical sensitivity of $0.45$~mJy~beam$^{-1}$. More recently, the Rapid ASKAP Continuum Survey \citep[RACS;][]{mcconnell2020} has mapped the entire southern sky at multiple frequency bands. RACS–Low \citep{hale2021}, centred at $887.5$\,MHz, provides imaging at $15$–$25\arcsec$ resolution with a typical rms sensitivity of $\sim250~\mu$Jy\,beam$^{-1}$, while RACS–Mid, centred at 1367.5\,MHz, delivers a median PSF of $\sim11\arcsec\times9\arcsec$ with rms in the few $\times10^{-4}$\,Jy\,beam$^{-1}$ range \citep{Duchesne2023}. Complementing the rapid RACS survey is the deeper Evolutionary Map of the Universe \citep[EMU;][]{norris2021, Hopkins2025}, another flagship ASKAP project. EMU is planned to survey the entire sky south of equator until 2028 at 900~MHz, aiming for $\sim 15\arcsec$ resolution and a much deeper RMS sensitivity of $\sim$20 -- 30~$\mu$Jy~beam$^{-1}$. The ongoing VLA Sky Survey \citep[VLASS;][]{lacy2020} offers all-sky coverage at 3~GHz with $2.5\arcsec$ resolution and a sensitivity of $\sim$120~$\mu$Jy~beam$^{-1}$ per epoch, enabling high-resolution comparisons. In parallel, the MeerKAT International GHz Tiered Extragalactic Exploration survey \citep[MIGHTEE;][]{jarvis2016, heywood2022}, conducted with MeerKAT in the L-band, targets deep imaging ($\sim$1~$\mu$Jy~beam$^{-1}$) at $\sim 6\arcsec$ resolution over a few well-studied extragalactic fields such as COSMOS and XMM-LSS.

MeerKLASS fills a unique niche in this landscape. It offers significantly better surface brightness sensitivity and image fidelity than RACS, vastly broader sky coverage than MIGHTEE, and complementary resolution and frequency to legacy surveys like NVSS and SUMSS. These characteristics make MeerKLASS well suited for source–population studies, and synergy with both low– and high–frequency radio data. The wide bandwidth enables precise in–band measurements of spectral indices and curvature, facilitating source classification (e.g. separating AGN from SFGs) and the identification of peaked–spectrum/compact sources \citep[e.g.][]{Mauch2007,Smolcic2017,ODEa1998,Callingham2017,Sinha_2023}. The resulting deep continuum images and catalogues will support studies of galaxy evolution and large–scale structure \citep[e.g.][]{Padovani2016, Blake2002}, and can be used to test the cosmic radio dipole with wide–area source counts \citep[e.g.][]{BlakeWall2002,Singal2011,RubartSchwarz2013,2025PhRvL.135t1001B}. In addition, polarised emission extracted from the same dataset enables studies of cosmic magnetism via Faraday rotation, using RM–synthesis and dense extragalactic RM grids to probe magnetic fields in galaxies, clusters, and the intergalactic medium \citep[e.g.][]{BrentjensDeBruyn2005,Taylor2009,Gaensler2025}. Finally, the survey’s large area and cadence provide discovery space for slow transients and variables in the radio sky \citep[e.g.][]{Bell2015,Mooley2016}. Collectively, these science outcomes align closely with several key goals of the SKA.

Beyond its standalone scientific value, MeerKLASS also functions as a proving ground for developing scalable data processing and analysis workflows required for next-generation wide-area radio surveys. The combination of large sky coverage, high-resolution continuum imaging, and OTF interferometric scanning presents a set of technical challenges that are directly relevant to the SKA’s survey operations. To date, approximately 380 hours of pilot observations have been conducted, demonstrating the viability of the OTF technique and enabling the refinement of data processing pipelines. Of this, roughly 270 hours have been acquired using the UHF system since the transition from L-band in 2022, with processing still ongoing. The full MeerKLASS programme aims to survey 10,000 deg$^2$ over 2500 hours of nighttime observing time. The data release presented in this paper is based on a small subset of this total -- eight independent OTF-mode data blocks amounting to just 12 hours of observing time. Despite its limited scope, this subset yields high-fidelity continuum images and a source catalogue containing approximately 95,483 radio sources. These results demonstrate the scientific potential of the MeerKLASS strategy even at an early stage, offering a valuable benchmark for testing calibration, imaging, and source extraction pipelines, and already opening several avenues for standalone science -- including source counts, spectral indices, and cross-matching with large-area optical surveys like DESI.

This paper presents the first public data release (DR1) of the MeerKLASS UHF survey. In Section~\ref{sec:survey}, we describe the OTF observing strategy and the specific observations included in this data release. In Section~3, we detail the data processing and calibration pipeline, including the crucial on-the-fly phase correction required for fast-scanning data. Our visibility-domain imaging strategy is outlined in Section~\ref{sec:imaging}. We assess the quality of the resulting images in Section~5, presenting noise maps and synthesised beam characteristics. The source extraction, catalogue generation, and validation are described in Section~\ref{sec:source_catalog}, where we also present our analysis of source compactness, survey completeness, and the differential source counts. We outline the released data products in Section~\ref{sec:public_release}. Finally, we summarise our findings and future plans in Section~\ref{sec:conclusions}.

\section{Survey Description}
\label{sec:survey}

In the OTF observing mode, MeerKAT performs continuous slewing across the sky at a fixed elevation, producing stable and repeatable beam tracks. This strategy not only minimises variations in system temperature due to ground spill and atmospheric fluctuations, but also ensures uniform sky coverage with well-behaved primary beam shapes—factors that are critical for high-fidelity mosaicking and accurate image reconstruction over wide areas. This scanning approach achieves a nominal survey speed of $\sim150$\,deg$^2$\,hr$^{-1}$, with each scan stripe covering $\sim18^{\circ}$ in azimuth over $\sim200$\,s of observing time \citep{Wang_2021, Cunnington_2023}. All observations are scheduled during night-time LSTs to suppress far-sidelobe solar contamination in the autocorrelation (H\,I intensity mapping, \cite{2025arXiv251027549C}) data; this also reduces solar/ionospheric systematics in the interferometric stream. Observations are typically conducted at a scan speed of $7$\,arcmin\,s$^{-1}$, corresponding to $\sim11.7^{\circ}$ of sky coverage in 100\,s, well within the gain stability timescale of the MeerKAT system. The visibilities are recorded at a time resolution of 2 seconds, which balances sufficient temporal sampling of the sky motion with manageable data volume, and ensures minimal time-average smearing across the field of view.

In the context of MeerKLASS, a “rising” scan refers to an OTF observation performed as a target field rises in the eastern sky during the early evening, while a “setting” scan is acquired as the same field sets in the west later in the night. In both cases, the telescope slews back and forth in azimuth along the same constant–elevation track. Due to Earth's rotation, the projected scan paths on the sky intersect at different angles, naturally producing a cross-hatched pattern in equatorial coordinates. Each field is typically observed twice per night—once during rising and once during setting—yielding orthogonal scan tracks that enhance coverage uniformity and suppress striping artefacts associated with single-direction scanning. This cross-linked strategy improves the sampling of the sky, averages down direction-dependent systematics (e.g. residual gain drifts or pointing errors), and increases the effective sensitivity in regions where multiple scan blocks overlap.

The present data release includes eight such OTF interferometric scan blocks, each approximately 1.5 hours in duration (excluding calibrator overheads), yielding a total of 12 hours of usable data. Of these, four blocks were taken during field rising and four during setting. Three of the rising scans overlap directly, while a fourth is offset to overlap with the setting scans, producing a cross-hatched region of enhanced sensitivity. All fields lie within the footprint of the Dark Energy Spectroscopic Instrument (DESI) survey, ensuring excellent multi-wavelength complementarity.

\begin{table*}
\centering
\begin{tabular}{ccccccc}
\hline
Block ID & Rising (R)/Setting (S) & Observation Window (SAST) & Bandpass/Flux Cal & Gain Cal & OTF Duration \\
\hline
1676657789 & R & 2023-02-17 20:17:27 -- 22:34:46 & J0408$-$6545 & J1051$-$2023 & 95.5 min \\
1678122565 & R & 2023-03-06 19:11:56 -- 21:28:33 & J0408$-$6545 & J1051$-$2023 & 95.1 min \\
1678295187 & R & 2023-03-08 19:07:35 -- 21:24:45 & J0408$-$6545 & J1051$-$2023 & 95.7 min \\
1680626188 & R & 2023-04-04 18:37:46 -- 20:59:58 & J0408$-$6545 & J1058$+$0133 & 97.6 min \\
1688399183 & S & 2023-07-03 17:48:36 -- 20:29:51 & J1331$+$3030 & J1058$+$0133 & 121.5 min \\
1689003684 & S & 2023-07-10 17:42:28 -- 20:14:36 & J1331$+$3030 & J1058$+$0133 & 112.5 min \\
1689090392 & S & 2023-07-11 17:48:57 -- 20:18:31 & J1331$+$3030 & J1058$+$0133 & 110.0 min \\
1689176790 & S & 2023-07-12 17:48:55 -- 20:16:52 & J1331$+$3030 & J1058$+$0133 & 108.5 min \\
\hline
\end{tabular}
\caption{Summary of the eight MeerKLASS UHF observing blocks included in this data release. Each entry lists the block identifier, scan direction, observation start and end times, calibrator sources, and effective duration of the OTF science scan.}
\label{tab:obs_blocks}
\end{table*}

\begin{figure*}
\centering
\includegraphics[width=\textwidth, trim=0 75 0 70, clip]{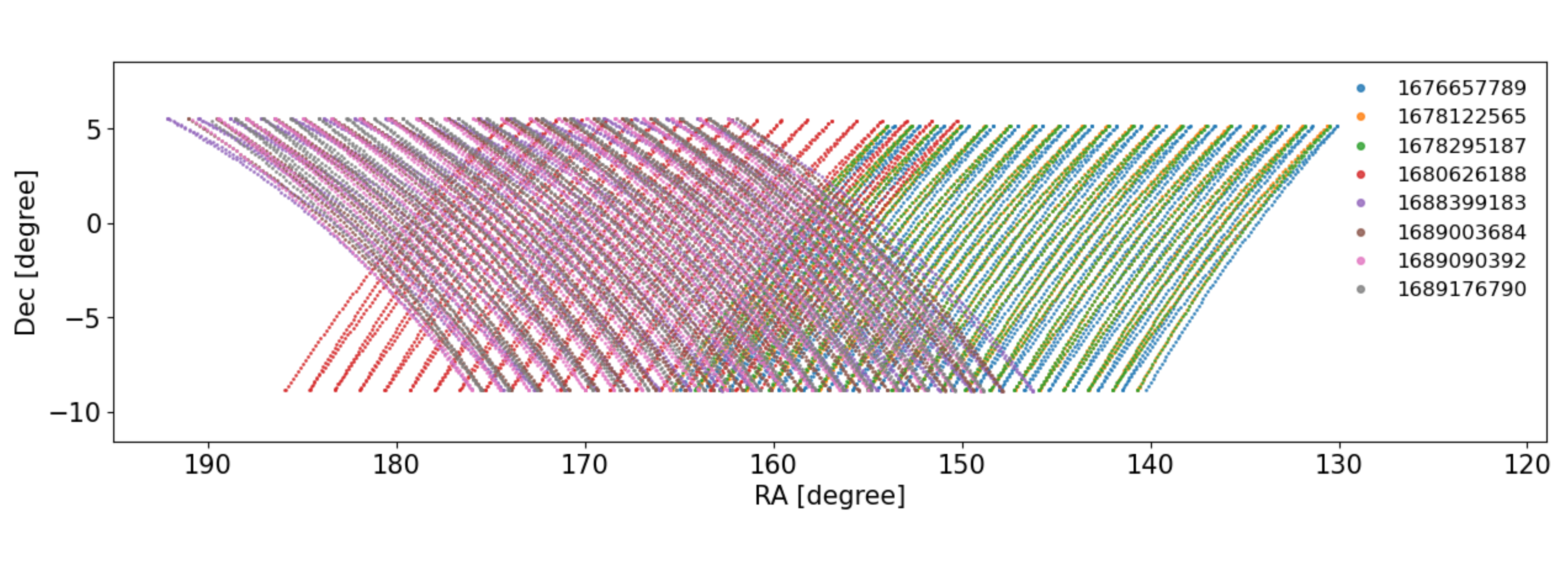}
\caption{Sky footprint of the eight MeerKLASS UHF observing blocks included in this data release. Each point marks the pointing centre of the reference antenna (m008) during the OTF scan. Rising and setting scans are shown with distinct orientations, producing a cross-hatched pattern that enhances coverage uniformity and sensitivity in overlapping regions. All fields lie within the DESI survey footprint.}
\label{fig:scan_footprint}
\end{figure*}

Each observing block is structured as follows. It begins with a short ($\sim 2$-minute) observation of a bright point-source gain calibrator, typically used to derive complex gain solutions. This calibrator is observed twice per block: once at the beginning and once at the end of the fast scanning sequence, bracketing the main science observation to allow interpolation of time-variable gains. The central science portion consists of $\sim$ 1.5 -- 2 hour OTF scans, during which the antennas continuously slew across the target field while recording visibilities. This portion of the schedule forms the basis of the continuum data products in this release. Several other calibrator scans are included in the observing schedule. These include single-dish calibrator fields (e.g., HydraA or PictorA and their angular offsets), which are used exclusively in the total power pipeline and are not used for interferometric calibration. After the main OTF scan, the telescope observes a bandpass and flux calibrator (J$0408-6545$ or J$1331+3030$) for approximately 5 minutes to calibrate frequency-dependent gains and set the absolute flux scale. In most blocks, this is followed by a $\sim 5$-minute observation of a dedicated polarisation calibrator to characterise leakage and cross-polarisation terms. Although polarisation data are available, polarimetric calibration and science analysis are not included in this data release and remain a work in progress. This calibration sequence ensures that each block is self-contained for interferometric processing, with the necessary observations for gain, bandpass, and flux calibration. However, due to the overlap structure and spatial layout of the eight scan blocks, sensitivity across the final 800\,deg$^2$ mosaics is inherently non-uniform -- peaking where multiple (rising and setting) blocks overlap, with maximum depth where all eight intersect, and declining toward survey edges covered by only a single block. Observations were conducted using MeerKAT's UHF receiver (544--1088\,MHz), and data were recorded with $4,096$ spectral channels across the band. The effective field of view ranges from $\sim2.8^{\circ}$ at the low end of the band to $\sim1.4^{\circ}$ at the high end, with an average interferometric beam size of $\sim15$\,arcsec after imaging.

Table~\ref{tab:obs_blocks} provides a summary of the eight interferometric blocks included in this data release. Their scan directions and approximate sky locations are shown in Figure~\ref{fig:scan_footprint}.

\begin{figure*}
\centering
\includegraphics[width=0.9\textwidth, trim=0 60 0 70, clip]{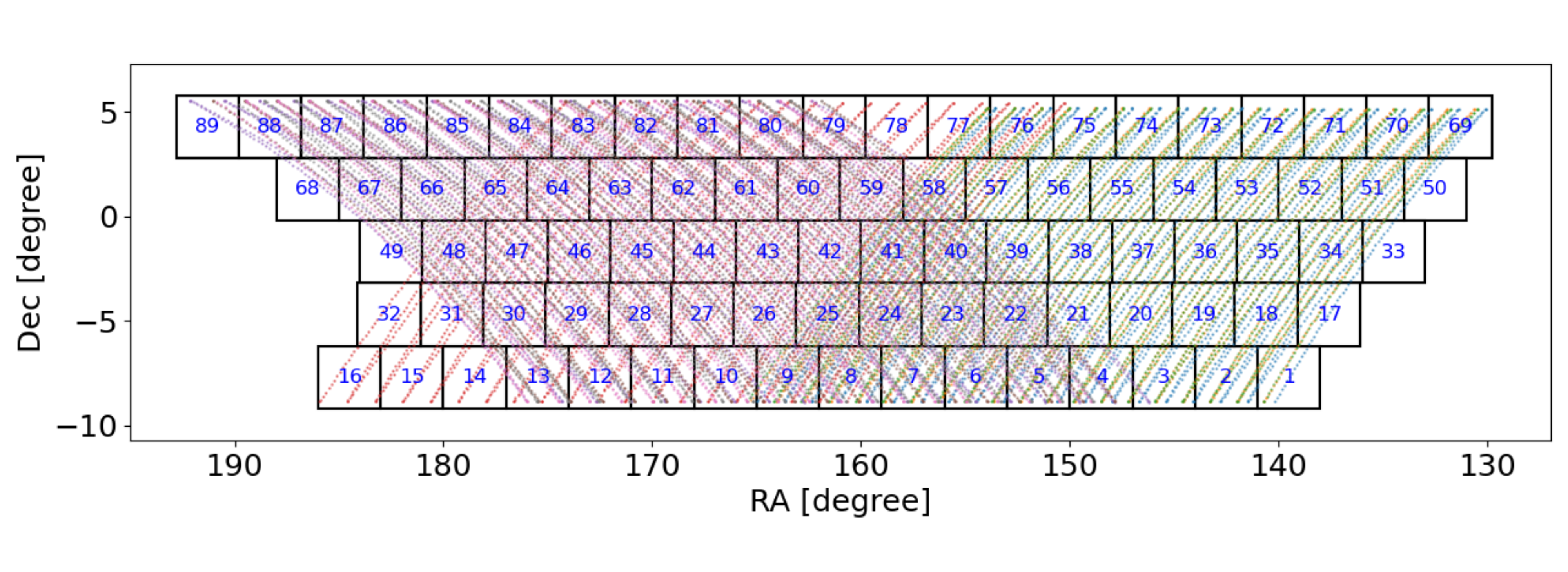}
\caption{Sky coverage of the MeerKLASS UHF imaging region, divided into $3.2^\circ \times 3.2^\circ$ tiles with $0.1^\circ$ overlap between adjacent tiles. Each square represents a tile, and the tile ID is indicated at its centre. The layout ensures full coverage of the $\sim$800\,deg$^2$ field. Background points denote the instantaneous pointing centres from OTF scan blocks.}
\label{fig:tile_layout}
\end{figure*}
\section{Data Processing and Calibration}

The data processing pipeline developed for this project transforms raw visibilities from fast-scanning OTF-mode observations into calibrated 2-second visibility datasets, which are subsequently imaged and combined in the visibility domain to produce science-ready, wide-field continuum mosaics. While a comprehensive treatment of the techniques is presented in a parallel OTF methodology paper (Chatterjee et al.), we summarize here the key steps relevant to this data release, including the flagging, calibration strategy and quality control.

\subsection{Initial Preprocessing and Flagging}

Each observing block is first subjected to a preprocessing stage that identifies and flags both conventional Radio-Frequency Interference (RFI) and instrument-specific systematics. RFI arising from terrestrial and satellite sources such as GSM, DTV, GPS, and Inmarsat, is mitigated using the \textsc{Tricolour} package \citep{2022ASPC..532..541H}, which is integrated into the \textsc{Caracal} pipeline \citep{2020ascl.soft06014J}. \textsc{Tricolour} implements a MeerKAT-optimised version of the SumThreshold algorithm \citep{Offringa2010}, and is applied uniformly to both calibrator and science scans. Flagging fractions are typically modest across the central part of the UHF band, but increase toward the band edges.

In addition to traditional RFI excision, we apply a dedicated purpose-built algorithm to identify and flag
``rogue'' antennas—those that fail to maintain synchronised motion during OTF scanning. These antennas may momentarily lag behind or deviate from the expected pointing trajectory, leading to decorrelated visibilities and image artefacts. The algorithm computes the pointing offsets of each antenna relative to the median pointing position per scan across the array. Antennas deviating by more than $\sim$0.1$^\circ$ from the array median are flagged either partially or fully, depending on the severity and duration of the deviation.

\subsection{Calibration}

Calibration is performed using the automated \textsc{Caracal} pipeline, configured to process each observing block independently using the standard set of calibrator scans, as described in \autoref{sec:survey}. The pipeline applies a sequence of model setting, primary calibration (bandpass and flux), and secondary calibration (time-dependent complex gains). The flux scale is anchored to standard models for the primary calibrators, and gain solutions are interpolated across the science scans to track instrumental and atmospheric variations.

The calibration strategy uses a primary calibrator sequence of \texttt{KGBAKGB}, where \texttt{K} solves for delays, \texttt{G} for complex gains, \texttt{B} for bandpass, and \texttt{A} for leakage. Calibration is performed over the full scan (solint = \texttt{inf}) for most terms, and at 60s intervals for the final gain steps. The gains are solved with a \texttt{combine} strategy that allows per-scan flexibility and are reused where possible to increase efficiency. The secondary calibrator is processed with a reduced sequence \texttt{KGAKF}, using solutions from the primary calibrator (\texttt{apply: B}) to anchor the flux scale.

The calibrated solutions are then applied to the science data using the \texttt{apply\_cal} step within \textsc{Caracal}, with calibration tables applied to the gain, bandpass, and cross-hand calibrator scans as appropriate. This approach enables fully self-contained calibration of each block, without requiring external calibration databases.

\subsection{On-the-Fly Phase Correction}
In OTF-mode observations, the MeerKAT correlator maintains the delay centre at a fixed azimuth (centre of the azimuth sweep) and elevation throughout the scan. However, as the antennas slew continuously across the sky, the actual pointing centre -- the centroid of the dishes' primary beam response -- drifts significantly from this static delay centre in azimuth.
When transformed to equatorial coordinates, this results in a dynamic offset between the phase centre of the correlation and the true sky position of the telescope beam. This offset introduces a varying geometric delay, which manifests as a time-dependent phase term in the visibilities. This time-dependent phase has two effects on our data. First, we must apply a fringe rotation to the visibilities in order to align their phase centre with the pointing centre. This rotation leads to no degradation of the final image. Second, during the integration period of a sample, the phase of each visibility is changing, and this is not compensated for in the telescope correlator as would normally occur for a standard tracking observation. The result of this is that the integration of visibilities within a sample is not perfectly coherent, leading to a signal loss which depends upon the fringe-rate of the visibility. Without any correction, this leads to image smearing, decorrelation, and astrometric distortions across the field, particularly at higher resolutions and frequencies.

To address the first effect, we apply a post-correlation correction using the \texttt{CHGCENTRE} task from the \textsc{WSClean} suite \citep{wsclean}. This task computes the required phase rotation for each visibility based on the time-varying difference between the fixed delay centre and the recorded pointing centre, and applies it to shift the effective phase centre to match the antenna pointing. The pointing centre is extracted per integration from the metadata of a designated reference antenna (typically m008). This correction is applied at the native time resolution of 2 seconds, ensuring minimal residual smearing even on long baselines. This step is essential for preserving astrometric accuracy and image fidelity across the wide field of view. It also enables effective mosaicking across snapshots with varying pointing centres, which is a key requirement for the final imaging strategy (see \autoref{sec:imaging}). Correction of the second effect is applied during imaging deconvolution as discussed at the end of \autoref{sec:imaging}.

\section{Imaging}
\label{sec:imaging}

Given the fast-scanning nature of MeerKLASS OTF observations, each $\sim$90-minute scan yields roughly 2,700 individual 2-second visibility datasets. One approach to imaging would be to process each snapshot independently and mosaic the resulting images in the image plane. Indeed, our prototype pipeline used this approach.  However, this method becomes computationally prohibitive due to the sheer number of datasets and the high overhead of managing image-plane mosaicking at this scale. Moreover, it would restrict deconvolution to the relatively shallow 2-second snapshots.

To overcome these limitations, we adopt a visibility-domain mosaicking strategy using a customised version of the \textsc{DDFacet} imaging pipeline \citep{tasse2018}, specifically modified to handle MeerKLASS OTF-mode data. This approach enables simultaneous imaging and deconvolution across multiple overlapping pointings while accounting for primary beam variation across the field.

\begin{figure}
\centering
\includegraphics[width=0.9\columnwidth, trim=150 15 150 15, clip]{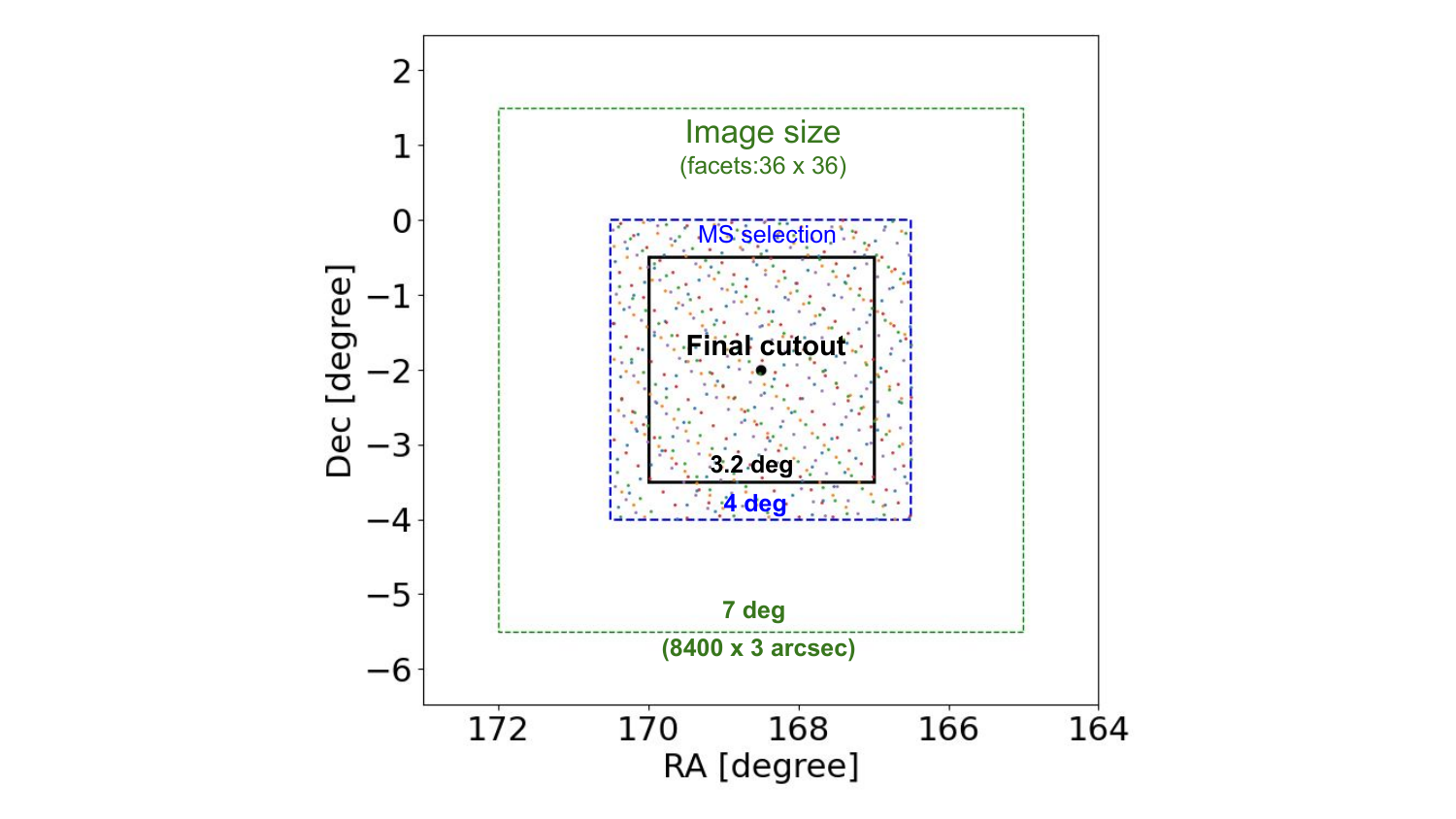}
\caption{Schematic of an individual tile used in the imaging pipeline. For each tile, all measurement sets with pointings falling inside a $4^\circ \times 4^\circ$ selection region (dashed blue box) are used as input for imaging with \textsc{DDFacet}. A large $7^\circ$ image is produced (green dashed box) to allow clean deconvolution of sources near the edges. The central $3.2^\circ \times 3.2^\circ$ science cutout (solid black square) is extracted for catalogue generation. This buffer strategy minimises edge artefacts and ensures uniform noise in the final mosaics.}
\label{fig:tile_design}
\end{figure}

Instead of imaging the full $\sim800$\,deg$^2$ field in a single run, the sky is divided into smaller overlapping tiles, as shown in Figure~\ref{fig:tile_layout}. Each tile is 3.2$^\circ \times$ 3.2$^\circ$ in size, with 0.1$^\circ$ overlap with adjacent tiles. For each tile, all measurement sets within a 4$^\circ \times$ 4$^\circ$ region are selected and imaged together. This larger input region ensures that sources outside the central cutout are properly deconvolved, minimising sidelobe contamination and producing a more uniform noise floor in the final cutout. Moreover, given the primary beam scale, all the measurement sets within the 4$^\circ \times$ 4$^\circ$ region contribute directly to the observations of the sources inside the 3.2$^\circ \times$ 3.2$^\circ$ region.

The structure of each tile is illustrated in Figure~\ref{fig:tile_design}. We perform imaging with 8,400$\times$8,400 pixels at 3\,arcsec resolution. This yields an effective image size of $\sim7^\circ$ across, large enough to accommodate beam roll-off and ensure clean deconvolution in the central regions. The central $3.2^\circ \times 3.2^\circ$ region is retained as the final science image, while the outer buffer region allows for accurate modelling of nearby sources during deconvolution. The use of consistent image geometry, pixel size, and facet configuration across tiles ensures smooth stitching for future mosaic combination and source cataloguing.

Imaging is carried out with \textsc{DDFacet} \citep{tasse2018}. For each $7^{\circ}\times7^{\circ}$ image per tile we use a $36\times36$ facet grid (1,296 facets total), so each facet spans $\approx7^{\circ}/36\simeq0.19^{\circ}$ on a side ($\sim11.7\arcmin$), over which the primary beam and instrumental response can be treated as approximately constant. This facet-based approach allows the visibilities to be gridded and deconvolved locally in each direction before being reassembled into a single wide-field image.

We adopt SSD2 deconvolution using Briggs weighting \citep{1999ASPC..180..127B} with a robust parameter of $0$. Imaging proceeds in two passes. In the first pass, we enable \textsc{DDFacet}'s automatic masking to pick up compact, high–S/N emission in each facet. We then construct an \emph{external} mask from the first–pass restored image using \texttt{MakeMask.py} with a $5\sigma$ threshold, which defines source islands across all facets. In the deeper second pass, we disable auto–masking and supply this external mask to \textsc{DDFacet} (\texttt{--Mask-Auto 0 --Mask-External <mask>}), while seeding the deconvolution with the first–pass sky model (\texttt{--Predict-InitDicoModel}). This scheme expands the clean region to include low–surface–brightness and extended emission that the initial auto–mask can miss.

Direction-independent self-calibration is performed using the \textsc{killMS} package \citep{2023ascl.soft05005T}, solving for per-antenna, full-Stokes complex gains at 60s intervals across the band. These solutions are derived using the CLEAN model from the second imaging run, smoothed in time and frequency, and then applied uniformly to all datasets. A third and final imaging run is performed using fixed CLEAN masks and restored source models, improving PSF stability and suppressing artefacts from calibration residuals. Residual images are also generated at this stage for quality control and noise characterisation.

Images are restored using frequency-dependent primary beam models. This includes corrections for the beam variation across the UHF band, ensuring accurate flux recovery and astrometric alignment across the wide field. The final outputs for each tile include a model image, a restored science image, a residual image, and beam information, all stored in standard FITS format and suitable for catalogue extraction and spectral index fitting.

\smallskip
\noindent
A summary of the imaging workflow is as follows:
\begin{itemize}
\item Facet layout and visibility selection for each tile.
\item First-round imaging with \textsc{DDFacet} using SSD2 deconvolution and auto-masking.
\item Generation of external masks based on first-pass source models.
\item Deeper second-round imaging with external masks to recover extended emission.
\item Direction-independent self-calibration with \textsc{killMS} in full-Stokes mode.
\item Final imaging with fixed masks and smoothed calibration solutions.
\item Restoration using UHF beam models; output of model, residual, and restored images.
\end{itemize}

One subtle but important effect in fast OTF scanning is residual smearing due to the continuous motion of antennas during correlation. While the \texttt{CHGCENTRE} correction accounts for the time-varying pointing centre at the visibility level, it does not fully eliminate baseline-dependent phase errors introduced by averaging visibilities over finite time and frequency intervals. These errors manifest as direction-dependent resolution loss and flux suppression. To address this, the customised \textsc{DDFacet} implementation used here includes an approximate smearing correction by constructing an effective point-spread-function (PSF) which accounts for the fringe-rate-dependent signal loss and implements this in both the minor and major cycles of deconvolutional {\sc Clean}ing. While this correction improves flux recovery and reduces artefacts, it broadens the effective PSF. A detailed description of this correction scheme and its performance is provided in the companion methodology paper (Chatterjee et al.).

\section{Image Quality and Assessment}
Before proceeding to source extraction and cataloguing, we evaluate the quality of the final restored images produced by the DDFacet pipeline. This includes visual inspection of representative fields, statistical analysis of the background noise, and assessment of the point spread function (PSF) and image artefacts. These diagnostics serve both as a validation of the imaging strategy and as a benchmark for interpreting catalogue reliability and completeness. 

\begin{figure*}
\centering
\includegraphics[width=\textwidth]{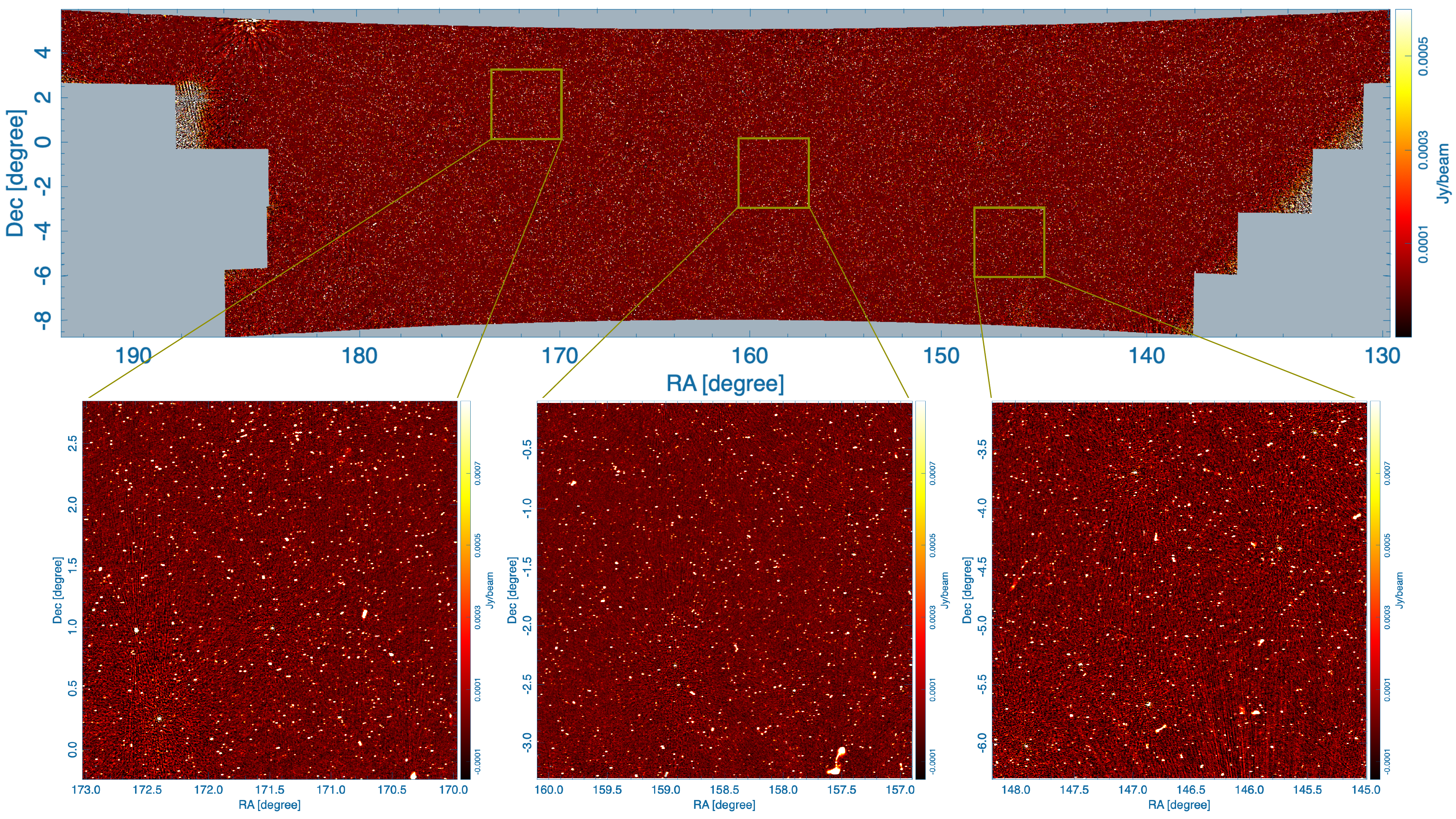}
\caption{
Overview of the mosaic and image quality across the MeerKLASS UHF survey area used in this work.
The top panel shows the full $\sim$800\,deg$^2$ continuum mosaic constructed from 89 tiles (\autoref{fig:tile_layout}) at the UHF band centre (816 MHz).
Three example tiles are highlighted on the mosaic and displayed below as zoomed-in cutouts, shown from left to right: Tile 63, Tile 41, and Tile 20.
These illustrate the high dynamic range and low noise floor achieved across a range of sky positions.
The selected fields span a variety of declinations and survey depths, and their structure reflects the uniformity and imaging fidelity delivered by the pipeline.
}
\label{fig:mosaic_zoomtiles}
\end{figure*}

\subsection{Full Mosaic}
\autoref{fig:mosaic_zoomtiles} provides a global view of the continuum image quality across the entire 800 deg$^2$ footprint. The top panel displays a composite mosaic constructed from the restored images of 89 individual tiles. This full-field image was assembled using \textsc{Montage}\footnote{\url{http://montage.ipac.caltech.edu}}\citep{Jacob2010} which reprojects each input FITS image to a common WCS frame before co-adding them into a seamless large-area mosaic. To produce this, we first collected the final restored images for all tiles, each corresponding to a $3.2^\circ \times 3.2^\circ$ field generated by a single DDFacet imaging run. All tiles share a common pixel scale of 3 arcsec and are aligned in equatorial coordinates. Overlap between adjacent tiles (typically 0.1$^\circ$) ensures that edge effects are suppressed and that the final mosaic remains visually continuous. \textsc{Montage} was used in its standard reprojection mode without any background matching or flux scaling applied, preserving the original image intensities produced by the pipeline.

To assess the fidelity of the imaging across the field, three representative tiles—Tile 20, Tile 41, and Tile 63—are highlighted in the top panel, and their zoomed-in views are shown below. These tiles span a range of declinations and coverage depths within the mosaic (see \autoref{fig:tile_layout}). The zoom panels demonstrate the high image quality achieved across the survey, revealing numerous compact and extended sources, clean background regions, and minimal residual artefacts. In Tile 41, where all eight OTF blocks overlap, the noise floor reaches 35 $\mu$Jy beam$^{-1}$, consistent with expectations based on overlapping block sensitivity and integration time.

These visual diagnostics confirm that the adopted imaging and mosaicking approach yields high-fidelity continuum maps over large sky areas. The absence of strong artefacts or residual sidelobe structure supports the robustness of the full processing pipeline in handling fast OTF interferometric data. This visual inspection forms the basis for the more quantitative image quality assessment steps that follow, including noise estimation and PSF characterization.

\subsection{RMS Noise Map}
To characterise the sensitivity achieved across the survey footprint, we constructed a spatially resolved RMS noise map using the residual images produced during the DDFacet imaging process. These residual maps, which contain no source model, provide an unbiased estimate of the thermal noise and imaging artefacts present in each tile.

For each of the 89 image tiles, the residual image was included in a full-sky mosaic using the \textsc{Montage} toolkit. This ensured a consistent astrometric frame across tiles and enabled pixel-wise comparison across the entire survey area. To estimate the local RMS, we applied a sigma-clipping approach using a sliding window. Specifically, a window of $100 \times 100$ pixels (corresponding to $5'\times5'$ at 3~arcsec resolution) was moved across the mosaic in $50$-pixel steps. Within each window, the standard deviation was estimated after iteratively clipping pixels beyond $3\sigma$ from the median, thereby mitigating contamination from residual sidelobes or unsubtracted sources. This process yielded a 2D noise map at reduced spatial resolution, capturing large-scale variations in image sensitivity.

\begin{figure*}
\centering
\includegraphics[width=\textwidth]{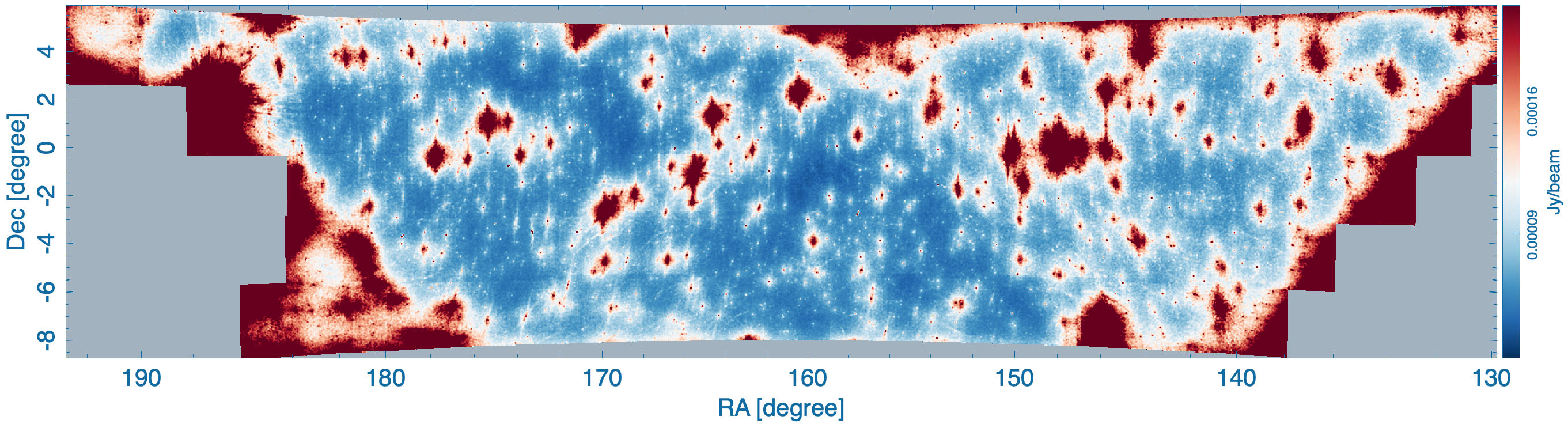}
\caption{Estimated background noise across the survey area, derived from the full residual mosaic at 816 MHz. The RMS map is computed by applying sigma-clipped statistics within overlapping sliding windows ($100\times100$ pixels, stepped every 50 pixels) directly on the combined residual image. The map reveals significant spatial variation in noise level due to differences in integration time, primary beam coverage, and residual imaging artefacts. The deepest regions, with RMS values as low as $\sim35\,\mu$Jy\,beam$^{-1}$, are located near the centre of the field where multiple blocks overlap.}
\label{fig:rms_map}
\end{figure*}

The resulting map, shown in Figure~\ref{fig:rms_map}, clearly reveals how the depth varies across the field. Noise levels are lowest in regions where multiple overlapping blocks contribute to the imaging, as evident from the darker blue patches, consistent with \autoref{fig:scan_footprint}. The tiles with the deepest coverage achieved an RMS of $\sim35$~$\mu$Jy~beam$^{-1}$, consistent with expectations based on cumulative integration time and system equivalent flux density. Conversely, edge regions and isolated tiles exhibit elevated noise due to lower sky redundancy and more pronounced primary beam attenuation. The prominent high-RMS regions (visible as ``red islands'' in \autoref{fig:rms_map}) correspond to areas where residual calibration artefacts or deconvolution imperfections persist. They coincide with fields containing very bright and/or extended sources; in these cases, imperfect deconvolution leaves residual PSF sidelobes around the sources, which increase the local RMS (i.e. dynamic–range–limited regions).  These higher noise regions do not reflect a higher thermal noise floor. Because the noise estimation is based on a sigma-clipped RMS from residual images, artefacts such as residual sidelobes and improper clean subtraction inflate the background estimate. These ``red islands'' are therefore valuable diagnostics: they highlight regions where local imaging fidelity is reduced and serve as a cautionary guide for interpreting faint source populations. Future data releases may benefit from improved direction-dependent calibration or peeling strategies to mitigate these effects. This noise map plays a key role in assessing survey uniformity, evaluating source detection completeness, and supporting accurate flux uncertainty estimates in the final catalogue. It also serves as an important input for simulations and matched-filter source finding in future work.

To quantify the dynamic range around bright, unresolved sources, we calculate $S_{\rm peak}/\sigma_{\rm local}$, where $S_{\rm peak}$ is the peak brightness measured on the restored image and $\sigma_{\rm local}$ is a sigma–clipped rms estimated in an annulus from $2$ to $5$ times the synthesized-beam FWHM on the corresponding residual image. Applying this to $99$ bright, compact sources ($S_{\rm Code}=\,$‘S’, $\mathrm{Peak\_flux}\!\gtrsim\!0.1$\,Jy\,beam$^{-1}$), we obtain a median residual-based dynamic range of $\sim\!5.42\times10^{2}$ with a 16–84th percentile range of $3.06\times10^{2}$–$7.18\times10^{2}$. These values indicate that a minority of fields near very bright or extended sources remain dynamic-range limited—i.e. elevated residual sidelobes and calibration imperfections locally inflate the rms—while the bulk of the survey approaches the thermal-noise regime away from such sources. Representative sources are listed in Table~\ref{tab:dr_residual}, which also illustrates the spread in $\sigma_{\rm local}$ at fixed $S_{\rm peak}$ and the corresponding variation in dynamic range across the footprint.

\begin{table*}
\centering
\begin{tabular}{cccccc}
\hline
Source & RA (deg) & Dec (deg) & $S_{\rm peak}$ (Jy\,beam$^{-1}$) & $\sigma_{\rm local}$ (Jy\,beam$^{-1}$) & Dynamic range \\
\hline
MeerKLASS-UHF\_DR1 J+115453.0 $-$082519.7 & 178.72093 & $-8.42213$ & 0.323 & 0.0005 & 656 \\
MeerKLASS-UHF\_DR1 J+124853.0 +024800.7   & 192.22066 &  2.80019  & 0.320 & 0.0013 & 251 \\
MeerKLASS-UHF\_DR1 J+120316.7 $-$053220.9 & 180.81947 & $-5.53914$ & 0.301 & 0.0005 & 604 \\
MeerKLASS-UHF\_DR1 J+121216.4 $-$033931.0 & 183.06849 & $-3.65861$ & 0.270 & 0.0004 & 645 \\
MeerKLASS-UHF\_DR1 J+121611.1 $-$025754.5 & 184.04605 & $-2.96514$ & 0.260 & 0.0010 & 257 \\
MeerKLASS-UHF\_DR1 J+085558.2 $-$030323.8 & 133.99263 & $-3.05661$ & 0.259 & 0.0017 & 152 \\
MeerKLASS-UHF\_DR1 J+121542.1 $-$071537.1 & 183.92554 & $-7.26030$ & 0.251 & 0.0003 & 718 \\
MeerKLASS-UHF\_DR1 J+094205.1 $-$081813.0 & 145.52121 & $-8.30362$ & 0.249 & 0.0006 & 398 \\
MeerKLASS-UHF\_DR1 J+095727.1 +030958.9   & 149.36303 &  3.16635  & 0.247 & 0.0003 & 935 \\
MeerKLASS-UHF\_DR1 J+092956.5 +014844.6   & 142.48543 &  1.81238  & 0.227 & 0.0003 & 704 \\
\hline
\end{tabular}
\caption{Residual-based dynamic range for a subset of bright, point-like sources, defined as $S_{\rm peak}/\sigma_{\rm local}$. Here $S_{\rm peak}$ is taken from the restored image and $\sigma_{\rm local}$ is the sigma–clipped rms in an annulus $[2,5]\times$FWHM measured on the residual image.}
\label{tab:dr_residual}
\end{table*}

\subsection{Synthesized Beam Characterisation}
\label{sec:beam_characterisation}

\begin{figure*}
    \centering
    \includegraphics[width=\textwidth, trim=0 80 0 30, clip]{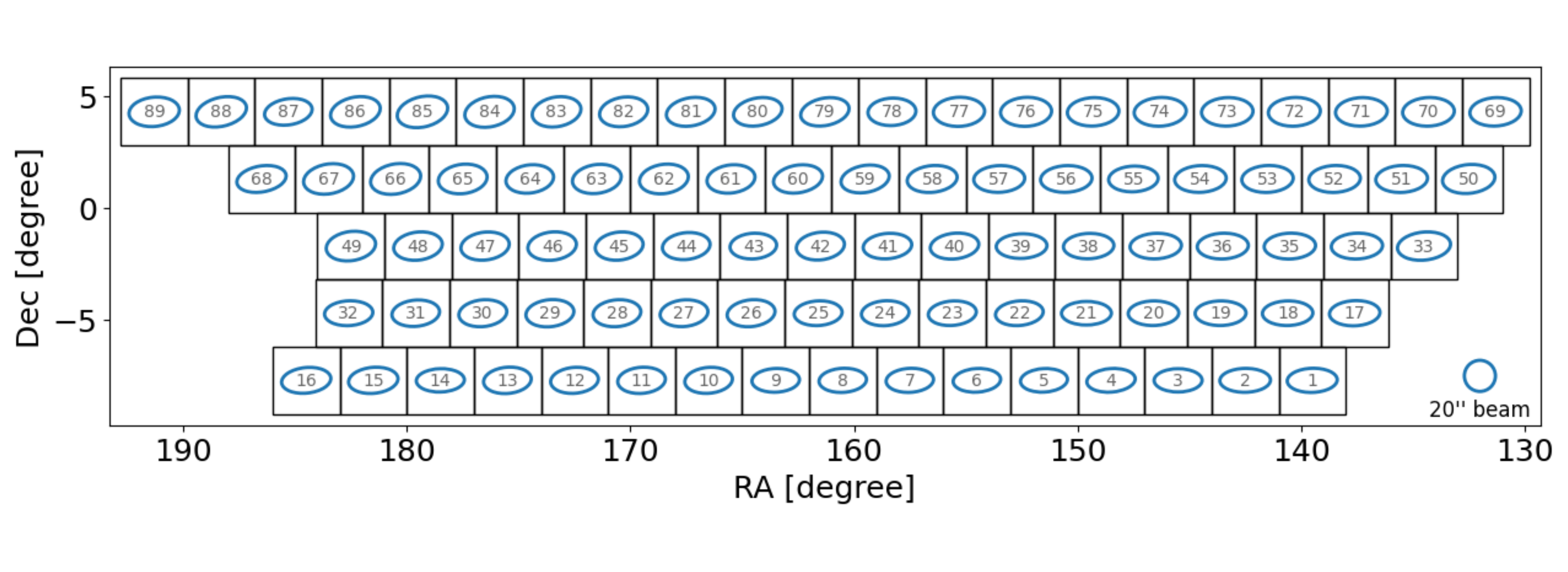}
    \caption{
        Spatial distribution of restoring beam ellipses across the MeerKLASS UHF tile grid. Each ellipse represents the major/minor axis and orientation of the synthesized beam for a given tile at 816 MHz. A reference $20''$ circular beam is shown in the bottom right for scale. The ellipses are plotted at the geometric centre of each tile. The consistent beam shape across the field validates the stability and uniformity of the imaging pipeline.
    }
    \label{fig:beam_size_map}
\end{figure*}

\begin{figure*}
    \centering
    \includegraphics[width=\textwidth, trim=0 57 0 60, clip]{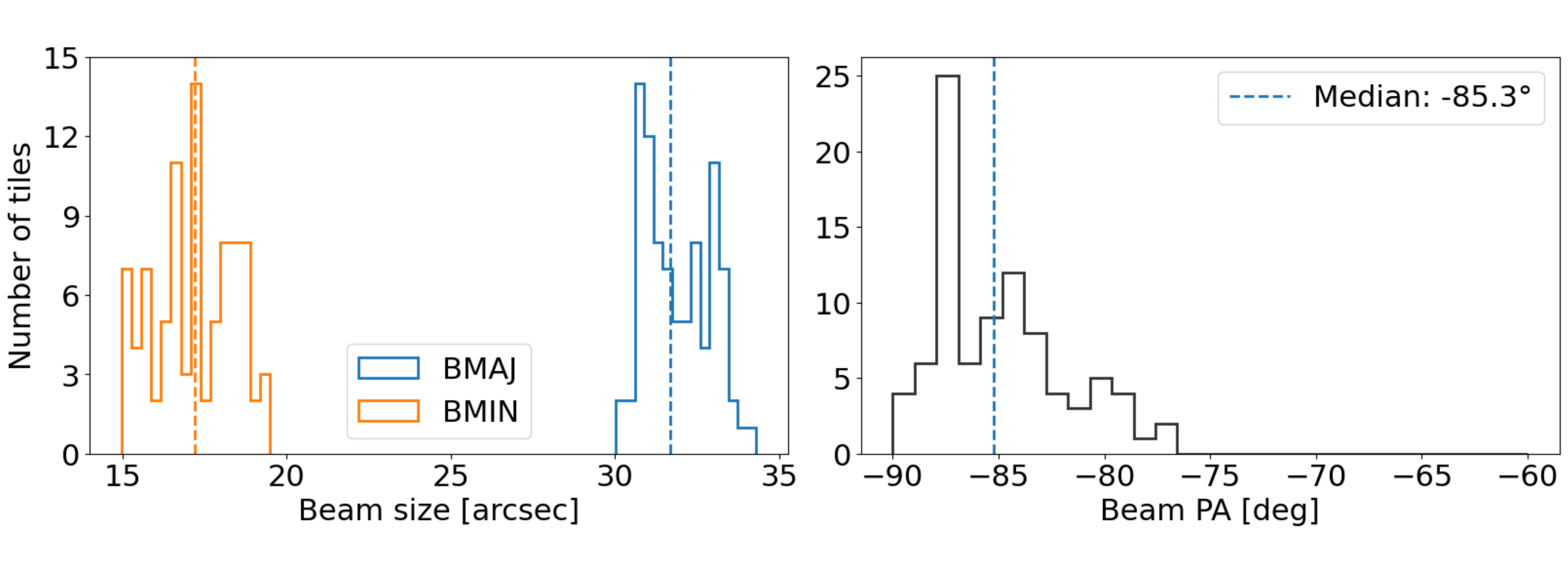}
    \caption{
        Histogram of restoring beam sizes across all 89 tiles at 816 MHz. \textbf{Left:} The distributions of \texttt{BMAJ} (major axis) and \texttt{BMIN} (minor axis) are tightly clustered around median values of $\sim32''$ and $\sim17''$ respectively (dashed vertical lines), indicating uniform angular resolution over the full survey footprint. \textbf{Right:} The distribution of \texttt{BPA} (beam position angle) of the major axis.
    }
    \label{fig:beam_size_hist}
\end{figure*}

A critical diagnostic of imaging fidelity is the structure of the synthesised beam (point–spread function, PSF), which sets the effective resolution and influences source morphology and completeness. For the MeerKLASS OTF data, the PSF varies across the footprint because the declination of the target field changes across the survey, altering the projected baselines; more importantly, the \emph{uv}–coverage differs from tile to tile. In practice, the dominant drivers of PSF variation are (i) the different numbers of passes for a given sky patch (rising/setting repeats and overlap regions) which change the density and azimuthal diversity of \emph{uv} tracks, and (ii) time–dependent flagging (e.g. RFI and data excision) that introduces non–uniformities in the \emph{uv} sampling. Imaging choices (e.g. Briggs robustness and any taper) then map these \emph{uv} variations into corresponding changes of the PSF width and sidelobe pattern.

To assess the PSF quality and spatial uniformity, we extracted the restoring beam parameters—major axis (\texttt{BMAJ}), minor axis (\texttt{BMIN}), and position angle (\texttt{BPA})—from the final restored images for all 89 tiles. These parameters represent the Gaussian restoring beam fitted to the synthesized PSF at the end of the deconvolution process using \textsc{DDFacet}. \autoref{fig:beam_size_map} shows a sky map of beam ellipses, plotted at the geometric centre of each tile. Each ellipse reflects the shape and orientation of the restoring beam, scaled up for better visualization. A reference $20''$ circular beam is overlaid in the lower right of the field for comparison. The beam shapes remain consistent and symmetric across most of the field. \autoref{fig:beam_size_hist} presents histograms of the major, minor axis lengths and position angle across the full survey. The median \texttt{BMAJ} is approximately $32''$, while the median \texttt{BMIN} is $\sim17''$. The narrow spread in beam sizes across all tiles confirms the uniform angular resolution achieved by the imaging pipeline within the context of the MeerKLASS OTF observing strategy. This stable beam is a prerequisite for reliable source detection, deblending, and morphological analyses. We note that the median beam size is broader than the theoretical resolution expected from uniform-weighted MeerKAT UHF data (typically $\sim$14--15$''$ at UHF band centre). This is due to the smearing correction implemented in \textsc{DDFacet} for fast-scanning OTF observations. As discussed in the companion methodology paper, this correction compensates for time-averaged smearing during correlation, but inherently broadens the effective PSF. The measured beam properties presented here therefore reflect this correction and represent the true image resolution in the final restored maps.

\subsection{Examples of Extended Sources and Multi-Survey Comparison}

\begin{figure*}
    \centering
    \includegraphics[width=\textwidth, trim=10 0 10 0, clip]{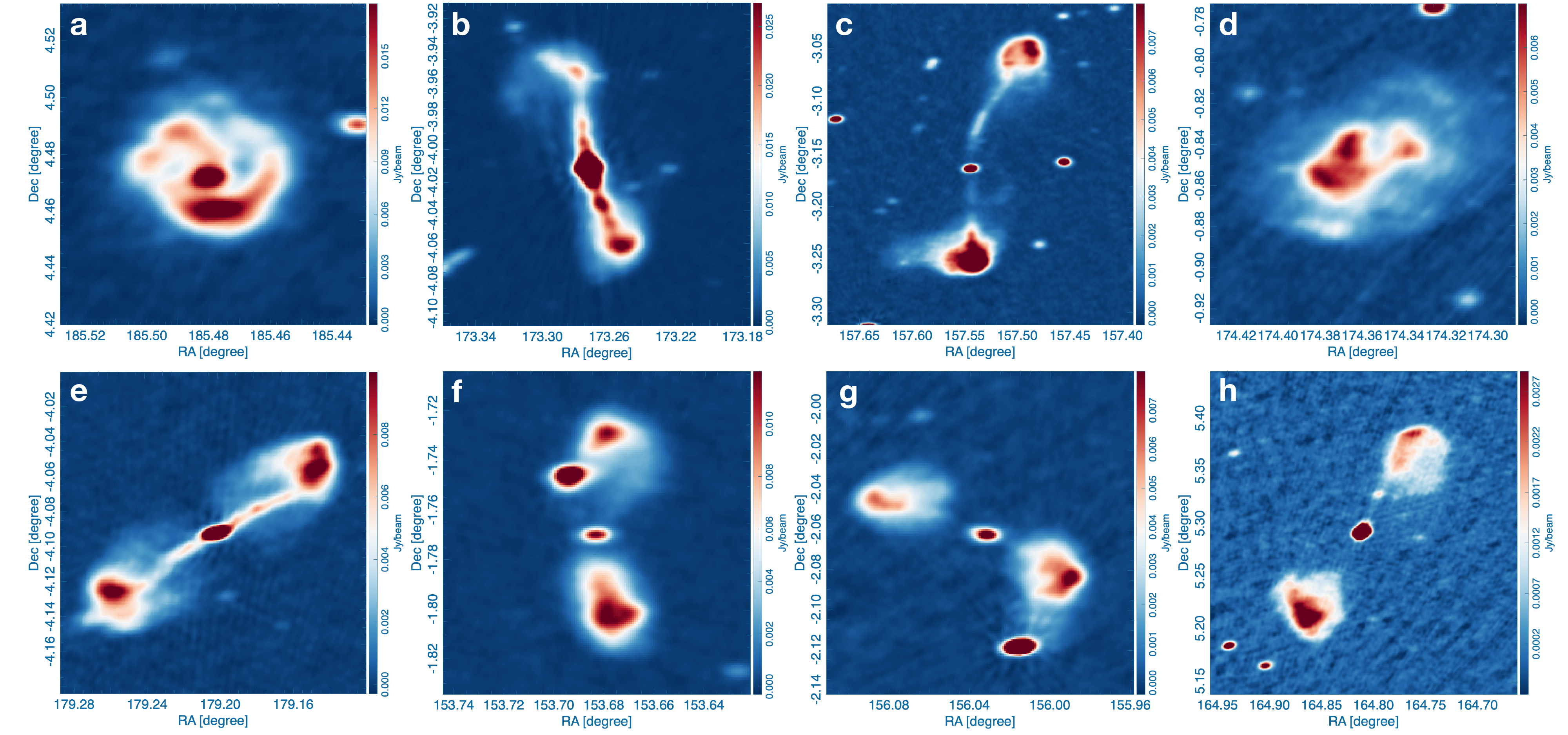}
    \caption{Demonstration of morphologically rich sources in the MeerKLASS OTF images at 816 MHz. 
    \textbf{a) NGC 4303:} a nearby barred spiral galaxy (Virgo Cluster), shows disk radio emission. 
    \textbf{b) 2dFGRS TGN171Z312:} a luminous galaxy with extended radio lobes, likely an FR I/II intermediate morphology. 
    \textbf{c) 2dFGRS TGN223Z101:} a complex structured galaxy exhibiting double-lobed structure with diffuse wings. 
    \textbf{d) 2dFGRS TGN307Z092:} extended radio galaxy with asymmetric lobes, indicative of environmental interaction. 
    \textbf{e) MCG–01–31–001:} clearly shows a prominent double-lobed radio galaxy (FR II), with two distinct bright lobes extending from a central compact core (likely the host galaxy)
    \textbf{f) PMN J1014-0146:} appears as a compact core with two relatively symmetric, elongated extensions, possibly indicating an FR I radio galaxy.
    \textbf{g) 2dFGRS TGN222Z318:} shows a source with a brighter, more compact core region and fainter, more diffuse extended emission.
    \textbf{h) UGC 6068:} reveals a complex, asymmetric radio morphology, with diffuse emission extending unevenly from a central brighter region, possibly indicative of a disturbed galaxy or jet-medium interaction.
    }
    \label{fig:extended_sources}
\end{figure*}

In addition to global image quality diagnostics, we present a selection of extended radio sources detected in the MeerKLASS UHF survey footprint to qualitatively demonstrate the survey’s high angular resolution, dynamic range, and sensitivity to diffuse emission. \autoref{fig:extended_sources} shows eight representative examples, each displayed as an enlarged cutout. The selected sources span a range of morphologies, including classical double-lobed AGN, asymmetric jets, and diffuse halo-like structures. Identified objects include well-known galaxies such as NGC~4303 and PMN~J1014$-$0146, as well as 2dFGRS and UGC galaxies associated with extended radio emission. These examples underscore the ability of MeerKLASS to resolve both compact cores and low-surface-brightness lobes, making it well suited for the discovery and characterization of giant radio galaxies, remnant AGN, and complex morphological systems.

\begin{figure*}
    \centering
    \includegraphics[width=\textwidth, trim=0 130 0 0, clip]{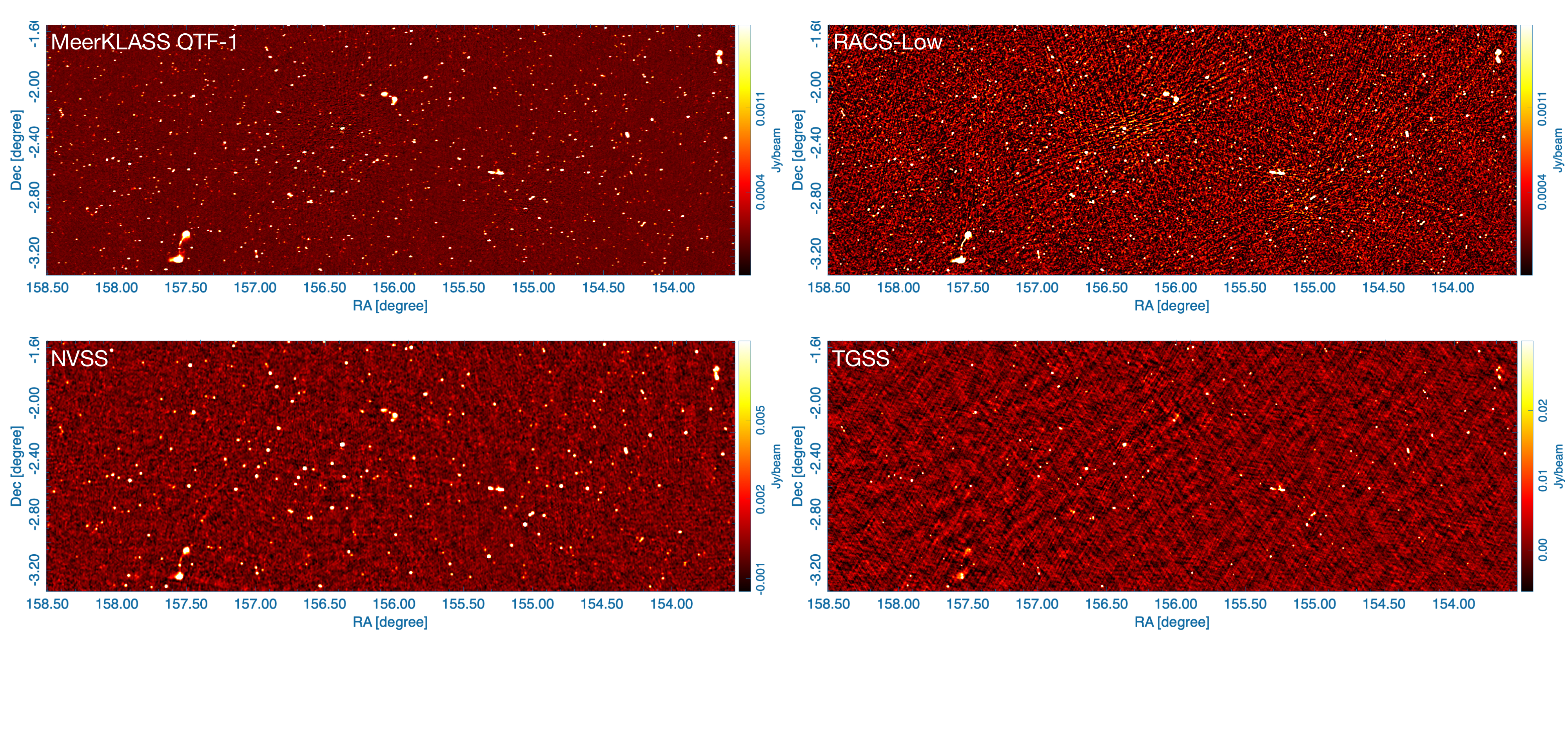}
    \caption{Multi-survey comparison of same sky patch, imaged by MeerKLASS OTF-1 (816\,MHz; top-left), RACS-Low (888\,MHz; top-right), NVSS (1.4\,GHz; bottom-left), and TGSS-ADR1 (150\,MHz; bottom-right). The MeerKLASS image exhibits the best combination of angular resolution and diffuse sensitivity, clearly resolving compact components and faint lobe structures that appear blended or suppressed in the other surveys. For a fair visual comparison, the MeerKLASS and RACS-Low panels—being at similar frequencies—have been displayed using the same colour scaling.
    }
    \label{fig:comparison}
\end{figure*}

\begin{figure*}
    \centering
    \includegraphics[width=\textwidth, trim=0 150 0 100, clip]{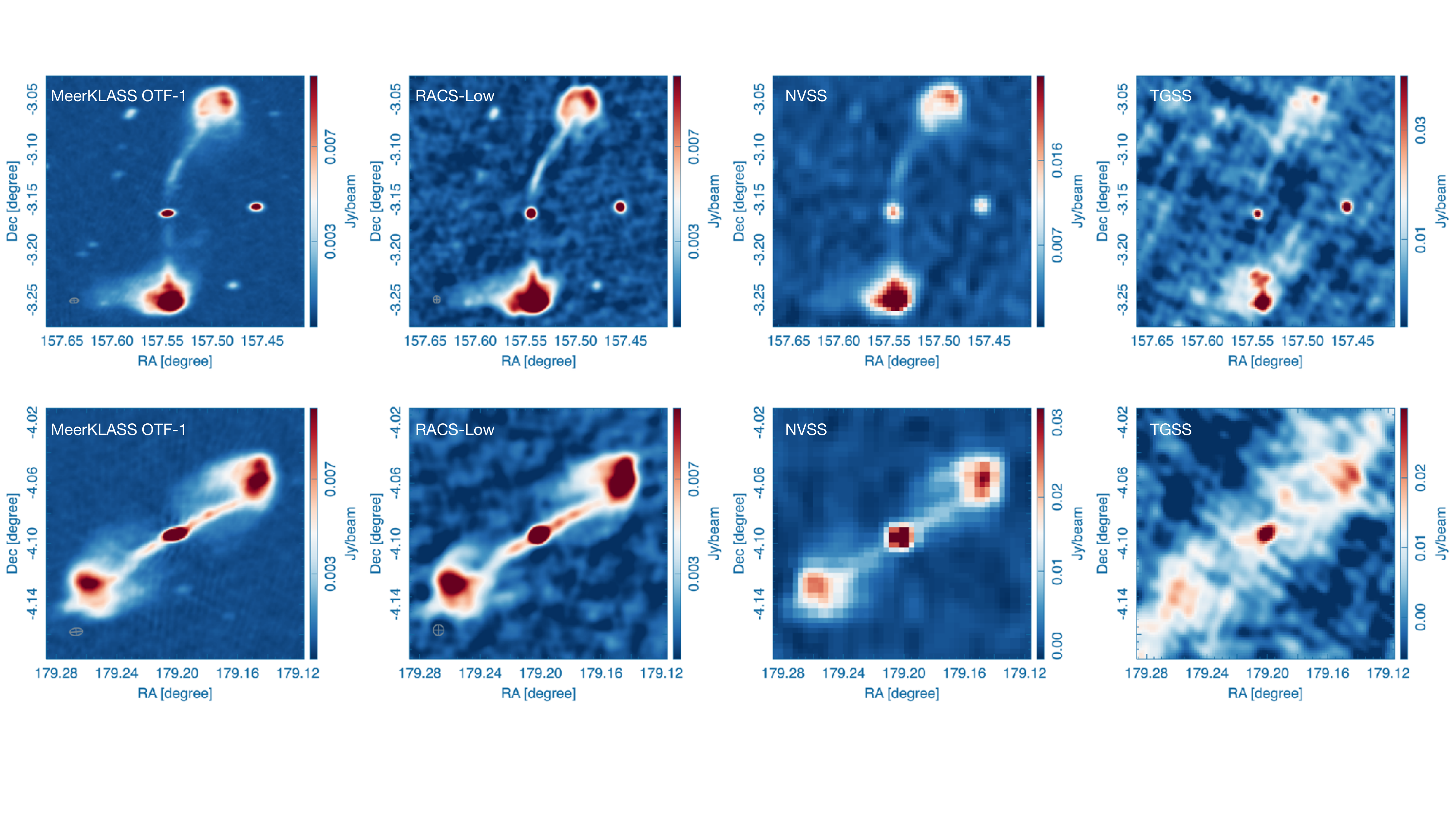}
    \caption{Comparison of MeerKLASS UHF OTF imaging with external surveys for two representative radio galaxies. Each panel shows a cut-out centred on the source in Stokes I. From left to right, the columns show images from MeerKLASS OTF-1 (816\,MHz), RACS-Low (888\,MHz), NVSS (1.4\,GHz), and TGSS-ADR1 (150\,MHz). The MeerKLASS image recovers more of the diffuse lobe emission and clearly separates multiple components, illustrating the improved surface-brightness sensitivity and imaging quality achieved with the MeerKAT OTF observations.
    }
    \label{fig:source_comparison}
\end{figure*}

To further benchmark the imaging performance, we compare MeerKLASS images with those from three widely used radio continuum surveys: RACS-Low, NVSS, and TGSS-ADR1. A representative sky region is shown in \autoref{fig:comparison}, with MeerKLASS at the top-left followed by RACS-Low, NVSS, and TGSS. All panels cover the same sky coordinates. The MeerKLASS cutout demonstrates improved angular resolution, cleaner background, and higher dynamic range than the comparison surveys. While NVSS (45\arcsec) is substantially coarser, TGSS (25\arcsec) offers comparable angular resolution to our effective beam (\(\sqrt{32\arcsec\times17\arcsec}\!\simeq\!23\arcsec\)) but at much lower frequency. RACS-Low, although closer in frequency, has a higher noise level and hence lower surface-brightness sensitivity. In contrast, the MeerKLASS image clearly resolves multiple emission components and diffuse lobes, demonstrating that our OTF mapping strategy, together with the sensitivity and uv coverage of MeerKAT, delivers high-quality imaging of complex radio sources (\autoref{fig:source_comparison}).

\section{Source Extraction and Catalogue Generation}
\label{sec:source_catalog}
Compact source detection was performed using the Python Blob Detection and Source Finder package \texttt{PyBDSF} \citep{mohan2015pybdsf}, applied independently to each restored continuum image tile. Prior to source finding, cutouts of size $3.2^\circ \times 3.2^\circ$ were generated from the original $7^\circ \times 7^\circ$ restored image tiles, as illustrated in \autoref{fig:tile_design}. These smaller subimages were chosen to avoid regions of higher noise and reduced fidelity near the tile edges, ensuring more uniform background characteristics for source detection.

The source detection parameters were optimized for the MeerKLASS UHF imaging characteristics, particularly the variation in background noise levels and beam shapes across the survey footprint. We adopted a pixel detection threshold of \texttt{thresh\_pix = 7.0} and an island threshold of \texttt{thresh\_isl = 5.0}, where the thresholds are expressed in units of the local RMS noise. Noise estimation was performed using \texttt{rms\_map = True} with a fixed \texttt{rms\_box = (150, 50)}, and the adaptive RMS mode (\texttt{adaptive\_rms\_box}) was disabled to maintain uniform sensitivity mapping across tiles. For bright regions, we additionally enabled \texttt{rms\_box\_bright = (20, 7)} to allow finer structure modeling without biasing the global RMS. To improve source characterization in regions with slowly varying PSF and complex backgrounds, we enabled wavelet-based detection using \texttt{atrous\_do = True} and allowed for spatially varying PSFs by setting \texttt{psf\_vary\_do = True}. These options helped recover diffuse features and improved deblending of nearby or overlapping components.

Each detected island was modeled as a combination of one or more 2D elliptical Gaussians, and fitted source properties (position, peak and integrated flux densities, size, and orientation) were stored in the resulting source (\texttt{srl}) catalogues. Residual, model, and RMS images were also exported for each tile to support quality assessment and image-level validation. These steps were executed uniformly across all 89 tiles in the survey, ensuring consistent source detection thresholds and modeling procedures. The process resulted in one PyBDSF source catalogue per tile.

Following source detection, the individual \texttt{srl} catalogues generated by \texttt{PyBDSF} for each tile were merged into a unified catalogue. To facilitate this, we read in the source lists from all available tiles, each stored in FITS format. For each catalogue entry, the originating tile number was recorded as a metadata field to preserve provenance and enable diagnostic tracing.

Because our tiling introduced a deliberate small overlap between adjacent pointings, a subset of sources was detected in more than one tile. We identified and removed such cross-tile duplicates using a graph-based approach. First, we performed a self-match of the concatenated catalogues with \texttt{Astropy}'s \texttt{SkyCoord.search\_around\_sky} using a maximum separation of $3\arcsec$. This value corresponds to one pixel width and balances the trade-off between completeness and spurious matching. To avoid suppressing genuine close pairs within a single image, we retained only links between entries from \emph{different} tiles. The resulting links defined a graph on the catalogue entries; we then computed connected components (union--find) so that each spatial cluster yielded exactly one surviving source. Within each component we selected the representative detection using a data-driven score: (i) highest peak signal-to-noise ratio, computed as \texttt{Peak\_flux/Isl\_rms}; if \texttt{Isl\_rms} was unavailable, we used \texttt{Total\_flux} as a proxy; (ii) if still tied, higher \texttt{Total\_flux}; (iii) if still tied, the smaller angular distance to the tile centre (final tie-breaker). We flagged and excluded all non-selected cross-tile duplicates from the final catalogue.

\subsection{Astrometric and Flux Density Accuracy}
\label{subsec:astrophoto}
\begin{figure*}
    \centering
    \includegraphics[width=\textwidth, trim=20 120 20 0, clip]{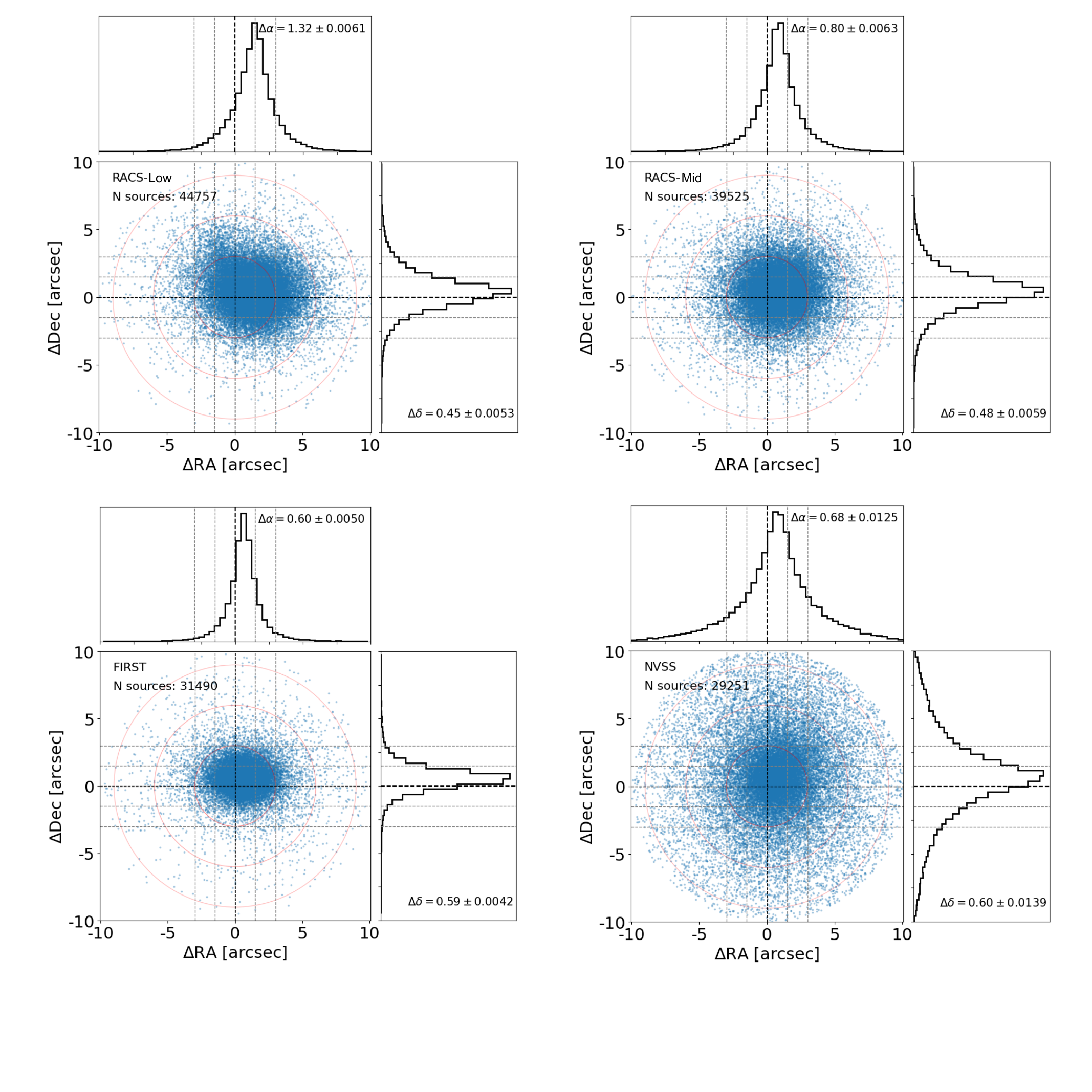}
    \caption{ Astrometric comparison between MeerKLASS OTF-1 and external surveys for cross-matched compact sources. The central panel shows the distribution of RA and Dec positional offsets, defined as MeerKLASS minus NVSS. Concentric red circles are spaced at intervals of $3\arcsec$, corresponding to the MeerKLASS image pixel size. Horizontal and vertical dashed lines are placed at $\pm1.5\arcsec$, i.e., half the pixel width. Marginal histograms summarize the mean offset and the associated uncertainty. The tight central clustering confirms high astrometric precision.
}
    \label{fig:astrometry}
\end{figure*}

\begin{figure*}
    \centering
    \includegraphics[width=\textwidth]{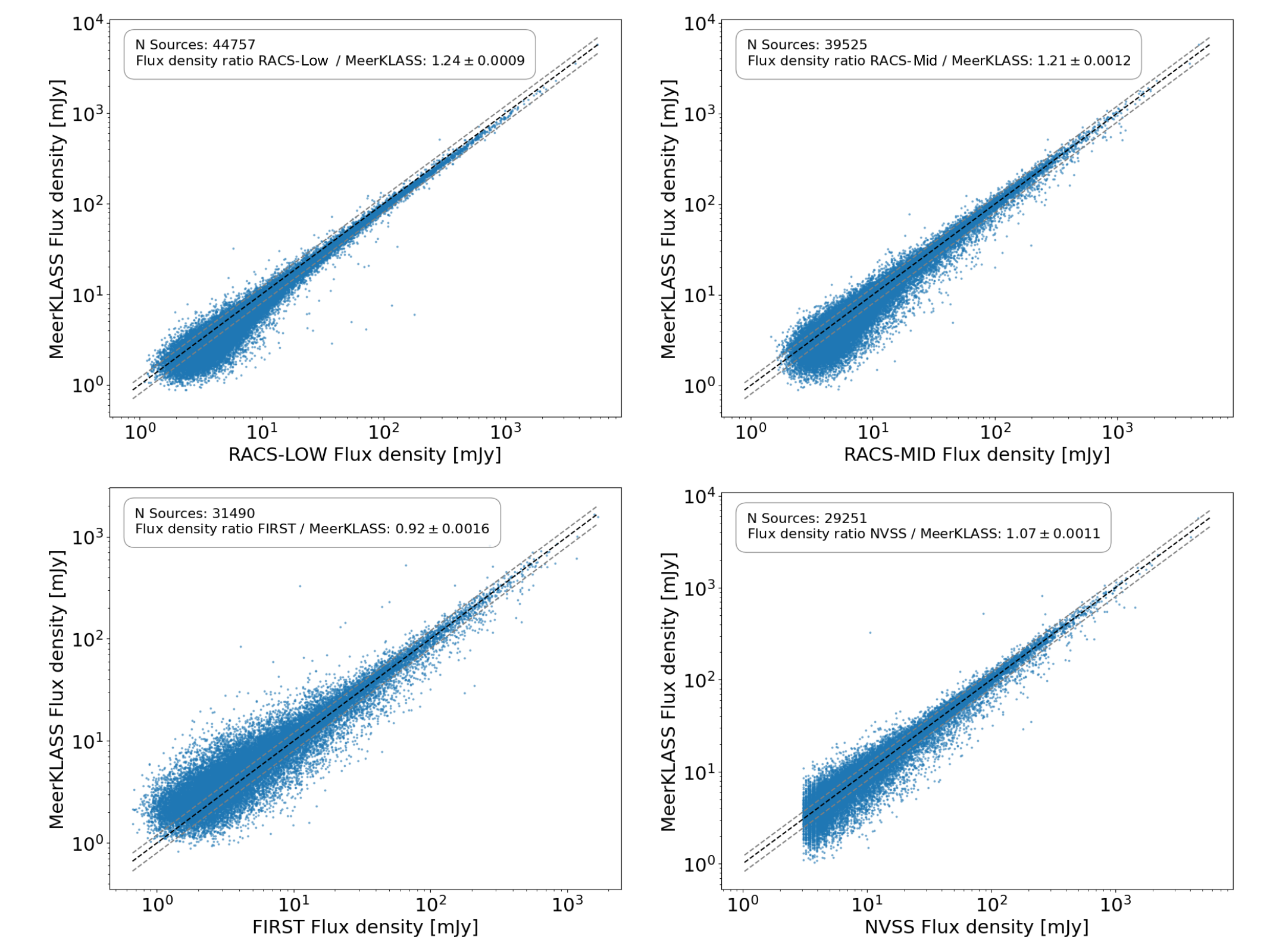}
    \caption{Flux density comparison between MeerKLASS OTF-1 and four external surveys: RACS-Low, RACS-Mid, NVSS, and FIRST. Each panel shows integrated flux density from the external survey versus MeerKLASS on log–log axes. The dashed black line represents the 1:1 relation, while the grey dashed lines denote $\pm$20\% deviations. Mean flux density ratios and the number of matched sources are annotated. The high degree of alignment confirms the robustness of the MeerKLASS flux scale.
}
    \label{fig:photometry}
\end{figure*}

We assessed the positional and flux density accuracy of the MeerKLASS UHF DR1 catalogue through cross-matching with four external interferometric surveys that overlap with the DR1 footprint: RACS-Low \citep{hale2021}, RACS-Mid \citep{Duchesne2023}, NVSS \citep{condon1998}, and FIRST \citep{becker1994}. These surveys span frequencies from 843\,MHz to 1.4\,GHz, and angular resolutions from 45\arcsec\ (NVSS) down to 5\arcsec\ (FIRST), making them ideal for benchmarking the astrometric and photometric quality of MeerKLASS data.

\medskip
\noindent
\textbf{Astrometric comparison.} Compact, isolated sources with $\rm{S/N}>10$ were selected from the MeerKLASS catalogue and matched to each reference survey using a $10\arcsec$ radius. Only sources with no neighbouring detection (i.e. no additional catalogue entries within $50\arcsec$ of the MeerKLASS position in either catalogue) were included to minimize misidentifications due to resolution differences or source blending.

For each matched pair, we computed the RA and Dec offsets as
$\Delta\alpha = \alpha_{\rm MeerKLASS} - \alpha_{\rm ext}$ and
$\Delta\delta = \delta_{\rm MeerKLASS} - \delta_{\rm ext}$, respectively.
The resulting offset distributions are shown in \autoref{fig:astrometry}, and
their summary statistics are listed in \autoref{tab:astrometry_offsets}.
For each reference survey we report both the median RA and Dec offsets with
asymmetric 68\% confidence intervals (from the 16th and 84th percentiles) and
the mean offsets with a robust estimate of the per–source scatter
(NMAD) and corresponding uncertainty on the mean ($\sigma_{\rm mean} =
{\rm NMAD}/\sqrt{N_{\rm offsets}}$). 

\begin{table*}
    \centering
    \begingroup
    \setlength{\tabcolsep}{3pt}
    \renewcommand{\arraystretch}{1.3} 

    \begin{tabular}{l c c c c c c c c c c c}
        \hline
        & & \multicolumn{5}{c}{RA ($\alpha$)} & \multicolumn{5}{c}{Dec ($\delta$)} \\
        \cline{3-7} \cline{8-12}
        Survey & $N_{\rm offsets}$ 
               & $\langle\Delta\alpha\rangle$ 
               & ${\rm NMAD}_{\Delta\alpha}$ 
               & $\sigma_{{\rm mean},\Delta\alpha}$ 
               & ${\rm median}(\Delta\alpha)$ 
               & $16$--$84\%$ range 
               & $\langle\Delta\delta\rangle$ 
               & ${\rm NMAD}_{\Delta\delta}$ 
               & $\sigma_{{\rm mean},\Delta\delta}$ 
               & ${\rm median}(\Delta\delta)$ 
               & $16$--$84\%$ range \\
        & & [arcsec] & [arcsec] & [arcsec] & [arcsec] & [arcsec]
          & [arcsec] & [arcsec] & [arcsec] & [arcsec] & [arcsec] \\
        \hline
        RACS-Low & 44\,757 
                 & 1.32 
                 & 1.29 
                 & 0.0061 
                 & 1.39 
                 & [$-1.53$, $+1.32$] 
                 & 0.45 
                 & 1.13 
                 & 0.0053 
                 & 0.41 
                 & [$-1.19$, $+1.24$] \\
        RACS-Mid & 39\,525 
                 & 0.80 
                 & 1.26 
                 & 0.0063 
                 & 0.82 
                 & [$-1.45$, $+1.39$] 
                 & 0.48 
                 & 1.18 
                 & 0.0059 
                 & 0.48 
                 & [$-1.27$, $+1.27$] \\
        FIRST    & 31\,490 
                 & 0.60 
                 & 0.89 
                 & 0.0050 
                 & 0.61 
                 & [$-1.02$, $+0.98$] 
                 & 0.60 
                 & 0.74 
                 & 0.0042 
                 & 0.59 
                 & [$-0.80$, $+0.79$] \\
        NVSS     & 29\,251 
                 & 0.68 
                 & 2.14 
                 & 0.0125 
                 & 0.74 
                 & [$-2.51$, $+2.40$] 
                 & 0.61 
                 & 2.37 
                 & 0.0139 
                 & 0.69 
                 & [$-2.79$, $+2.64$] \\
        \hline
    \end{tabular}
    \endgroup
    \caption{Astrometric offsets between the MeerKLASS UHF DR1 catalogue and external reference surveys. 
    $\langle\Delta\alpha\rangle$ and $\langle\Delta\delta\rangle$ are the mean positional offsets, 
    ${\rm NMAD}$ is a robust estimate of the per-source scatter, and 
    $\sigma_{\rm mean} = {\rm NMAD}/\sqrt{N_{\rm offsets}}$ gives the uncertainty on the mean offset.
    We also list the median offsets and the central 68\% ranges (16th--84th percentiles) of the offset distributions,
    expressed as relative deviations from the median.}
    \label{tab:astrometry_offsets}
\end{table*}

All offsets are smaller than the MeerKLASS image pixel size (3\arcsec), and the scatter is well within expectations for the image resolution and S/N regime. Concentric red circles in \autoref{fig:astrometry} mark radial steps of 3\arcsec, and the vertical and horizontal dashed lines are spaced at 1.5\arcsec, corresponding to half the pixel width. Notably, the comparison to FIRST (the highest–resolution survey in our set) yields median offsets $\lesssim 0.6$–$0.7$\,arcsec in both coordinates, implying any residual systematic positional error in MeerKLASS is $<1$\arcsec.

\medskip
\noindent
\textbf{Flux density comparison.} \autoref{fig:photometry} compares the integrated flux densities of compact cross-matched sources between MeerKLASS and the four external surveys. 
Fluxes from the reference surveys were scaled to 816\,MHz assuming a synchrotron spectrum $S \propto \nu^{\alpha}$ with $\alpha=-0.7$. 
Each panel shows a log--log scatter plot of $S_{\rm ext}$ versus $S_{\rm MeerKLASS}$ together with the corresponding flux–density–ratio histogram. 
The black dashed line marks the 1:1 relation, while grey dashed lines indicate the $\pm 20\%$ envelope around unity. 
The distributions of the flux density ratios $R = S_{\rm ext}/S_{\rm MeerKLASS}$ are summarised in \autoref{tab:flux_comparison}, where we report both the median ratios with 68\% confidence intervals (from the 16th--84th percentiles) and the mean ratios with their robust scatter (NMAD) and uncertainties on the mean.

\begin{table*}
    \centering
    \begingroup
    \renewcommand{\arraystretch}{1.3}
    \footnotesize                   

    \begin{tabular}{l c c c c c c}
        \hline
        Survey & $N_{\rm matches}$ 
               & $\langle R \rangle$ 
               & ${\rm NMAD}_R$ 
               & $\sigma_{{\rm mean},R}$ 
               & ${\rm median}(R)$ 
               & 16--84\% range \\
        & & \multicolumn{5}{c}{$R = S_{\rm ext} / S_{\rm MeerKLASS}$} \\
        \hline
        RACS-Low & 44\,757
                 & 1.24
                 & 0.18
                 & 0.0009
                 & 1.16
                 & [$-0.15$, $+0.29$] \\
        RACS-Mid & 39\,525
                 & 1.21
                 & 0.25
                 & 0.0012
                 & 1.10
                 & [$-0.19$, $+0.41$] \\
        FIRST    & 31\,490
                 & 0.92
                 & 0.28
                 & 0.0016
                 & 0.87
                 & [$-0.28$, $+0.30$] \\
        NVSS     & 29\,251
                 & 1.07
                 & 0.20
                 & 0.0011
                 & 0.99
                 & [$-0.18$, $+0.28$] \\
        \hline
    \end{tabular}
    \endgroup
    \caption{Flux density comparison between the MeerKLASS UHF DR1 catalogue and external reference surveys.
    For each survey we compute the flux density ratio
    $R = S_{\rm ext} / S_{\rm MeerKLASS}$ using matched sources with positive flux densities.
    $\langle R \rangle$ is the mean ratio, ${\rm NMAD}_R$ provides a robust estimate of the per-source scatter,
    and $\sigma_{\rm mean} = {\rm NMAD}_R/\sqrt{N_{\rm matches}}$ is the corresponding uncertainty on the mean.
    We also list the median ratios and the central 68\% ranges (16th--84th percentiles) of the $R$ distributions.}
    \label{tab:flux_comparison}
\end{table*}

The close agreement with the 1:1 line in all cases confirms the accuracy of the MeerKLASS flux scale across a range of observing frequencies and angular resolutions. Slight deviations from unity are expected due to spectral index variations, beam differences, and potential resolution-related effects. 

\medskip
\noindent
Together, these comparisons demonstrate that the MeerKLASS DR1 UHF catalogue achieves sub-arcsecond positional accuracy and maintains a flux density scale consistent with existing external interferometric benchmarks.

\subsection{Source compactness}
\label{sec:source_compactness}
\begin{figure*}
    \centering
    \includegraphics[width=\textwidth, trim=0 40 0 30, clip]{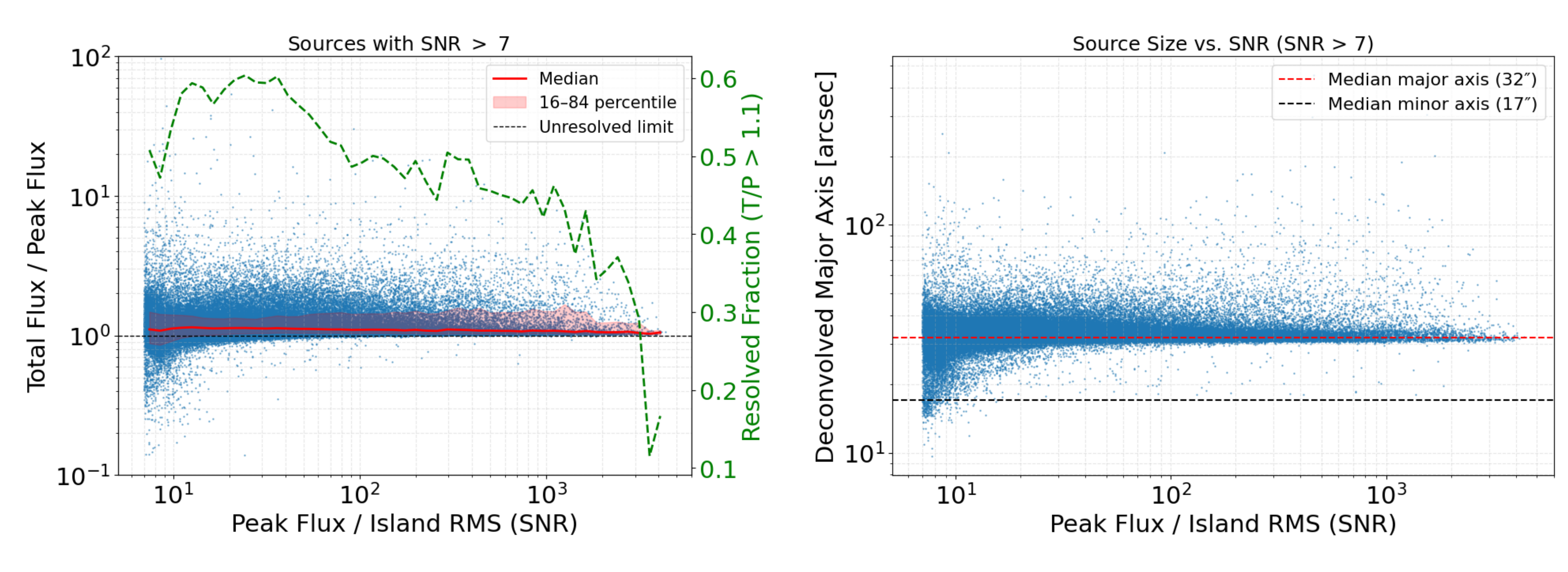}
    \caption{{\bf Left:} Flux density ratio $T/P$ as a function of SNR (defined as peak flux over island RMS) for sources with SNR~$>7$. The red curve shows the binned median, and the shaded region spans the 16th to 84th percentile range. The horizontal dashed line at $T/P=1$ indicates the unresolved limit. The green dashed curve (right axis) denotes the fraction of resolved sources, defined as those with $T/P > 1.1$. {\bf Right:} Deconvolved major axis of sources with SNR~$>7$ plotted against SNR. The dashed black line shows the median minor axis of the restoring beam ($\sim 17''$), while the dashed red line shows the median fitted major axis ($\sim 32''$) across all imaging tiles.}
    \label{fig:source_snr_dist}
\end{figure*}

Understanding the resolved nature of radio sources is crucial for accurate source characterization and statistical analysis. A common proxy to assess whether a source is resolved is the ratio of total to peak flux density ($T/P$), which should ideally equal unity for point-like sources \citep{Bondi2003,Hales2014}. Values significantly greater than one typically indicate extended emission, though modest deviations can also result from noise-driven uncertainties, especially at low signal-to-noise ratio (SNR). In this analysis, the SNR is computed as the ratio between the peak flux density and the island rms noise for each source, both of which are extracted directly from the PyBDSF catalog. This quantity provides a robust estimate of the detection significance, independent of the integrated flux measurement. Analyzing compactness as a function of SNR helps assess the survey’s resolving power across a wide range of source brightness and structure.

The left panel of \autoref{fig:source_snr_dist} presents the $T/P$ ratio as a function of SNR for all sources in the final catalogue with SNR $>7$. Each point represents a single source, and the red curve traces the median trend in log-spaced bins, with the shaded pink region marking the 16th to 84th percentile range. A horizontal dashed line at $T/P = 1$ indicates the expected value for unresolved sources. Overplotted on the right axis in green is the resolved fraction, defined as the fraction of sources within each bin that satisfy $T/P > 1.1$.

The median compactness ratio, $T/P$, remains close to unity across the full $\mathrm{S/N}$ range, indicating that the catalogue is predominantly composed of compact sources. A mild decline is visible, from $T/P \simeq 1.12$ at $\mathrm{S/N} \sim 10$ to $T/P \simeq 1.04$ at $\mathrm{S/N} \sim 3000$. Part of this trend is expected from the $\mathrm{S/N}$-dependent scatter and skewness of the flux--ratio estimator $T/P$: at low $\mathrm{S/N}$, noise in both the integrated and peak fluxes broadens and skews the ratio distribution, naturally pushing the median above unity even for intrinsically unresolved sources. The resolved fraction (defined via $T/P > 1.1$) likewise decreases from $\sim 58\%$ at $\mathrm{S/N} \sim 10$ to $\sim 30\%$ at $\mathrm{S/N} \sim 3000$, a behaviour that is also influenced by the shrinking scatter in $T/P$ with increasing $\mathrm{S/N}$. Taken together, these trends are consistent with a source population that is largely unresolved or marginally resolved at the effective survey resolution, with truly extended emission contributing primarily to the tail of high $T/P$ values.

To further investigate the deconvolved angular sizes, the right panel of \autoref{fig:source_snr_dist} shows the major axis of each source (from PyBDSF Gaussian fits) against its SNR, and only sources with SNR $>7$ are shown. The dashed black line marks the median minor axis of the restoring beam ($\sim17''$), and the dashed red line indicates the median fitted major axis ($\sim32''$) from all imaging tiles (\autoref{fig:beam_size_hist}). Most sources have deconvolved major axes clustered around the median synthesized beam size $32''$, consistent with the resolution limit of the survey. A smaller population of sources appear marginally resolved with sizes just above the median beam minor axis, while a minority exhibit significantly larger extents. The weak dependence of apparent size on SNR suggests that beam convolution dominates over intrinsic size variations.

Together, these plots provide consistent evidence that the majority of catalogued sources are unresolved or marginally resolved, with the resolving power limited by the effective PSF. The decreasing trend in $T/P$ and deconvolved angular size with $\mathrm{S/N}$ is qualitatively
consistent with resolved structure being more readily identified in lower--$\mathrm{S/N}$ sources, where marginal resolution effects and low--surface--brightness components can contribute to elevated $T/P$ values. However, we stress that part of this behaviour is also driven by the $\mathrm{S/N}$ dependence of the size and flux--ratio measurements themselves: even intrinsically point-like Gaussian sources will be classified as marginally resolved a non-negligible fraction of the time at low $\mathrm{S/N}$, whereas at $\mathrm{S/N} \gtrsim 10^3$ the fitted sizes converge to the beam and such sources almost never appear extended. Thus, the observed trends in $T/P$ and ``resolved fraction'' as a function of $\mathrm{S/N}$ should be interpreted as the combined result of intrinsic morphology and $\mathrm{S/N}$-dependent measurement effects, rather than as a purely population-driven change in source structure.

\subsection{Survey Completeness}
\label{sec:completeness}
\begin{figure*}
    \centering
    \includegraphics[width=\textwidth]{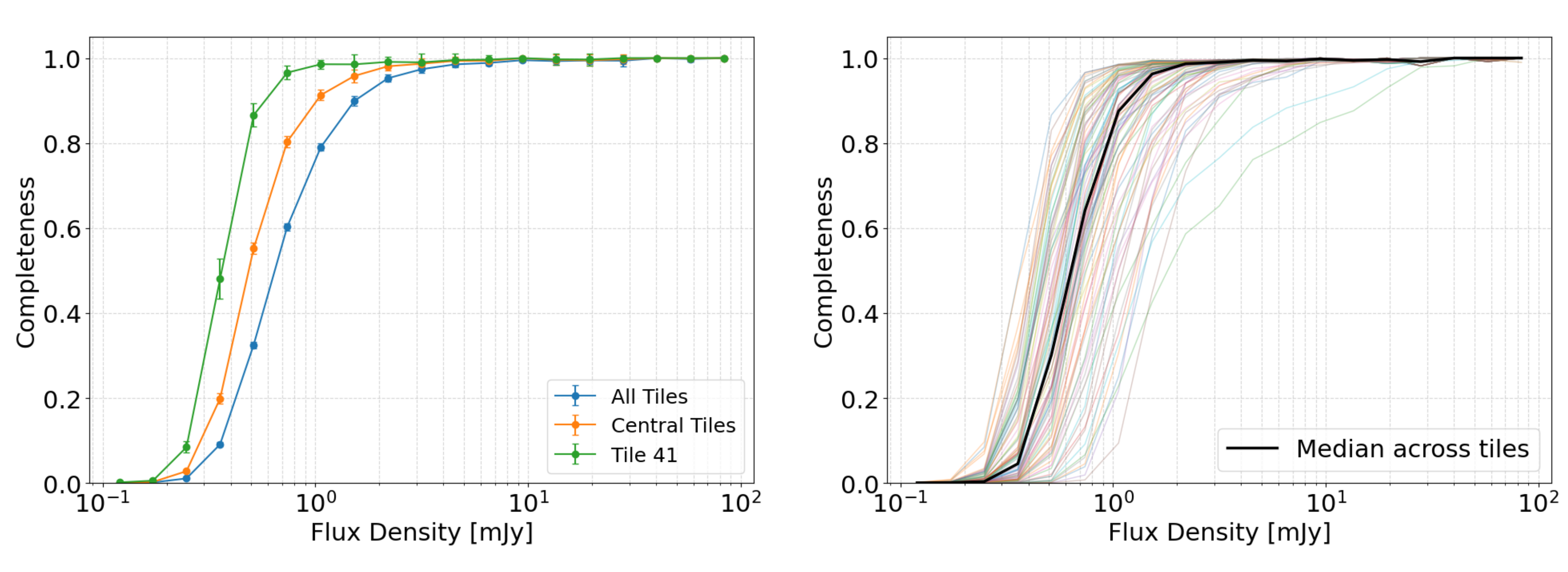}
    \caption{
        {\bf Left:} Completeness curves for the full survey (blue), central high-sensitivity tiles (orange), and a representative tile (Tile 41, green), with error bars showing standard deviation across realizations. 
        {\bf Right:} Completeness curves for individual tiles (thin lines), and the median curve across all tiles (solid black).
    }
    \label{fig:completeness}
\end{figure*}
To quantify the detection efficiency of our source catalog, we conducted a suite of injection-recovery simulations across the entire survey footprint. The goal of this analysis is to estimate the probability of detecting a source as a function of flux density, i.e., the survey completeness.

For each tile, we generated artificial point sources with flux densities drawn from a power-law distribution of the form $\mathrm{d}N/\mathrm{d}S \propto S^{-1.6}$, spanning the range $0.1-100$~mJy. These sources were assigned a fixed elliptical Gaussian shape consistent with the synthesized beam parameters recorded in the restoring image of that tile, and injected into the PyBDSF residual maps (in units of Jy/beam). Each tile was simulated over 20 independent realizations, with 1500 sources injected per realization. To avoid artificial crowding, sources were randomly placed across the image with no overlap constraints. We then re-ran PyBDSF using the same configuration applied to the original data. A source was considered successfully recovered if a detection was found within $5''$ of the injected position. Completeness in each flux bin was defined as the fraction of injected sources that were recovered under this criterion.

\autoref{fig:completeness} presents the results of the injection-recovery analysis. The left panel displays the mean completeness curves for three representative cases: (i) the entire set of tiles included in the data release (blue), (ii) a subset of central tiles—specifically Tiles 24, 25, 26, 41, 42, 43, 59, 60, and 61—corresponding to a high-sensitivity region of the mosaic (orange), and (iii) a single representative tile (Tile 41, green), which benefits from overlap across all eight observing blocks and is free of bright sources, resulting in a particularly low RMS. A clear trend is observed across these curves: the completeness improves systematically from the full set to the central subset, and is highest for the individual deep tile. The survey-wide average (blue) reaches 50\% completeness at approximately 0.6~mJy and approaches 90\% completeness by $1.5$~mJy. The central tiles (orange), benefiting from deeper integration and lower noise -- though still affected by the presence of bright sources -- reach 50\% completeness at slightly lower flux densities, around 0.5~mJy. Tile 41 (green), representative of the deepest and cleanest regions of the mosaic, attains 50\% completeness near 0.35~mJy and rapidly approches 90\% at $\sim$0.6~mJy. Error bars represent the standard deviation across 20 realizations, capturing both statistical noise and variation in source recovery performance. These comparisons highlight the impact of depth and local noise properties on detection efficiency and motivate the use of tile-dependent completeness corrections in downstream analyses.

\autoref{fig:completeness} right panel displays the completeness curves for all individual tiles (thin colored lines), highlighting the tile-to-tile variation, especially at low flux densities. The bold black line denotes the median completeness curve across the full survey. While the completeness rises sharply with increasing flux, it asymptotes to unity at different levels for different tiles. 

These completeness curves are subsequently used to correct the differential source counts presented in \autoref{sec:dnds}.

\begin{figure*}
    \centering
    \includegraphics[width=0.75\textwidth]{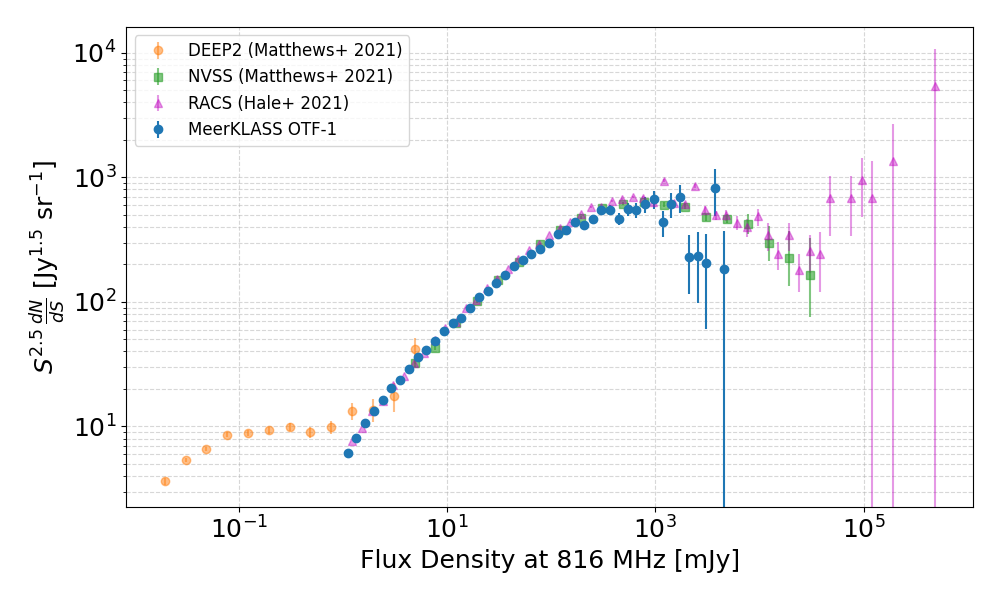}
    \caption{Euclidean-normalised differential source counts, $S^{2.5}dN/dS$, at 816\,MHz for the MeerKLASS OTF DR1 catalogue (blue circles: completeness-corrected) compared with DEEP2 and NVSS counts from \citet{matthews2021} and RACS counts from \citet{hale2021}, all scaled to a common reference frequency assuming $\alpha = -0.7$.}
    \label{fig:dnds}
\end{figure*}

\subsection{Differential Source Counts}
\label{sec:dnds}
The differential source count, $dN/dS$, which measures the number of sources per unit solid angle per unit flux density, is a fundamental statistical tool in extragalactic astronomy. It provides a direct probe of the cosmic evolution of radio source populations, including active galactic nuclei (AGN) and star-forming galaxies (SFGs). In a static, non-evolving Euclidean universe, these counts follow $dN/dS \propto S^{-2.5}$; hence, it is customary to present the \emph{Euclidean-normalised} form, $S^{2.5}\,dN/dS$, which is expected to be constant with flux density under such conditions. Departures from a flat distribution are therefore sensitive diagnostics of cosmological evolution and the changing demographics of radio emitters over cosmic time. 

We constructed the source counts in this work using sources with an integrated flux density greater than 1\,mJy and a signal-to-noise ratio greater than 7 in each of the 89 survey tiles. To ensure robust statistics, the sources were binned logarithmically by their integrated flux density. A critical step in this analysis is correcting for completeness, which becomes significant at faint flux densities close to the survey's detection limit. We applied a completeness correction to the raw counts in each bin by dividing them by the corresponding completeness fraction derived from our extensive injection-recovery simulations (\autoref{sec:completeness}). The uncertainty in each bin was calculated assuming Poisson statistics on the globally aggregated, completeness-corrected source counts. 

The resulting Euclidean-normalised differential source counts at a reference frequency of 816\,MHz are presented in \autoref{fig:dnds}. 
 
For context and validation, we compare our measurements with established results from the literature, including deep counts from the DEEP2 field \citep{matthews2021}, and wider-area survey counts from NVSS \citep{matthews2021} and RACS \citep{hale2021}. To facilitate a direct comparison, all literature counts were scaled to the MeerKLASS reference frequency of 816\,MHz assuming a mean spectral index of $\alpha = -0.7$ ($S_\nu \propto \nu^\alpha$). 

The MeerKLASS counts are in excellent agreement with the general trends established by previous work. The overall distribution, primarily traced by RACS and NVSS, shows the expected rise from faint fluxes to a broad peak at higher flux values, before declining towards brighter sources. Our corrected MeerKLASS counts are consistent with this picture.The large error bars on the high-flux MeerKLASS data points are likely due to statistical fluctuations from the small number of sources in those bins, rather than a true feature of the source population. 

The ability of MeerKLASS to reliably trace the source counts across this wide flux range validates the survey's calibration and completeness corrections. Overall, the DR1 counts successfully bridge the gap between wide, shallow surveys and deep, narrow fields by providing a statistically powerful measurement of the source population down to the flux density regime where the contribution from star-forming galaxies and radio-quiet AGN becomes dominant.

\begin{table}
\centering
\begin{tabular}{ccc}
\hline
Flux $S$ & Count & Corrected $S^{2.5}\,dN/dS$ \\
(mJy) & & (Jy$^{1.5}$ sr$^{-1}$) \\
\hline
1.107 & 5797 & $6.147 \pm 0.074$ \\
1.343 & 6563 & $8.022 \pm 0.094$ \\
1.630 & 7209 & $10.666 \pm 0.121$ \\
1.978 & 7139 & $13.286 \pm 0.153$ \\
2.401 & 6811 & $16.355 \pm 0.195$ \\
2.913 & 6498 & $20.397 \pm 0.250$ \\
3.536 & 5715 & $23.680 \pm 0.310$ \\
4.291 & 5321 & $29.174 \pm 0.397$ \\
5.207 & 4978 & $36.284 \pm 0.511$ \\
6.319 & 4281 & $41.487 \pm 0.631$ \\
7.669 & 3765 & $48.550 \pm 0.788$ \\
9.307 & 3411 & $58.559 \pm 1.000$ \\
11.295 & 2948 & $67.678 \pm 1.243$ \\
13.707 & 2409 & $73.913 \pm 1.501$ \\
16.634 & 2200 & $90.166 \pm 1.916$ \\
20.187 & 1983 & $108.567 \pm 2.431$ \\
24.499 & 1683 & $123.144 \pm 2.993$ \\
29.731 & 1461 & $142.741 \pm 3.725$ \\
36.081 & 1257 & $163.644 \pm 4.611$ \\
43.787 & 1118 & $194.293 \pm 5.809$ \\
53.139 & 930  & $216.280 \pm 7.086$ \\
64.488 & 777  & $241.537 \pm 8.658$ \\
78.262 & 643  & $266.927 \pm 10.523$ \\
94.977 & 531  & $294.581 \pm 12.782$ \\
115.262 & 470 & $348.578 \pm 16.077$ \\
139.879 & 378 & $374.993 \pm 19.280$ \\
169.754 & 330 & $437.427 \pm 24.077$ \\
206.009 & 235 & $416.639 \pm 27.170$ \\
250.008 & 196 & $464.678 \pm 33.176$ \\
303.405 & 174 & $551.110 \pm 41.776$ \\
368.205 & 130 & $551.291 \pm 48.311$ \\
446.845 & 82  & $464.564 \pm 51.278$ \\
542.281 & 73  & $553.361 \pm 64.708$ \\
658.101 & 54  & $546.267 \pm 74.338$ \\
798.656 & 45  & $609.962 \pm 90.825$ \\
969.231 & 37  & $668.984 \pm 109.980$ \\
1176.237 & 18  & $436.500 \pm 102.719$ \\
1427.454 & 19  & $614.002 \pm 140.862$ \\
1732.326 & 16  & $691.254 \pm 172.814$ \\
2102.312 & 4   & $231.036 \pm 115.518$ \\
2551.319 & 3   & $231.655 \pm 133.746$ \\
3096.223 & 2   & $206.467 \pm 145.995$ \\
3757.506 & 6   & $828.083 \pm 338.064$ \\
4560.025 & 1   & $184.512 \pm 184.512$ \\
\hline
\end{tabular}
\caption{Euclidean-normalised differential source counts at 816\,MHz. Flux bins are the bin-centre values in mJy; counts are raw source counts in each bin. The third column lists $S^{2.5}\,dN/dS$ in Jy$^{1.5}$\,sr$^{-1}$, shown as value~$\pm$~Poisson error (completeness-corrected).}
\label{tab:source_counts}
\end{table}

\section{Public data release}
\label{sec:public_release}
We release a suite of data products to facilitate downstream science. The package comprises tile–level imaging (continuum and sub–band cubes), source and component catalogues, and survey–scale summary maps. All images are provided as FITS files with standard WCS; flux densities are in Jy\,beam$^{-1}$, and the restoring beam is recorded in the headers via \texttt{BMAJ/BMIN/BPA}. Astrometry is ICRS (J2000) throughout.

\begin{enumerate}
\item {\bf Tile–level imaging:} For every tile we provide:
\begin{itemize}
  \item a Stokes~$I$ \emph{continuum} image at the central UHF frequency of 816\,MHz with a $3\arcsec$ pixel scale, and the corresponding \emph{residual} image.
  \item a 9–plane \emph{sub–band cube} sampled at 574, 635, 695, 755, 816, 876, 937, 997, and 1058\,MHz (one image per frequency plane).
\end{itemize}
The beam for each tile is stored in the image headers and mirrors the values reported alongside the catalogues.

\item {\bf Catalogues:} We release two FITS binary tables produced with \textsc{PyBDSF}: the SRL (source) catalogue, containing \textbf{95\,483} radio sources after cross–tile de–duplication, and the GAUL (Gaussian–component) catalogue, containing \textbf{115\,328} components. The columns for both catalogues are detailed in Appendix~A. Researchers are encouraged to use \texttt{Source\_id} as the primary key when joining GAUL to SRL, and \texttt{Tile\_ID} when computing tile–level statistics.

\item {\bf Survey–scale maps:} In addition to tile products, we release the full survey mosaic (shown in \autoref{fig:mosaic_zoomtiles}) and the RMS noise map (shown in \autoref{fig:rms_map}) as FITS images suitable for global analyses and quick–look visualisation.
\end{enumerate}

\section{Conclusions}
\label{sec:conclusions}
In this paper, we have presented the first public data release (DR1) of the MeerKLASS UHF survey, a legacy program utilising MeerKAT's 
scanning mode to support single-dish intensity mapping while enabling commensal on-the-fly interferometric imaging.
Based on just 12 hours of early science observations, we have demonstrated the viability and power of this fast-scanning technique for efficiently mapping large areas of the sky. We have developed and validated a dedicated data processing pipeline to handle the unique challenges of OTF 
interferometric imaging,
producing high-fidelity continuum images and a comprehensive source catalogue for an $\sim$800\,deg$^2$ field within the DESI footprint. We characterised the survey performance in terms of image fidelity, angular resolution, flux density accuracy, astrometric precision, and completeness. Injection–recovery simulations were used to quantify the completeness as a function of flux density for each tile, enabling accurate corrections to the measured source counts.

Our main findings and data products are summarised as follows:

\begin{enumerate}
    \item We have produced deep, high‐resolution continuum images at a central frequency of 816~MHz. The deepest regions of our mosaic reach an RMS sensitivity of $\sim35\,\mu$Jy\,beam$^{-1}$ with a typical angular resolution of $\sim32''\times17''$. As expected for thermal–noise–dominated imaging at fixed bandwidth, the local RMS decreases approximately as $\sigma \propto t^{-1/2}$ across the mosaic, closely tracking the spatial variation in integration time from the 
    scan pattern (equivalently, $\sigma \propto N_{\rm hit}^{-1/2}$ for repeated passes). Departures from this scaling are confined to areas around the brightest sources where dynamic-range limitations elevate the noise. Overall image quality is excellent, revealing a rich variety of compact and extended radio morphologies. 

    \item From these images, we have extracted and validated a source catalogue containing approximately 100,000 unique radio sources. Through cross-matching with external surveys (RACS, NVSS, FIRST), we have confirmed the catalogue's high astrometric precision (any systematic positional error is $<1''$) and a robust flux density scale consistent with established benchmarks.

    \item We have computed the differential source counts for sources above 1\,mJy, finding excellent agreement with previous surveys. This result validates our entire processing chain, from calibration and imaging to source extraction and completeness correction, and demonstrates that even this initial, limited dataset can produce scientifically robust 
    measurements.
\end{enumerate}

This first data release represents only a small fraction of the full MeerKLASS survey, which aims to cover 10,000\,deg$^2$. The success of this pilot study serves as a crucial proof of concept for 
commensal scanning and OTF interferometric imaging,
highlighting
the associated data reduction techniques, which will be essential for future large-scale radio surveys with MeerKAT and the SKA. The public release of these images and the source catalogue provides a valuable resource for the community, enabling a wide range of scientific investigations, including studies of AGN and galaxy evolution, spectral index mapping, and cross-identification with optical surveys like DESI. Future MeerKLASS data releases will expand the sky coverage, improve sensitivity, and incorporate polarisation data, further enhancing the legacy value of this unique survey. In addition, forthcoming enhancements to the MeerKAT correlator -- specifically, sidereal tracking of the delay centre during OTF scans -- will remove the current smearing of the synthesised beam due to the fixed delay centre in azimuth, thereby improving the effective resolution and point-source flux recovery and sensitivity.

\section*{Acknowledgments}
SP thanks Steve Cunnington, Aishrila Mazumder, and Benedict Bahr-Kalus for their helpful feedback on the manuscript. 
SP acknowledges support from the Science and Technology Facilities Council (STFC) through the Consolidated Grant ST/X001229/1 at the Jodrell Bank Centre for Astrophysics, University of Manchester. SC acknowledges financial support from the South African National Research Foundation (Grant No. 84156) and the Inter-University Institute for Data Intensive Astronomy (IDIA). IDIA is a partnership of the University of Cape Town, the University of Pretoria and the University of the Western Cape. IDIA is registered on the Research Organization Registry with ROR ID 01edhwb26, and on Open Funder Registry with funder ID 100031500. SM and JM acknowledge the support provided by the German Federal Ministry of Education and Research (BMBF) through the BMBF D-MeerKAT III award (number 05A23WM2). OMS's research is supported by the South African Research Chairs Initiative of the Department of Science and Technology and National Research Foundation (grant No. 81737). The MeerKAT telescope is operated by the South African Radio Astronomy Observatory, which is a facility of the National Research Foundation, an agency of the Department of Science and Innovation. We acknowledge the use of the ilifu cloud computing facility – www.ilifu.ac.za, a partnership between the University of Cape Town, the University of the Western Cape, Stellenbosch University, Sol Plaatje University and the Cape Peninsula University of Technology. The ilifu facility is supported by contributions from the Inter-University Institute for Data Intensive Astronomy (IDIA – a partnership between the University of Cape Town, the University of Pretoria and the University of the Western Cape), the Computational Biology division at UCT and the Data Intensive Research Initiative of South Africa (DIRISA). This work made use of the CARTA (Cube Analysis and Rendering Tool for Astronomy) software (DOI: 10.5281/zenodo.3377984 – https://cartavis.github.io). We thank the developers of open-source Python libraries \textsc{NumPy} \citep{Numpy}, \textsc{SciPy} \citep{Scipy}, \textsc{Matplotlib} \citep{Matplotlib}, and \textsc{Astropy} \citep{astropy:2022}.

\section*{Data availability}
The data used in this study are available in the SARAO Online Archive \href{https://archive.sarao.ac.za}{(https://archive.sarao.ac.za)} with proposal ID SCI-20220822-MS-01. All data products described in \autoref{sec:public_release} will be made available via the SARAO-hosted MeerKLASS data release at: \url{https://doi.datacite.org/dois/10.48479/h9k3-7294}

\bibliographystyle{mnras}
\bibliography{ref}

@ARTICLE{jonas2009,
  author={Jonas, Justin L.},
  journal={Proceedings of the IEEE}, 
  title={MeerKAT—The South African Array With Composite Dishes and Wide-Band Single Pixel Feeds}, 
  year={2009},
  volume={97},
  number={8},
  pages={1522-1530},
  doi={10.1109/JPROC.2009.2020713}}

@INPROCEEDINGS{meerklass,
       author = {{Santos}, M. and {Bull}, P. and {Camera}, S. and {Chen}, S. and {Fonseca}, J. and {Heywood}, I. and {Hilton}, M. and {Jarvis}, M. and {Jozsa}, G.~I.~G. and {Knowles}, K. and {Leeuw}, L. and {Maartens}, R. and {Malefahlo}, E. and {McAlpine}, K. and {Moodley}, K. and {Patel}, P. and {Pourtsidou}, A. and {Prescott}, M. and {Spekkens}, K. and {Taylor}, R. and {Witzemann}, A. and {Whittam}, I.~H.},
        title = "{A Large Sky Survey with MeerKAT}",
    booktitle = {MeerKAT Science: On the Pathway to the SKA},
         year = 2016,
        month = jan,
          eid = {32},
        pages = {32},
          doi = {10.22323/1.277.0032},
archivePrefix = {arXiv},
       eprint = {1709.06099},
 primaryClass = {astro-ph.CO},
       adsurl = {https://ui.adsabs.harvard.edu/abs/2016mks..confE..32S}
}

@article{Chang_2010,
    author = "Chang, Tzu-Ching and Pen, Ue-Li and Bandura, Kevin and Peterson, Jeffrey B.",
    archivePrefix = "arXiv",
    doi = "10.1038/nature09187",
    eprint = "1007.3709",
    journal = "Nature",
    pages = "463--465",
    primaryClass = "astro-ph.CO",
    title = "{Hydrogen 21-cm Intensity Mapping at redshift 0.8}",
    volume = "466",
    year = "2010"
}

@ARTICLE{Masui_2013,
       author = {{Masui}, K.~W. and {Switzer}, E.~R. and {Banavar}, N. and {Bandura}, K. and {Blake}, C. and {Calin}, L. -M. and {Chang}, T. -C. and {Chen}, X. and {Li}, Y. -C. and {Liao}, Y. -W. and {Natarajan}, A. and {Pen}, U. -L. and {Peterson}, J.~B. and {Shaw}, J.~R. and {Voytek}, T.~C.},
        title = "{Measurement of 21 cm Brightness Fluctuations at z \raisebox{-0.5ex}\textasciitilde 0.8 in Cross-correlation}",
      journal = {\apjl},
         year = 2013,
        month = jan,
       volume = {763},
       number = {1},
          eid = {L20},
        pages = {L20},
          doi = {10.1088/2041-8205/763/1/L20},
archivePrefix = {arXiv},
       eprint = {1208.0331},
 primaryClass = {astro-ph.CO},
       adsurl = {https://ui.adsabs.harvard.edu/abs/2013ApJ...763L..20M}
}

@ARTICLE{Wolz_2022,
       author = {{Wolz}, Laura and {Pourtsidou}, Alkistis and {Masui}, Kiyoshi W. and {Chang}, Tzu-Ching and {Bautista}, Julian E. and {M{\"u}ller}, Eva-Maria and {Avila}, Santiago and {Bacon}, David and {Percival}, Will J. and {Cunnington}, Steven and {Anderson}, Chris and {Chen}, Xuelei and {Kneib}, Jean-Paul and {Li}, Yi-Chao and {Liao}, Yu-Wei and {Pen}, Ue-Li and {Peterson}, Jeffrey B. and {Rossi}, Graziano and {Schneider}, Donald P. and {Yadav}, Jaswant and {Zhao}, Gong-Bo},
        title = "{H I constraints from the cross-correlation of eBOSS galaxies and Green Bank Telescope intensity maps}",
      journal = {\mnras},
         year = 2022,
        month = mar,
       volume = {510},
       number = {3},
        pages = {3495-3511},
          doi = {10.1093/mnras/stab3621},
archivePrefix = {arXiv},
       eprint = {2102.04946},
 primaryClass = {astro-ph.CO},
       adsurl = {https://ui.adsabs.harvard.edu/abs/2022MNRAS.510.3495W}
}

@ARTICLE{Anderson_2018,
       author = {{Anderson}, C.~J. and {Luciw}, N.~J. and {Li}, Y. -C. and {Kuo}, C.~Y. and {Yadav}, J. and {Masui}, K.~W. and {Chang}, T. -C. and {Chen}, X. and {Oppermann}, N. and {Liao}, Y. -W. and {Pen}, U. -L. and {Price}, D.~C. and {Staveley-Smith}, L. and {Switzer}, E.~R. and {Timbie}, P.~T. and {Wolz}, L.},
        title = "{Low-amplitude clustering in low-redshift 21-cm intensity maps cross-correlated with 2dF galaxy densities}",
      journal = {\mnras},
         year = 2018,
        month = may,
       volume = {476},
       number = {3},
        pages = {3382-3392},
          doi = {10.1093/mnras/sty346},
archivePrefix = {arXiv},
       eprint = {1710.00424},
 primaryClass = {astro-ph.CO},
       adsurl = {https://ui.adsabs.harvard.edu/abs/2018MNRAS.476.3382A}
}

@ARTICLE{CHIME_2022,
       author = {{CHIME Collaboration} and {Amiri}, Mandana and {Bandura}, Kevin and {Chen}, Tianyue and {Deng}, Meiling and {Dobbs}, Matt and {Fandino}, Mateus and {Foreman}, Simon and {Halpern}, Mark and {Hill}, Alex S. and {Hinshaw}, Gary and {H{\"o}fer}, Carolin and {Kania}, Joseph and {Landecker}, T.~L. and {MacEachern}, Joshua and {Masui}, Kiyoshi and {Mena-Parra}, Juan and {Milutinovic}, Nikola and {Mirhosseini}, Arash and {Newburgh}, Laura and {Ordog}, Anna and {Pen}, Ue-Li and {Pinsonneault-Marotte}, Tristan and {Polzin}, Ava and {Reda}, Alex and {Renard}, Andre and {Shaw}, J. Richard and {Siegel}, Seth R. and {Singh}, Saurabh and {Vanderlinde}, Keith and {Wang}, Haochen and {Wiebe}, Donald V. and {Wulf}, Dallas},
        title = "{Detection of Cosmological 21 cm Emission with the Canadian Hydrogen Intensity Mapping Experiment}",
      journal = {arXiv e-prints},
         year = 2022,
        month = feb,
          eid = {arXiv:2202.01242},
        pages = {arXiv:2202.01242},
archivePrefix = {arXiv},
       eprint = {2202.01242},
 primaryClass = {astro-ph.CO},
       adsurl = {https://ui.adsabs.harvard.edu/abs/2022arXiv220201242C}
}

@ARTICLE{Cunnington_2023,
       author = {{Cunnington}, Steven and {Li}, Yichao and {Santos}, Mario G. and {Wang}, Jingying and {Carucci}, Isabella P. and {Irfan}, Melis O. and {Pourtsidou}, Alkistis and {Spinelli}, Marta and {Wolz}, Laura and {Soares}, Paula S. and {Blake}, Chris and {Bull}, Philip and {Engelbrecht}, Brandon and {Fonseca}, Jos{\'e} and {Grainge}, Keith and {Ma}, Yin-Zhe},
        title = "{H I intensity mapping with MeerKAT: power spectrum detection in cross-correlation with WiggleZ galaxies}",
      journal = {\mnras},
         year = 2023,
        month = feb,
       volume = {518},
       number = {4},
        pages = {6262-6272},
          doi = {10.1093/mnras/stac3060},
archivePrefix = {arXiv},
       eprint = {2206.01579},
 primaryClass = {astro-ph.CO},
       adsurl = {https://ui.adsabs.harvard.edu/abs/2023MNRAS.518.6262C}
}

@ARTICLE{Paul_2023,
       author = {{Paul}, Sourabh and {Santos}, Mario G. and {Chen}, Zhaoting and {Wolz}, Laura},
        title = "{A first detection of neutral hydrogen intensity mapping on Mpc scales at $z\approx 0.32$ and $z\approx 0.44$}",
      journal = {arXiv e-prints},
         year = 2023,
        month = jan,
          eid = {arXiv:2301.11943},
        pages = {arXiv:2301.11943},
          doi = {10.48550/arXiv.2301.11943},
archivePrefix = {arXiv},
       eprint = {2301.11943},
 primaryClass = {astro-ph.CO},
       adsurl = {https://ui.adsabs.harvard.edu/abs/2023arXiv230111943P}
}

@ARTICLE{Wang_2021,
       author = {{Wang}, Jingying and {Santos}, Mario G. and {Bull}, Philip and {Grainge}, Keith and {Cunnington}, Steven and {Fonseca}, Jos{\'e} and {Irfan}, Melis O. and {Li}, Yichao and {Pourtsidou}, Alkistis and {Soares}, Paula S. and {Spinelli}, Marta and {Bernardi}, Gianni and {Engelbrecht}, Brandon},
        title = "{H I intensity mapping with MeerKAT: calibration pipeline for multidish autocorrelation observations}",
      journal = {\mnras},
         year = 2021,
        month = aug,
       volume = {505},
       number = {3},
        pages = {3698-3721},
          doi = {10.1093/mnras/stab1365},
archivePrefix = {arXiv},
       eprint = {2011.13789},
 primaryClass = {astro-ph.CO},
       adsurl = {https://ui.adsabs.harvard.edu/abs/2021MNRAS.505.3698W}
}

@ARTICLE{shimwell2022,
       author = {{Shimwell}, T.~W. and {Hardcastle}, M.~J. and {Tasse}, C. and {Best}, P.~N. and {R{\"o}ttgering}, H.~J.~A. and {Williams}, W.~L. and {Botteon}, A. and {Drabent}, A. and {Mechev}, A. and {Shulevski}, A. and {van Weeren}, R.~J. and {Bester}, L. and {Br{\"u}ggen}, M. and {Brunetti}, G. and {Callingham}, J.~R. and {Chy{\.z}y}, K.~T. and {Conway}, J.~E. and {Dijkema}, T.~J. and {Duncan}, K. and {de Gasperin}, F. and {Hale}, C.~L. and {Haverkorn}, M. and {Hugo}, B. and {Jackson}, N. and {Mevius}, M. and {Miley}, G.~K. and {Morabito}, L.~K. and {Morganti}, R. and {Offringa}, A. and {Oonk}, J.~B.~R. and {Rafferty}, D. and {Sabater}, J. and {Smith}, D.~J.~B. and {Schwarz}, D.~J. and {Smirnov}, O. and {O'Sullivan}, S.~P. and {Vedantham}, H. and {White}, G.~J. and {Albert}, J.~G. and {Alegre}, L. and {Asabere}, B. and {Bacon}, D.~J. and {Bonafede}, A. and {Bonnassieux}, E. and {Brienza}, M. and {Bilicki}, M. and {Bonato}, M. and {Calistro Rivera}, G. and {Cassano}, R. and {Cochrane}, R. and {Croston}, J.~H. and {Cuciti}, V. and {Dallacasa}, D. and {Danezi}, A. and {Dettmar}, R.~J. and {Di Gennaro}, G. and {Edler}, H.~W. and {En{\ss}lin}, T.~A. and {Emig}, K.~L. and {Franzen}, T.~M.~O. and {Garc{\'\i}a-Vergara}, C. and {Grange}, Y.~G. and {G{\"u}rkan}, G. and {Hajduk}, M. and {Heald}, G. and {Heesen}, V. and {Hoang}, D.~N. and {Hoeft}, M. and {Horellou}, C. and {Iacobelli}, M. and {Jamrozy}, M. and {Jeli{\'c}}, V. and {Kondapally}, R. and {Kukreti}, P. and {Kunert-Bajraszewska}, M. and {Magliocchetti}, M. and {Mahatma}, V. and {Ma{\l}ek}, K. and {Mandal}, S. and {Massaro}, F. and {Meyer-Zhao}, Z. and {Mingo}, B. and {Mostert}, R.~I.~J. and {Nair}, D.~G. and {Nakoneczny}, S.~J. and {Nikiel-Wroczy{\'n}ski}, B. and {Orr{\'u}}, E. and {Pajdosz-{\'S}mierciak}, U. and {Pasini}, T. and {Prandoni}, I. and {van Piggelen}, H.~E. and {Rajpurohit}, K. and {Retana-Montenegro}, E. and {Riseley}, C.~J. and {Rowlinson}, A. and {Saxena}, A. and {Schrijvers}, C. and {Sweijen}, F. and {Siewert}, T.~M. and {Timmerman}, R. and {Vaccari}, M. and {Vink}, J. and {West}, J.~L. and {Wo{\l}owska}, A. and {Zhang}, X. and {Zheng}, J.},
        title = "{The LOFAR Two-metre Sky Survey. V. Second data release}",
      journal = {\aap},
         year = 2022,
        month = mar,
       volume = {659},
          eid = {A1},
        pages = {A1},
          doi = {10.1051/0004-6361/202142484},
archivePrefix = {arXiv},
       eprint = {2202.11733},
 primaryClass = {astro-ph.GA},
       adsurl = {https://ui.adsabs.harvard.edu/abs/2022A&A...659A...1S}
}

@ARTICLE{mcconnell2020,
       author = {{McConnell}, D. and {Hale}, C.~L. and {Lenc}, E. and {Banfield}, J.~K. and {Heald}, George and {Hotan}, A.~W. and {Leung}, James K. and {Moss}, Vanessa A. and {Murphy}, Tara and {O'Brien}, Andrew and {Pritchard}, Joshua and {Raja}, Wasim and {Sadler}, Elaine M. and {Stewart}, Adam and {Thomson}, Alec J.~M. and {Whiting}, M. and {Allison}, James R. and {Amy}, S.~W. and {Anderson}, C. and {Ball}, Lewis and {Bannister}, Keith W. and {Bell}, Martin and {Bock}, Douglas C. -J. and {Bolton}, Russ and {Bunton}, J.~D. and {Chippendale}, A.~P. and {Collier}, J.~D. and {Cooray}, F.~R. and {Cornwell}, T.~J. and {Diamond}, P.~J. and {Edwards}, P.~G. and {Gupta}, N. and {Hayman}, Douglas B. and {Heywood}, Ian and {Jackson}, C.~A. and {Koribalski}, B{\"a}rbel S. and {Lee-Waddell}, Karen and {McClure-Griffiths}, N.~M. and {Ng}, Alan and {Norris}, Ray P. and {Phillips}, Chris and {Reynolds}, John E. and {Roxby}, Daniel N. and {Schinckel}, Antony E.~T. and {Shields}, Matt and {Tremblay}, Chenoa and {Tzioumis}, A. and {Voronkov}, M.~A. and {Westmeier}, Tobias},
        title = "{The Rapid ASKAP Continuum Survey I: Design and first results}",
      journal = {\pasa},
         year = 2020,
        month = nov,
       volume = {37},
          eid = {e048},
        pages = {e048},
          doi = {10.1017/pasa.2020.41},
archivePrefix = {arXiv},
       eprint = {2012.00747},
 primaryClass = {astro-ph.IM},
       adsurl = {https://ui.adsabs.harvard.edu/abs/2020PASA...37...48M}
}

@ARTICLE{hale2021,
       author = {{Hale}, Catherine L. and {McConnell}, D. and {Thomson}, A.~J.~M. and {Lenc}, E. and {Heald}, G.~H. and {Hotan}, A.~W. and {Leung}, J.~K. and {Moss}, V.~A. and {Murphy}, T. and {Pritchard}, J. and {Sadler}, E.~M. and {Stewart}, A.~J. and {Whiting}, M.~T.},
        title = "{The Rapid ASKAP Continuum Survey Paper II: First Stokes I Source Catalogue Data Release}",
      journal = {\pasa},
         year = 2021,
        month = nov,
       volume = {38},
          eid = {e058},
        pages = {e058},
          doi = {10.1017/pasa.2021.47},
archivePrefix = {arXiv},
       eprint = {2109.00956},
 primaryClass = {astro-ph.GA},
       adsurl = {https://ui.adsabs.harvard.edu/abs/2021PASA...38...58H}
}

@INPROCEEDINGS{jarvis2016,
       author = {{Jarvis}, M. and {Taylor}, R. and {Agudo}, I. and {Allison}, J.~R. and {Deane}, R.~P. and {Frank}, B. and {Gupta}, N. and {Heywood}, I. and {Maddox}, N. and {McAlpine}, K. and {Santos}, M. and {Scaife}, A.~M.~M. and {Vaccari}, M. and {Zwart}, J.~T.~L. and {Adams}, E. and {Bacon}, D.~J. and {Baker}, A.~J. and {Bassett}, B.~A. and {Best}, P.~N. and {Beswick}, R. and {Blyth}, S. and {Brown}, M.~L. and {Bruggen}, M. and {Cluver}, M. and {Colafrancesco}, S. and {Cotter}, G. and {Cress}, C. and {Dav{\'e}}, R. and {Ferrari}, C. and {Hardcastle}, M.~J. and {Hale}, C.~L. and {Harrison}, I. and {Hatfield}, P.~W. and {Klockner}, H.~R. and {Kolwa}, S. and {Malefahlo}, E. and {Marubini}, T. and {Mauch}, T. and {Moodley}, K. and {Morganti}, R. and {Norris}, R.~P. and {Peters}, J.~A. and {Prandoni}, I. and {Prescott}, M. and {Oliver}, S. and {Oozeer}, N. and {Rottgering}, H.~J.~A. and {Seymour}, N. and {Simpson}, C. and {Smirnov}, O. and {Smith}, D.~J.~B.},
        title = "{The MeerKAT International GHz Tiered Extragalactic Exploration (MIGHTEE) Survey}",
    booktitle = {MeerKAT Science: On the Pathway to the SKA},
         year = 2016,
        month = jan,
          eid = {6},
        pages = {6},
          doi = {10.22323/1.277.0006},
archivePrefix = {arXiv},
       eprint = {1709.01901},
primaryClass = {astro-ph.GA},
       adsurl = {https://ui.adsabs.harvard.edu/abs/2016mks..confE...6J}
}

@ARTICLE{heywood2022,
       author = {{Heywood}, I. and {Jarvis}, M.~J. and {Hale}, C.~L. and {Whittam}, I.~H. and {Bester}, H.~L. and {Hugo}, B. and {Kenyon}, J.~S. and {Prescott}, M. and {Smirnov}, O.~M. and {Tasse}, C. and {Afonso}, J.~M. and {Best}, P.~N. and {Collier}, J.~D. and {Deane}, R.~P. and {Frank}, B.~S. and {Hardcastle}, M.~J. and {Knowles}, K. and {Maddox}, N. and {Murphy}, E.~J. and {Prandoni}, I. and {Randriamampandry}, S.~M. and {Santos}, M.~G. and {Sekhar}, S. and {Tabatabaei}, F. and {Taylor}, A.~R. and {Thorat}, K.},
        title = "{MIGHTEE: total intensity radio continuum imaging and the COSMOS/XMM-LSS Early Science fields}",
      journal = {\mnras},
         year = 2022,
        month = jan,
       volume = {509},
       number = {2},
        pages = {2150-2168},
          doi = {10.1093/mnras/stab3021},
archivePrefix = {arXiv},
       eprint = {2110.00347},
 primaryClass = {astro-ph.GA},
       adsurl = {https://ui.adsabs.harvard.edu/abs/2022MNRAS.509.2150H}
}

@article{intema2017,
	author = {{Intema, H. T.} and {Jagannathan, P.} and {Mooley, K. P.} and {Frail, D. A.}},
	title = {The GMRT 150 MHz all-sky radio survey⋆ - First alternative data release TGSS ADR1},
	DOI= "10.1051/0004-6361/201628536",
	url= "https://doi.org/10.1051/0004-6361/201628536",
	journal = {A&A},
	year = 2017,
	volume = 598,
	pages = "A78",
}

@ARTICLE{wayth2015,
       author = {{Wayth}, R.~B. and {Lenc}, E. and {Bell}, M.~E. and {Callingham}, J.~R. and {Dwarakanath}, K.~S. and {Franzen}, T.~M.~O. and {For}, B. -Q. and {Gaensler}, B. and {Hancock}, P. and {Hindson}, L. and {Hurley-Walker}, N. and {Jackson}, C.~A. and {Johnston-Hollitt}, M. and {Kapi{\'n}ska}, A.~D. and {McKinley}, B. and {Morgan}, J. and {Offringa}, A.~R. and {Procopio}, P. and {Staveley-Smith}, L. and {Wu}, C. and {Zheng}, Q. and {Trott}, C.~M. and {Bernardi}, G. and {Bowman}, J.~D. and {Briggs}, F. and {Cappallo}, R.~J. and {Corey}, B.~E. and {Deshpande}, A.~A. and {Emrich}, D. and {Goeke}, R. and {Greenhill}, L.~J. and {Hazelton}, B.~J. and {Kaplan}, D.~L. and {Kasper}, J.~C. and {Kratzenberg}, E. and {Lonsdale}, C.~J. and {Lynch}, M.~J. and {McWhirter}, S.~R. and {Mitchell}, D.~A. and {Morales}, M.~F. and {Morgan}, E. and {Oberoi}, D. and {Ord}, S.~M. and {Prabu}, T. and {Rogers}, A.~E.~E. and {Roshi}, A. and {Shankar}, N. Udaya and {Srivani}, K.~S. and {Subrahmanyan}, R. and {Tingay}, S.~J. and {Waterson}, M. and {Webster}, R.~L. and {Whitney}, A.~R. and {Williams}, A. and {Williams}, C.~L.},
        title = "{GLEAM: The GaLactic and Extragalactic All-Sky MWA Survey}",
      journal = {\pasa},
         year = 2015,
        month = jun,
       volume = {32},
          eid = {e025},
        pages = {e025},
          doi = {10.1017/pasa.2015.26},
archivePrefix = {arXiv},
       eprint = {1505.06041},
 primaryClass = {astro-ph.IM},
       adsurl = {https://ui.adsabs.harvard.edu/abs/2015PASA...32...25W}
}

@ARTICLE{mauch2003,
       author = {{Mauch}, T. and {Murphy}, T. and {Buttery}, H.~J. and {Curran}, J. and {Hunstead}, R.~W. and {Piestrzynski}, B. and {Robertson}, J.~G. and {Sadler}, E.~M.},
        title = "{SUMSS: a wide-field radio imaging survey of the southern sky - II. The source catalogue}",
      journal = {\mnras},
         year = 2003,
        month = jul,
       volume = {342},
       number = {4},
        pages = {1117-1130},
          doi = {10.1046/j.1365-8711.2003.06605.x},
archivePrefix = {arXiv},
       eprint = {astro-ph/0303188},
 primaryClass = {astro-ph},
       adsurl = {https://ui.adsabs.harvard.edu/abs/2003MNRAS.342.1117M}
}

@ARTICLE{condon1998,
       author = {{Condon}, J.~J. and {Cotton}, W.~D. and {Greisen}, E.~W. and {Yin}, Q.~F. and {Perley}, R.~A. and {Taylor}, G.~B. and {Broderick}, J.~J.},
        title = "{The NRAO VLA Sky Survey}",
      journal = {\aj},
         year = 1998,
        month = may,
       volume = {115},
       number = {5},
        pages = {1693-1716},
          doi = {10.1086/300337},
       adsurl = {https://ui.adsabs.harvard.edu/abs/1998AJ....115.1693C}
}

@ARTICLE{lacy2020,
       author = {{Lacy}, M. and {Baum}, S.~A. and {Chandler}, C.~J. and {Chatterjee}, S. and {Clarke}, T.~E. and {Deustua}, S. and {English}, J. and {Farnes}, J. and {Gaensler}, B.~M. and {Gugliucci}, N. and {Hallinan}, G. and {Kent}, B.~R. and {Kimball}, A. and {Law}, C.~J. and {Lazio}, T.~J.~W. and {Marvil}, J. and {Mao}, S.~A. and {Medlin}, D. and {Mooley}, K. and {Murphy}, E.~J. and {Myers}, S. and {Osten}, R. and {Richards}, G.~T. and {Rosolowsky}, E. and {Rudnick}, L. and {Schinzel}, F. and {Sivakoff}, G.~R. and {Sjouwerman}, L.~O. and {Taylor}, R. and {White}, R.~L. and {Wrobel}, J. and {Andernach}, H. and {Beasley}, A.~J. and {Berger}, E. and {Bhatnager}, S. and {Birkinshaw}, M. and {Bower}, G.~C. and {Brandt}, W.~N. and {Brown}, S. and {Burke-Spolaor}, S. and {Butler}, B.~J. and {Comerford}, J. and {Demorest}, P.~B. and {Fu}, H. and {Giacintucci}, S. and {Golap}, K. and {G{\"u}th}, T. and {Hales}, C.~A. and {Hiriart}, R. and {Hodge}, J. and {Horesh}, A. and {Ivezi{\'c}}, {\v{Z}}. and {Jarvis}, M.~J. and {Kamble}, A. and {Kassim}, N. and {Liu}, X. and {Loinard}, L. and {Lyons}, D.~K. and {Masters}, J. and {Mezcua}, M. and {Moellenbrock}, G.~A. and {Mroczkowski}, T. and {Nyland}, K. and {O'Dea}, C.~P. and {O'Sullivan}, S.~P. and {Peters}, W.~M. and {Radford}, K. and {Rao}, U. and {Robnett}, J. and {Salcido}, J. and {Shen}, Y. and {Sobotka}, A. and {Witz}, S. and {Vaccari}, M. and {van Weeren}, R.~J. and {Vargas}, A. and {Williams}, P.~K.~G. and {Yoon}, I.},
        title = "{The Karl G. Jansky Very Large Array Sky Survey (VLASS). Science Case and Survey Design}",
      journal = {\pasp},
         year = 2020,
        month = mar,
       volume = {132},
       number = {1009},
          eid = {035001},
        pages = {035001},
          doi = {10.1088/1538-3873/ab63eb},
archivePrefix = {arXiv},
       eprint = {1907.01981},
 primaryClass = {astro-ph.IM},
       adsurl = {https://ui.adsabs.harvard.edu/abs/2020PASP..132c5001L}
}

@ARTICLE{Perrott2013,
       author = {{Perrott}, Yvette C. and {Scaife}, Anna M.~M. and {Green}, David A. and {Davies}, Matthew L. and {Franzen}, Thomas M.~O. and {Grainge}, Keith J.~B. and {Hobson}, Michael P. and {Hurley-Walker}, Natasha and {Lasenby}, Anthony N. and {Olamaie}, Malak and {Pooley}, Guy G. and {Rodr{\'\i}guez-Gonz{\'a}lvez}, Carmen and {Rumsey}, Clare and {Saunders}, Richard D.~E. and {Schammel}, Michel P. and {Scott}, Paul F. and {Shimwell}, Timothy W. and {Titterington}, David J. and {Waldram}, Elizabeth M. and {AMI Consortium}},
        title = "{AMI Galactic Plane Survey at 16 GHz - I. Observing, mapping and source extraction}",
      journal = {\mnras},
         year = 2013,
        month = mar,
       volume = {429},
       number = {4},
        pages = {3330-3340},
          doi = {10.1093/mnras/sts589},
archivePrefix = {arXiv},
       eprint = {1208.5343},
 primaryClass = {astro-ph.GA},
       adsurl = {https://ui.adsabs.harvard.edu/abs/2013MNRAS.429.3330P}
}

@ARTICLE{norris2021,
       author = {{Norris}, Ray P. and {Marvil}, Joshua and {Collier}, J.~D. and {Kapi{\'n}ska}, Anna D. and {O'Brien}, Andrew N. and {Rudnick}, L. and {Andernach}, Heinz and {Asorey}, Jacobo and {Brown}, Michael J.~I. and {Br{\"u}ggen}, Marcus and {Crawford}, Evan and {English}, Jayanne and {Rahman}, Syed Faisal ur and {Filipovi{\'c}}, Miroslav D. and {Gordon}, Yjan and {G{\"u}rkan}, G{\"u}lay and {Hale}, Catherine and {Hopkins}, Andrew M. and {Huynh}, Minh T. and {HyeongHan}, Kim and {James Jee}, M. and {Koribalski}, B{\"a}rbel S. and {Lenc}, Emil and {Luken}, Kieran and {Parkinson}, David and {Prandoni}, Isabella and {Raja}, Wasim and {Reiprich}, Thomas H. and {Riseley}, Christopher J. and {Shabala}, Stanislav S. and {Sheil}, Jaimie R. and {Vernstrom}, Tessa and {Whiting}, Matthew T. and {Allison}, James R. and {Anderson}, C.~S. and {Ball}, Lewis and {Bell}, Martin and {Bunton}, John and {Galvin}, T.~J. and {Gupta}, Neeraj and {Hotan}, Aidan and {Jacka}, Colin and {Macgregor}, Peter J. and {Mahony}, Elizabeth K. and {Maio}, Umberto and {Moss}, Vanessa and {Pandey-Pommier}, M. and {Voronkov}, Maxim A.},
        title = "{The Evolutionary Map of the Universe pilot survey}",
      journal = {\pasa},
         year = 2021,
        month = sep,
       volume = {38},
          eid = {e046},
        pages = {e046},
          doi = {10.1017/pasa.2021.42},
archivePrefix = {arXiv},
       eprint = {2108.00569},
 primaryClass = {astro-ph.CO},
       adsurl = {https://ui.adsabs.harvard.edu/abs/2021PASA...38...46N}
}

@ARTICLE{Duchesne2023,
       author = {{Duchesne}, S.~W. and {Thomson}, A.~J.~M. and {Pritchard}, J. and {Lenc}, E. and {Moss}, V.~A. and {McConnell}, D. and {Wieringa}, M.~H. and {Whiting}, M.~T. and {Wang}, Z. and {Wang}, Y. and {Rose}, K. and {Raja}, W. and {Murphy}, Tara and {Leung}, J.~K. and {Huynh}, M.~T. and {Hotan}, A.~W. and {Hodgson}, T. and {Heald}, G.~H.},
        title = "{The Rapid ASKAP Continuum Survey IV: continuum imaging at 1367.5 MHz and the first data release of RACS-mid}",
      journal = {\pasa},
         year = 2023,
        month = aug,
       volume = {40},
          eid = {e034},
        pages = {e034},
          doi = {10.1017/pasa.2023.31},
archivePrefix = {arXiv},
       eprint = {2306.07194},
 primaryClass = {astro-ph.IM},
       adsurl = {https://ui.adsabs.harvard.edu/abs/2023PASA...40...34D}
}

@ARTICLE{Mauch2007,
       author = {{Mauch}, Tom and {Sadler}, Elaine M.},
        title = "{Radio sources in the 6dFGS: local luminosity functions at 1.4 GHz for star-forming galaxies and radio-loud AGN}",
      journal = {\mnras},
         year = 2007,
        month = mar,
       volume = {375},
       number = {3},
        pages = {931-950},
          doi = {10.1111/j.1365-2966.2006.11353.x},
archivePrefix = {arXiv},
       eprint = {astro-ph/0612018},
 primaryClass = {astro-ph},
       adsurl = {https://ui.adsabs.harvard.edu/abs/2007MNRAS.375..931M}
}

@ARTICLE{Smolcic2017,
       author = {{Smol{\v{c}}i{\'c}}, V. and {Novak}, M. and {Bondi}, M. and {Ciliegi}, P. and {Mooley}, K.~P. and {Schinnerer}, E. and {Zamorani}, G. and {Navarrete}, F. and {Bourke}, S. and {Karim}, A. and {Vardoulaki}, E. and {Leslie}, S. and {Delhaize}, J. and {Carilli}, C.~L. and {Myers}, S.~T. and {Baran}, N. and {Delvecchio}, I. and {Miettinen}, O. and {Banfield}, J. and {Balokovi{\'c}}, M. and {Bertoldi}, F. and {Capak}, P. and {Frail}, D.~A. and {Hallinan}, G. and {Hao}, H. and {Herrera Ruiz}, N. and {Horesh}, A. and {Ilbert}, O. and {Intema}, H. and {Jeli{\'c}}, V. and {Kl{\"o}ckner}, H. -R. and {Krpan}, J. and {Kulkarni}, S.~R. and {McCracken}, H. and {Laigle}, C. and {Middleberg}, E. and {Murphy}, E.~J. and {Sargent}, M. and {Scoville}, N.~Z. and {Sheth}, K.},
        title = "{The VLA-COSMOS 3 GHz Large Project: Continuum data and source catalog release}",
      journal = {\aap},
         year = 2017,
        month = jun,
       volume = {602},
          eid = {A1},
        pages = {A1},
          doi = {10.1051/0004-6361/201628704},
archivePrefix = {arXiv},
       eprint = {1703.09713},
 primaryClass = {astro-ph.GA},
       adsurl = {https://ui.adsabs.harvard.edu/abs/2017A&A...602A...1S}
}

@ARTICLE{ODEa1998,
       author = {{O'Dea}, Christopher P.},
        title = "{The Compact Steep-Spectrum and Gigahertz Peaked-Spectrum Radio Sources}",
      journal = {\pasp},
         year = 1998,
        month = may,
       volume = {110},
       number = {747},
        pages = {493-532},
          doi = {10.1086/316162},
       adsurl = {https://ui.adsabs.harvard.edu/abs/1998PASP..110..493O}
}

@ARTICLE{Callingham2017,
       author = {{Callingham}, J.~R. and {Ekers}, R.~D. and {Gaensler}, B.~M. and {Line}, J.~L.~B. and {Hurley-Walker}, N. and {Sadler}, E.~M. and {Tingay}, S.~J. and {Hancock}, P.~J. and {Bell}, M.~E. and {Dwarakanath}, K.~S. and {For}, B. -Q. and {Franzen}, T.~M.~O. and {Hindson}, L. and {Johnston-Hollitt}, M. and {Kapi{\'n}ska}, A.~D. and {Lenc}, E. and {McKinley}, B. and {Morgan}, J. and {Offringa}, A.~R. and {Procopio}, P. and {Staveley-Smith}, L. and {Wayth}, R.~B. and {Wu}, C. and {Zheng}, Q.},
        title = "{Extragalactic Peaked-spectrum Radio Sources at Low Frequencies}",
      journal = {\apj},
         year = 2017,
        month = feb,
       volume = {836},
       number = {2},
          eid = {174},
        pages = {174},
          doi = {10.3847/1538-4357/836/2/174},
archivePrefix = {arXiv},
       eprint = {1701.02771},
 primaryClass = {astro-ph.GA},
       adsurl = {https://ui.adsabs.harvard.edu/abs/2017ApJ...836..174C}
}

@ARTICLE{BlakeWall2002,
       author = {{Blake}, Chris and {Wall}, Jasper},
        title = "{A velocity dipole in the distribution of radio galaxies}",
      journal = {\nat},
         year = 2002,
        month = mar,
       volume = {416},
       number = {6877},
        pages = {150-152},
          doi = {10.1038/416150a},
archivePrefix = {arXiv},
       eprint = {astro-ph/0203385},
 primaryClass = {astro-ph},
       adsurl = {https://ui.adsabs.harvard.edu/abs/2002Natur.416..150B}
}

@ARTICLE{Singal2011,
       author = {{Singal}, Ashok K.},
        title = "{Large Peculiar Motion of the Solar System from the Dipole Anisotropy in Sky Brightness due to Distant Radio Sources}",
      journal = {\apjl},
         year = 2011,
        month = dec,
       volume = {742},
       number = {2},
          eid = {L23},
        pages = {L23},
          doi = {10.1088/2041-8205/742/2/L23},
archivePrefix = {arXiv},
       eprint = {1110.6260},
 primaryClass = {astro-ph.CO},
       adsurl = {https://ui.adsabs.harvard.edu/abs/2011ApJ...742L..23S}
}

@ARTICLE{RubartSchwarz2013,
       author = {{Rubart}, M. and {Schwarz}, D.~J.},
        title = "{Cosmic radio dipole from NVSS and WENSS}",
      journal = {\aap},
         year = 2013,
        month = jul,
       volume = {555},
          eid = {A117},
        pages = {A117},
          doi = {10.1051/0004-6361/201321215},
archivePrefix = {arXiv},
       eprint = {1301.5559},
 primaryClass = {astro-ph.CO},
       adsurl = {https://ui.adsabs.harvard.edu/abs/2013A&A...555A.117R}
}

@ARTICLE{BrentjensDeBruyn2005,
       author = {{Brentjens}, M.~A. and {de Bruyn}, A.~G.},
        title = "{Faraday rotation measure synthesis}",
      journal = {\aap},
         year = 2005,
        month = oct,
       volume = {441},
       number = {3},
        pages = {1217-1228},
          doi = {10.1051/0004-6361:20052990},
archivePrefix = {arXiv},
       eprint = {astro-ph/0507349},
 primaryClass = {astro-ph},
       adsurl = {https://ui.adsabs.harvard.edu/abs/2005A&A...441.1217B}
}

@ARTICLE{Taylor2009,
       author = {{Taylor}, A.~R. and {Stil}, J.~M. and {Sunstrum}, C.},
        title = "{A Rotation Measure Image of the Sky}",
      journal = {\apj},
         year = 2009,
        month = sep,
       volume = {702},
       number = {2},
        pages = {1230-1236},
          doi = {10.1088/0004-637X/702/2/1230},
       adsurl = {https://ui.adsabs.harvard.edu/abs/2009ApJ...702.1230T}
}

@ARTICLE{Gaensler2025,
       author = {{Gaensler}, B.~M. and {Heald}, G.~H. and {McClure-Griffiths}, N.~M. and {Anderson}, C.~S. and {Van Eck}, C.~L. and {West}, J.~L. and {Thomson}, A.~J.~M. and {Leahy}, J.~P. and {Rudnick}, L. and {Ma}, Y.~K. and {Akahori}, Takuya and {G{\"u}rkan}, G. and {Landecker}, T.~L. and {Mao}, S.~A. and {O'Sullivan}, S.~P. and {Raja}, W. and {Sun}, X. and {Vernstrom}, T. and {Baidoo}, Lerato and {Carretti}, Ettore and {Taylor}, A.~R. and {Willis}, A.~G. and {Osinga}, Erik and {Livingston}, J.~D. and {Alexander}, E.~L. and {Alonso-L{\'o}pez}, David and {Amaral}, A.~D. and {An}, T. and {Bracco}, Andrea and {Bradbury}, S. and {Br{\"u}ggen}, Marcus and {Eswaraiah}, Chakali and {En{\ss}lin}, Torsten and {Galvin}, T.~J. and {Haverkorn}, Marijke and {Hopkins}, A.~M. and {Hutschenreuter}, Sebastian and {Ideguchi}, Shinsuke and {Jaswanth}, S. and {Jung}, S. Lyla and {Kaczmarek}, J.~F. and {Kothes}, Roland and {Lazarevi{\'c}}, Sanja and {Leahy}, Denis and {Loi}, Francesca and {Marvil}, Joshua R. and {Norris}, Ray and {Pandhi}, Ayush and {Price}, Jason M. and {Riseley}, C.~J. and {Ryder}, P. and {Seta}, Amit and {Shaw}, Vasundhara and {Shen}, A.~X. and {Sobey}, C. and {Stil}, J. and {Stuardi}, Chiara and {Upasana}, Gupta and {Vanderwoude}, Shannon and {Velovi{\'c}}, Velibor},
        title = "{The Polarisation Sky Survey of the Universe's Magnetism (POSSUM): Science goals and survey description}",
      journal = {\pasa},
         year = 2025,
        month = jun,
       volume = {42},
          eid = {e091},
        pages = {e091},
          doi = {10.1017/pasa.2025.10031},
archivePrefix = {arXiv},
       eprint = {2505.08272},
 primaryClass = {astro-ph.GA},
       adsurl = {https://ui.adsabs.harvard.edu/abs/2025PASA...42...91G}
}

@ARTICLE{Bell2015,
       author = {{Bell}, M.~E. and {Huynh}, M.~T. and {Hancock}, P. and {Murphy}, Tara and {Gaensler}, B.~M. and {Burlon}, D. and {Trott}, C. and {Bannister}, K.},
        title = "{A search for variable and transient radio sources in the extended Chandra Deep Field South at 5.5 GHz}",
      journal = {\mnras},
         year = 2015,
        month = jul,
       volume = {450},
       number = {4},
        pages = {4221-4232},
          doi = {10.1093/mnras/stv882},
archivePrefix = {arXiv},
       eprint = {1504.06371},
 primaryClass = {astro-ph.GA},
       adsurl = {https://ui.adsabs.harvard.edu/abs/2015MNRAS.450.4221B}
}

@ARTICLE{Mooley2016,
       author = {{Mooley}, K.~P. and {Hallinan}, G. and {Bourke}, S. and {Horesh}, A. and {Myers}, S.~T. and {Frail}, D.~A. and {Kulkarni}, S.~R. and {Levitan}, D.~B. and {Kasliwal}, M.~M. and {Cenko}, S.~B. and {Cao}, Y. and {Bellm}, E. and {Laher}, R.~R.},
        title = "{The Caltech-NRAO Stripe 82 Survey (CNSS). I. The Pilot Radio Transient Survey In 50 deg$^{2}$}",
      journal = {\apj},
         year = 2016,
        month = feb,
       volume = {818},
       number = {2},
          eid = {105},
        pages = {105},
          doi = {10.3847/0004-637X/818/2/105},
archivePrefix = {arXiv},
       eprint = {1601.01693},
 primaryClass = {astro-ph.HE},
       adsurl = {https://ui.adsabs.harvard.edu/abs/2016ApJ...818..105M}
}

@ARTICLE{Padovani2016,
       author = {{Padovani}, Paolo},
        title = "{The faint radio sky: radio astronomy becomes mainstream}",
      journal = {\aapr},
         year = 2016,
        month = sep,
       volume = {24},
       number = {1},
          eid = {13},
        pages = {13},
          doi = {10.1007/s00159-016-0098-6},
archivePrefix = {arXiv},
       eprint = {1609.00499},
 primaryClass = {astro-ph.GA},
       adsurl = {https://ui.adsabs.harvard.edu/abs/2016A&ARv..24...13P}
}

@ARTICLE{Blake2002,
       author = {{Blake}, Chris and {Wall}, Jasper},
        title = "{Quantifying angular clustering in wide-area radio surveys}",
      journal = {\mnras},
         year = 2002,
        month = dec,
       volume = {337},
       number = {3},
        pages = {993-1003},
          doi = {10.1046/j.1365-8711.2002.05979.x},
archivePrefix = {arXiv},
       eprint = {astro-ph/0208350},
 primaryClass = {astro-ph},
       adsurl = {https://ui.adsabs.harvard.edu/abs/2002MNRAS.337..993B}
}

@ARTICLE{Sinha_2023,
       author = {{Sinha}, Akriti and {Mangla}, Sarvesh and {Datta}, Abhirup},
        title = "{Spectral study of faint radio sources in ELAIS N1 field}",
      journal = {Journal of Astrophysics and Astronomy},
         year = 2023,
        month = dec,
       volume = {44},
       number = {2},
          eid = {88},
        pages = {88},
          doi = {10.1007/s12036-023-09978-0},
archivePrefix = {arXiv},
       eprint = {2308.05192},
 primaryClass = {astro-ph.GA},
       adsurl = {https://ui.adsabs.harvard.edu/abs/2023JApA...44...88S}
}

@ARTICLE{tasse2018,
       author = {{Tasse}, C. and {Hugo}, B. and {Mirmont}, M. and {Smirnov}, O. and {Atemkeng}, M. and {Bester}, L. and {Hardcastle}, M.~J. and {Lakhoo}, R. and {Perkins}, S. and {Shimwell}, T.},
        title = "{Faceting for direction-dependent spectral deconvolution}",
      journal = {\aap},
         year = 2018,
        month = apr,
       volume = {611},
          eid = {A87},
        pages = {A87},
          doi = {10.1051/0004-6361/201731474},
archivePrefix = {arXiv},
       eprint = {1712.02078},
 primaryClass = {astro-ph.IM},
       adsurl = {https://ui.adsabs.harvard.edu/abs/2018A&A...611A..87T}
}

@ARTICLE{DESI,
       author = {{DESI Collaboration} and {Aghamousa}, Amir and {Aguilar}, Jessica and {Ahlen}, Steve and {Alam}, Shadab and {Allen}, Lori E. and {Allende Prieto}, Carlos and {Annis}, James and {Bailey}, Stephen and {Balland}, Christophe and {Ballester}, Otger and {Baltay}, Charles and {Beaufore}, Lucas and {Bebek}, Chris and {Beers}, Timothy C. and {Bell}, Eric F. and {Bernal}, Jos{\'e} Luis and {Besuner}, Robert and {Beutler}, Florian and {Blake}, Chris and {Bleuler}, Hannes and {Blomqvist}, Michael and {Blum}, Robert and {Bolton}, Adam S. and {Briceno}, Cesar and {Brooks}, David and {Brownstein}, Joel R. and {Buckley-Geer}, Elizabeth and {Burden}, Angela and {Burtin}, Etienne and {Busca}, Nicolas G. and {Cahn}, Robert N. and {Cai}, Yan-Chuan and {Cardiel-Sas}, Laia and {Carlberg}, Raymond G. and {Carton}, Pierre-Henri and {Casas}, Ricard and {Castander}, Francisco J. and {Cervantes-Cota}, Jorge L. and {Claybaugh}, Todd M. and {Close}, Madeline and {Coker}, Carl T. and {Cole}, Shaun and {Comparat}, Johan and {Cooper}, Andrew P. and {Cousinou}, M. -C. and {Crocce}, Martin and {Cuby}, Jean-Gabriel and {Cunningham}, Daniel P. and {Davis}, Tamara M. and {Dawson}, Kyle S. and {de la Macorra}, Axel and {De Vicente}, Juan and {Delubac}, Timoth{\'e}e and {Derwent}, Mark and {Dey}, Arjun and {Dhungana}, Govinda and {Ding}, Zhejie and {Doel}, Peter and {Duan}, Yutong T. and {Ealet}, Anne and {Edelstein}, Jerry and {Eftekharzadeh}, Sarah and {Eisenstein}, Daniel J. and {Elliott}, Ann and {Escoffier}, St{\'e}phanie and {Evatt}, Matthew and {Fagrelius}, Parker and {Fan}, Xiaohui and {Fanning}, Kevin and {Farahi}, Arya and {Farihi}, Jay and {Favole}, Ginevra and {Feng}, Yu and {Fernandez}, Enrique and {Findlay}, Joseph R. and {Finkbeiner}, Douglas P. and {Fitzpatrick}, Michael J. and {Flaugher}, Brenna and {Flender}, Samuel and {Font-Ribera}, Andreu and {Forero-Romero}, Jaime E. and {Fosalba}, Pablo and {Frenk}, Carlos S. and {Fumagalli}, Michele and {Gaensicke}, Boris T. and {Gallo}, Giuseppe and {Garcia-Bellido}, Juan and {Gaztanaga}, Enrique and {Pietro Gentile Fusillo}, Nicola and {Gerard}, Terry and {Gershkovich}, Irena and {Giannantonio}, Tommaso and {Gillet}, Denis and {Gonzalez-de-Rivera}, Guillermo and {Gonzalez-Perez}, Violeta and {Gott}, Shelby and {Graur}, Or and {Gutierrez}, Gaston and {Guy}, Julien and {Habib}, Salman and {Heetderks}, Henry and {Heetderks}, Ian and {Heitmann}, Katrin and {Hellwing}, Wojciech A. and {Herrera}, David A. and {Ho}, Shirley and {Holland}, Stephen and {Honscheid}, Klaus and {Huff}, Eric and {Hutchinson}, Timothy A. and {Huterer}, Dragan and {Hwang}, Ho Seong and {Illa Laguna}, Joseph Maria and {Ishikawa}, Yuzo and {Jacobs}, Dianna and {Jeffrey}, Niall and {Jelinsky}, Patrick and {Jennings}, Elise and {Jiang}, Linhua and {Jimenez}, Jorge and {Johnson}, Jennifer and {Joyce}, Richard and {Jullo}, Eric and {Juneau}, St{\'e}phanie and {Kama}, Sami and {Karcher}, Armin and {Karkar}, Sonia and {Kehoe}, Robert and {Kennamer}, Noble and {Kent}, Stephen and {Kilbinger}, Martin and {Kim}, Alex G. and {Kirkby}, David and {Kisner}, Theodore and {Kitanidis}, Ellie and {Kneib}, Jean-Paul and {Koposov}, Sergey and {Kovacs}, Eve and {Koyama}, Kazuya and {Kremin}, Anthony and {Kron}, Richard and {Kronig}, Luzius and {Kueter-Young}, Andrea and {Lacey}, Cedric G. and {Lafever}, Robin and {Lahav}, Ofer and {Lambert}, Andrew and {Lampton}, Michael and {Landriau}, Martin and {Lang}, Dustin and {Lauer}, Tod R. and {Le Goff}, Jean-Marc and {Le Guillou}, Laurent and {Le Van Suu}, Auguste and {Lee}, Jae Hyeon and {Lee}, Su-Jeong and {Leitner}, Daniela and {Lesser}, Michael and {Levi}, Michael E. and {L'Huillier}, Benjamin and {Li}, Baojiu and {Liang}, Ming and {Lin}, Huan and {Linder}, Eric and {Loebman}, Sarah R. and {Luki{\'c}}, Zarija and {Ma}, Jun and {MacCrann}, Niall and {Magneville}, Christophe and {Makarem}, Laleh and {Manera}, Marc and {Manser}, Christopher J. and {Marshall}, Robert and {Martini}, Paul and {Massey}, Richard and {Matheson}, Thomas and {McCauley}, Jeremy and {McDonald}, Patrick and {McGreer}, Ian D. and {Meisner}, Aaron and {Metcalfe}, Nigel and {Miller}, Timothy N. and {Miquel}, Ramon and {Moustakas}, John and {Myers}, Adam and {Naik}, Milind and {Newman}, Jeffrey A. and {Nichol}, Robert C. and {Nicola}, Andrina and {Nicolati da Costa}, Luiz and {Nie}, Jundan and {Niz}, Gustavo and {Norberg}, Peder and {Nord}, Brian and {Norman}, Dara and {Nugent}, Peter and {O'Brien}, Thomas and {Oh}, Minji and {Olsen}, Knut A.~G.},
        title = "{The DESI Experiment Part I: Science,Targeting, and Survey Design}",
      journal = {arXiv e-prints},
         year = 2016,
        month = oct,
          eid = {arXiv:1611.00036},
        pages = {arXiv:1611.00036},
          doi = {10.48550/arXiv.1611.00036},
archivePrefix = {arXiv},
       eprint = {1611.00036},
 primaryClass = {astro-ph.IM},
       adsurl = {https://ui.adsabs.harvard.edu/abs/2016arXiv161100036D}
}

@software{2023ascl.soft05005T,
       author = {{Tasse}, Cyril},
        title = "{killMS: Direction-dependent radio interferometric calibration package}",
 howpublished = {Astrophysics Source Code Library, record ascl:2305.005},
         year = 2023,
        month = may,
          eid = {ascl:2305.005},
archivePrefix = {ascl},
       eprint = {2305.005},
       adsurl = {https://ui.adsabs.harvard.edu/abs/2023ascl.soft05005T}
}

@INPROCEEDINGS{becker1994,
       author = {{Becker}, Robert H. and {White}, Richard L. and {Helfand}, David J.},
        title = "{The VLA's FIRST Survey}",
    booktitle = {Astronomical Data Analysis Software and Systems III},
         year = 1994,
       editor = {{Crabtree}, D.~R. and {Hanisch}, R.~J. and {Barnes}, J.},
       series = {Astronomical Society of the Pacific Conference Series},
       volume = {61},
        month = jan,
        pages = {165},
       adsurl = {https://ui.adsabs.harvard.edu/abs/1994ASPC...61..165B}
}

@ARTICLE{Numpy,
       author = {{Harris}, Charles R. and {Millman}, K. Jarrod and {van der Walt}, St{\'e}fan J. and {Gommers}, Ralf and {Virtanen}, Pauli and {Cournapeau}, David and {Wieser}, Eric and {Taylor}, Julian and {Berg}, Sebastian and {Smith}, Nathaniel J. and {Kern}, Robert and {Picus}, Matti and {Hoyer}, Stephan and {van Kerkwijk}, Marten H. and {Brett}, Matthew and {Haldane}, Allan and {del R{\'\i}o}, Jaime Fern{\'a}ndez and {Wiebe}, Mark and {Peterson}, Pearu and {G{\'e}rard-Marchant}, Pierre and {Sheppard}, Kevin and {Reddy}, Tyler and {Weckesser}, Warren and {Abbasi}, Hameer and {Gohlke}, Christoph and {Oliphant}, Travis E.},
        title = "{Array programming with NumPy}",
      journal = {\nat},
         year = 2020,
        month = sep,
       volume = {585},
       number = {7825},
        pages = {357-362},
          doi = {10.1038/s41586-020-2649-2},
archivePrefix = {arXiv},
       eprint = {2006.10256},
 primaryClass = {cs.MS},
       adsurl = {https://ui.adsabs.harvard.edu/abs/2020Natur.585..357H}
}

@ARTICLE{Scipy,
       author = {{Virtanen}, Pauli and {Gommers}, Ralf and {Oliphant}, Travis E. and {Haberland}, Matt and {Reddy}, Tyler and {Cournapeau}, David and {Burovski}, Evgeni and {Peterson}, Pearu and {Weckesser}, Warren and {Bright}, Jonathan and {van der Walt}, St{\'e}fan J. and {Brett}, Matthew and {Wilson}, Joshua and {Millman}, K. Jarrod and {Mayorov}, Nikolay and {Nelson}, Andrew R.~J. and {Jones}, Eric and {Kern}, Robert and {Larson}, Eric and {Carey}, C.~J. and {Polat}, {\.I}lhan and {Feng}, Yu and {Moore}, Eric W. and {VanderPlas}, Jake and {Laxalde}, Denis and {Perktold}, Josef and {Cimrman}, Robert and {Henriksen}, Ian and {Quintero}, E.~A. and {Harris}, Charles R. and {Archibald}, Anne M. and {Ribeiro}, Ant{\^o}nio H. and {Pedregosa}, Fabian and {van Mulbregt}, Paul and {SciPy 1. 0 Contributors}},
        title = "{SciPy 1.0: fundamental algorithms for scientific computing in Python}",
      journal = {Nature Methods},
         year = 2020,
        month = feb,
       volume = {17},
        pages = {261-272},
          doi = {10.1038/s41592-019-0686-2},
archivePrefix = {arXiv},
       eprint = {1907.10121},
 primaryClass = {cs.MS},
       adsurl = {https://ui.adsabs.harvard.edu/abs/2020NatMe..17..261V}
}

@ARTICLE{Matplotlib,
       author = {{Hunter}, John D.},
        title = "{Matplotlib: A 2D Graphics Environment}",
      journal = {Computing in Science and Engineering},
         year = 2007,
        month = may,
       volume = {9},
       number = {3},
        pages = {90-95},
          doi = {10.1109/MCSE.2007.55},
       adsurl = {https://ui.adsabs.harvard.edu/abs/2007CSE.....9...90H}
}

@ARTICLE{astropy:2022,
       author = {{Astropy Collaboration} and {Price-Whelan}, Adrian M. and {Lim}, Pey Lian and {Earl}, Nicholas and {Starkman}, Nathaniel and {Bradley}, Larry and {Shupe}, David L. and {Patil}, Aarya A. and {Corrales}, Lia and {Brasseur}, C.~E. and {N{"o}the}, Maximilian and {Donath}, Axel and {Tollerud}, Erik and {Morris}, Brett M. and {Ginsburg}, Adam and {Vaher}, Eero and {Weaver}, Benjamin A. and {Tocknell}, James and {Jamieson}, William and {van Kerkwijk}, Marten H. and {Robitaille}, Thomas P. and {Merry}, Bruce and {Bachetti}, Matteo and {G{"u}nther}, H. Moritz and {Aldcroft}, Thomas L. and {Alvarado-Montes}, Jaime A. and {Archibald}, Anne M. and {B{'o}di}, Attila and {Bapat}, Shreyas and {Barentsen}, Geert and {Baz{'a}n}, Juanjo and {Biswas}, Manish and {Boquien}, M{'e}d{'e}ric and {Burke}, D.~J. and {Cara}, Daria and {Cara}, Mihai and {Conroy}, Kyle E. and {Conseil}, Simon and {Craig}, Matthew W. and {Cross}, Robert M. and {Cruz}, Kelle L. and {D'Eugenio}, Francesco and {Dencheva}, Nadia and {Devillepoix}, Hadrien A.~R. and {Dietrich}, J{"o}rg P. and {Eigenbrot}, Arthur Davis and {Erben}, Thomas and {Ferreira}, Leonardo and {Foreman-Mackey}, Daniel and {Fox}, Ryan and {Freij}, Nabil and {Garg}, Suyog and {Geda}, Robel and {Glattly}, Lauren and {Gondhalekar}, Yash and {Gordon}, Karl D. and {Grant}, David and {Greenfield}, Perry and {Groener}, Austen M. and {Guest}, Steve and {Gurovich}, Sebastian and {Handberg}, Rasmus and {Hart}, Akeem and {Hatfield-Dodds}, Zac and {Homeier}, Derek and {Hosseinzadeh}, Griffin and {Jenness}, Tim and {Jones}, Craig K. and {Joseph}, Prajwel and {Kalmbach}, J. Bryce and {Karamehmetoglu}, Emir and {Ka{l}uszy{'n}ski}, Miko{l}aj and {Kelley}, Michael S.~P. and {Kern}, Nicholas and {Kerzendorf}, Wolfgang E. and {Koch}, Eric W. and {Kulumani}, Shankar and {Lee}, Antony and {Ly}, Chun and {Ma}, Zhiyuan and {MacBride}, Conor and {Maljaars}, Jakob M. and {Muna}, Demitri and {Murphy}, N.~A. and {Norman}, Henrik and {O'Steen}, Richard and {Oman}, Kyle A. and {Pacifici}, Camilla and {Pascual}, Sergio and {Pascual-Granado}, J. and {Patil}, Rohit R. and {Perren}, Gabriel I. and {Pickering}, Timothy E. and {Rastogi}, Tanuj and {Roulston}, Benjamin R. and {Ryan}, Daniel F. and {Rykoff}, Eli S. and {Sabater}, Jose and {Sakurikar}, Parikshit and {Salgado}, Jes{'u}s and {Sanghi}, Aniket and {Saunders}, Nicholas and {Savchenko}, Volodymyr and {Schwardt}, Ludwig and {Seifert-Eckert}, Michael and {Shih}, Albert Y. and {Jain}, Anany Shrey and {Shukla}, Gyanendra and {Sick}, Jonathan and {Simpson}, Chris and {Singanamalla}, Sudheesh and {Singer}, Leo P. and {Singhal}, Jaladh and {Sinha}, Manodeep and {Sip{H{o}}cz}, Brigitta M. and {Spitler}, Lee R. and {Stansby}, David and {Streicher}, Ole and {{{S}}umak}, Jani and {Swinbank}, John D. and {Taranu}, Dan S. and {Tewary}, Nikita and {Tremblay}, Grant R. and {Val-Borro}, Miguel de and {Van Kooten}, Samuel J. and {Vasovi{'c}}, Zlatan and {Verma}, Shresth and {de Miranda Cardoso}, Jos{'e} Vin{'i}cius and {Williams}, Peter K.~G. and {Wilson}, Tom J. and {Winkel}, Benjamin and {Wood-Vasey}, W.~M. and {Xue}, Rui and {Yoachim}, Peter and {Zhang}, Chen and {Zonca}, Andrea and {Astropy Project Contributors}},
        title = "{The Astropy Project: Sustaining and Growing a Community-oriented Open-source Project and the Latest Major Release (v5.0) of the Core Package}",
      journal = {\apj},
         year = 2022,
        month = aug,
       volume = {935},
       number = {2},
          eid = {167},
        pages = {167},
          doi = {10.3847/1538-4357/ac7c74},
archivePrefix = {arXiv},
       eprint = {2206.14220},
 primaryClass = {astro-ph.IM},
       adsurl = {https://ui.adsabs.harvard.edu/abs/2022ApJ...935..167A}
}

@ARTICLE{matthews2021,
       author = {{Matthews}, A.~M. and {Condon}, J.~J. and {Cotton}, W.~D. and {Mauch}, T.},
        title = "{Source Counts Spanning Eight Decades of Flux Density at 1.4 GHz}",
      journal = {\apj},
         year = 2021,
        month = mar,
       volume = {909},
       number = {2},
          eid = {193},
        pages = {193},
          doi = {10.3847/1538-4357/abdd37},
archivePrefix = {arXiv},
       eprint = {2101.07827},
 primaryClass = {astro-ph.GA},
       adsurl = {https://ui.adsabs.harvard.edu/abs/2021ApJ...909..193M}
}

@ARTICLE{wsclean,
       author = {{Offringa}, A.~R. and {McKinley}, B. and {Hurley-Walker}, N. and {Briggs}, F.~H. and {Wayth}, R.~B. and {Kaplan}, D.~L. and {Bell}, M.~E. and {Feng}, L. and {Neben}, A.~R. and {Hughes}, J.~D. and {Rhee}, J. and {Murphy}, T. and {Bhat}, N.~D.~R. and {Bernardi}, G. and {Bowman}, J.~D. and {Cappallo}, R.~J. and {Corey}, B.~E. and {Deshpande}, A.~A. and {Emrich}, D. and {Ewall-Wice}, A. and {Gaensler}, B.~M. and {Goeke}, R. and {Greenhill}, L.~J. and {Hazelton}, B.~J. and {Hindson}, L. and {Johnston-Hollitt}, M. and {Jacobs}, D.~C. and {Kasper}, J.~C. and {Kratzenberg}, E. and {Lenc}, E. and {Lonsdale}, C.~J. and {Lynch}, M.~J. and {McWhirter}, S.~R. and {Mitchell}, D.~A. and {Morales}, M.~F. and {Morgan}, E. and {Kudryavtseva}, N. and {Oberoi}, D. and {Ord}, S.~M. and {Pindor}, B. and {Procopio}, P. and {Prabu}, T. and {Riding}, J. and {Roshi}, D.~A. and {Shankar}, N. Udaya and {Srivani}, K.~S. and {Subrahmanyan}, R. and {Tingay}, S.~J. and {Waterson}, M. and {Webster}, R.~L. and {Whitney}, A.~R. and {Williams}, A. and {Williams}, C.~L.},
        title = "{WSCLEAN: an implementation of a fast, generic wide-field imager for radio astronomy}",
      journal = {\mnras},
         year = 2014,
        month = oct,
       volume = {444},
       number = {1},
        pages = {606-619},
          doi = {10.1093/mnras/stu1368},
archivePrefix = {arXiv},
       eprint = {1407.1943},
 primaryClass = {astro-ph.IM},
       adsurl = {https://ui.adsabs.harvard.edu/abs/2014MNRAS.444..606O}
}

@software{mohan2015pybdsf,
       author = {{Mohan}, Niruj and {Rafferty}, David},
        title = "{PyBDSF: Python Blob Detection and Source Finder}",
 howpublished = {Astrophysics Source Code Library, record ascl:1502.007},
         year = 2015,
        month = feb,
          eid = {ascl:1502.007},
archivePrefix = {ascl},
       eprint = {1502.007},
       adsurl = {https://ui.adsabs.harvard.edu/abs/2015ascl.soft02007M}
}

@ARTICLE{Offringa2010,
       author = {{Offringa}, A.~R. and {de Bruyn}, A.~G. and {Biehl}, M. and {Zaroubi}, S. and {Bernardi}, G. and {Pandey}, V.~N.},
        title = "{Post-correlation radio frequency interference classification methods}",
      journal = {\mnras},
         year = 2010,
        month = jun,
       volume = {405},
       number = {1},
        pages = {155-167},
          doi = {10.1111/j.1365-2966.2010.16471.x},
archivePrefix = {arXiv},
       eprint = {1002.1957},
 primaryClass = {astro-ph.IM},
       adsurl = {https://ui.adsabs.harvard.edu/abs/2010MNRAS.405..155O}
}

@software{Jacob2010,
       author = {{Jacob}, Joseph C. and {Katz}, Daniel S. and {Berriman}, G. Bruce and {Good}, John and {Laity}, Anastasia C. and {Deelman}, Ewa and {Kesselman}, Carl and {Singh}, Gurmeet and {Su}, Mei-Hui and {Prince}, Thomas A. and {Williams}, Roy},
        title = "{Montage: An Astronomical Image Mosaicking Toolkit}",
 howpublished = {Astrophysics Source Code Library, record ascl:1010.036},
         year = 2010,
        month = oct,
          eid = {ascl:1010.036},
archivePrefix = {ascl},
       eprint = {1010.036},
       adsurl = {https://ui.adsabs.harvard.edu/abs/2010ascl.soft10036J}
}

@ARTICLE{Mangum2007,
       author = {{Mangum}, J.~G. and {Emerson}, D.~T. and {Greisen}, E.~W.},
        title = "{The On The Fly imaging technique}",
      journal = {\aap},
         year = 2007,
        month = nov,
       volume = {474},
       number = {2},
        pages = {679-687},
          doi = {10.1051/0004-6361:20077811},
archivePrefix = {arXiv},
       eprint = {0709.0553},
 primaryClass = {astro-ph},
       adsurl = {https://ui.adsabs.harvard.edu/abs/2007A&A...474..679M}
}

@ARTICLE{2025arXiv251027549C,
       author = {{Cunnington}, Steven and {Barberi-Squarotti}, Matilde and {Bernal}, Jos{\'e} Luis and {Camera}, Stefano and {Carucci}, Isabella P. and {Chen}, Zhaoting and {Fonseca}, Jos{\'e} and {Santos}, Mario and {Spinelli}, Marta and {Wang}, Jingying and {Wolz}, Laura},
        title = "{Revealing cosmological fluctuations in 21cm intensity maps with MeerKLASS: from maps to power spectra}",
      journal = {arXiv e-prints},
         year = 2025,
        month = oct,
          eid = {arXiv:2510.27549},
        pages = {arXiv:2510.27549},
          doi = {10.48550/arXiv.2510.27549},
archivePrefix = {arXiv},
       eprint = {2510.27549},
 primaryClass = {astro-ph.CO},
       adsurl = {https://ui.adsabs.harvard.edu/abs/2025arXiv251027549C}
}

@ARTICLE{2025PhRvL.135t1001B,
       author = {{B{\"o}hme}, Lukas and {Schwarz}, Dominik J. and {Tiwari}, Prabhakar and {Pashapour-Ahmadabadi}, Morteza and {Bahr-Kalus}, Benedict and {Bilicki}, Maciej and {Hale}, Catherine L. and {Heneka}, Caroline S. and {Siewert}, Thilo M.},
        title = "{Overdispersed Radio Source Counts and Excess Radio Dipole Detection}",
      journal = {\prl},
         year = 2025,
        month = nov,
       volume = {135},
       number = {20},
          eid = {201001},
        pages = {201001},
          doi = {10.1103/6z32-3zf4},
archivePrefix = {arXiv},
       eprint = {2509.16732},
 primaryClass = {astro-ph.CO},
       adsurl = {https://ui.adsabs.harvard.edu/abs/2025PhRvL.135t1001B}
}

@ARTICLE{Hopkins2025,
       author = {{Hopkins}, Andrew and {Kapinska}, Anna and {Marvil}, Joshua and {Vernstrom}, Tessa and {Collier}, Jordan and {Norris}, Ray and {Gordon}, Yjan and {Duchesne}, Stefan and {Rudnick}, Lawrence and {Gupta}, Nikhel and {Carretti}, Ettore and {Anderson}, Craig and {Dai}, Shi and {G{\"u}rkan}, Gulay and {Parkinson}, David and {Prandoni}, Isabella and {Riggi}, Simone and {Shekhar Saraf}, Chandra and {Ma}, Yik Ki and {Filipovi{\'c}}, Miroslav D. and {Umana}, Grazia and {Bahr-Kalus}, Benedict and {Koribalski}, B{\"a}rbel Silvia and {Lenc}, Emil and {Ingallinera}, Adriano and {Afonso}, Jos{\'e} and {Ahmad}, Adeel and {Ahmed}, Ummee Tania and {Alexander}, Emma and {Andernach}, Heinz and {Asorey}, Jacobo and {Battisti}, Andrew J. and {Bilicki}, Maciej and {Botteon}, Andrea and {Brown}, Michael and {Br{\"u}ggen}, Marcus and {Cowley}, Michael and {Dage}, Kristen and {Hale}, Catherine Laura and {Hardcastle}, Martin and {Kothes}, Roland and {Lazarevi{\'c}}, Sanja and {Lin}, Yen-Ting and {Luken}, Kieran and {Moss}, Jeremy and {Prathap}, P.~K. Jahang and {ur Rahman}, Syed Faisal and {Reiprich}, Thomas and {Riseley}, Christopher and {Salvato}, Mara and {Seymour}, Nicholas and {Shabala}, Stanislav and {Smith}, Daniel and {Vaccari}, Mattia and {van Loon}, Jacco Th. and {Wong}, O. Ivy Ivy and {Zainal Alsaberi}, Rami and {Asher}, Albany and {Ball}, Brianna and {Barbosa}, Davi and {Biava}, Nadia and {Bradley}, Aaron and {Carvajal}, Rodrigo and {Crawford}, Evan J. and {Galvin}, Timothy James and {Huynh}, Minh and {Leahy}, Denis and {Matute}, Israel and {Moss}, Vanessa and {Pappalardo}, Ciro and {Smeaton}, Zachary and {Velovi{\'c}}, Velibor and {Zafar}, Tayyaba},
        title = "{The Evolutionary Map of the Universe: A new radio atlas for the southern hemisphere sky}",
      journal = {\pasa},
         year = 2025,
        month = may,
       volume = {42},
          eid = {e071},
        pages = {e071},
          doi = {10.1017/pasa.2025.10042},
archivePrefix = {arXiv},
       eprint = {2505.08271},
 primaryClass = {astro-ph.GA},
       adsurl = {https://ui.adsabs.harvard.edu/abs/2025PASA...42...71H}
}

@software{2020ascl.soft06014J,
       author = {{J{\'o}zsa}, Gyula I.~G. and {White}, Sarah V. and {Thorat}, Kshitij and {Smirnov}, Oleg M. and {Serra}, Paolo and {Ramatsoku}, Mpati and {Ramaila}, Athanaseus J.~T. and {Perkins}, Simon J. and {Moln{\'a}r}, D{\'a}niel Cs. and {Makhathini}, Sphesihle and {Maccagni}, Filippo M. and {Kleiner}, Dane and {Kamphuis}, Peter and {Hugo}, Benjamin V. and {de Blok}, W.~J.~G. and {Andati}, Lexy A.~L.},
        title = "{CARACal: Containerized Automated Radio Astronomy Calibration pipeline}",
 howpublished = {Astrophysics Source Code Library, record ascl:2006.014},
         year = 2020,
        month = jun,
          eid = {ascl:2006.014},
archivePrefix = {ascl},
       eprint = {2006.014},
       adsurl = {https://ui.adsabs.harvard.edu/abs/2020ascl.soft06014J}
}

@INPROCEEDINGS{2022ASPC..532..541H,
       author = {{Hugo}, Benjamin V. and {Perkins}, S. and {Merry}, B. and {Mauch}, T. and {Smirnov}, O.~M.},
        title = "{Tricolour: An Optimized SumThreshold Flagger for MeerKAT}",
    booktitle = {Astronomical Data Analysis Software and Systems XXX},
         year = 2022,
       editor = {{Ruiz}, Jose Enrique and {Pierfedereci}, Francesco and {Teuben}, Peter},
       series = {Astronomical Society of the Pacific Conference Series},
       volume = {532},
        month = jul,
        pages = {541},
          doi = {10.48550/arXiv.2206.09179},
archivePrefix = {arXiv},
       eprint = {2206.09179},
 primaryClass = {astro-ph.IM},
       adsurl = {https://ui.adsabs.harvard.edu/abs/2022ASPC..532..541H}
}

@INPROCEEDINGS{1999ASPC..180..127B,
       author = {{Briggs}, Daniel S. and {Schwab}, Frederic R. and {Sramek}, Richard A.},
        title = "{Imaging}",
    booktitle = {Synthesis Imaging in Radio Astronomy II},
         year = 1999,
       editor = {{Taylor}, G.~B. and {Carilli}, C.~L. and {Perley}, R.~A.},
       series = {Astronomical Society of the Pacific Conference Series},
       volume = {180},
        month = jan,
        pages = {127},
       adsurl = {https://ui.adsabs.harvard.edu/abs/1999ASPC..180..127B}
}

@ARTICLE{Bondi2003,
       author = {{Bondi}, M. and {Ciliegi}, P. and {Zamorani}, G. and {Gregorini}, L. and {Vettolani}, G. and {Parma}, P. and {de Ruiter}, H. and {Le Fevre}, O. and {Arnaboldi}, M. and {Guzzo}, L. and {Maccagni}, D. and {Scaramella}, R. and {Adami}, C. and {Bardelli}, S. and {Bolzonella}, M. and {Bottini}, D. and {Cappi}, A. and {Foucaud}, S. and {Franzetti}, P. and {Garilli}, B. and {Gwyn}, S. and {Ilbert}, O. and {Iovino}, A. and {Le Brun}, V. and {Marano}, B. and {Marinoni}, C. and {McCracken}, H.~J. and {Meneux}, B. and {Pollo}, A. and {Pozzetti}, L. and {Radovich}, M. and {Ripepi}, V. and {Rizzo}, D. and {Scodeggio}, M. and {Tresse}, L. and {Zanichelli}, A. and {Zucca}, E.},
        title = "{The VLA-VIRMOS Deep Field. I. Radio observations probing the mu Jy source population}",
      journal = {\aap},
         year = 2003,
        month = jun,
       volume = {403},
        pages = {857-867},
          doi = {10.1051/0004-6361:20030382},
archivePrefix = {arXiv},
       eprint = {astro-ph/0303364},
 primaryClass = {astro-ph},
       adsurl = {https://ui.adsabs.harvard.edu/abs/2003A&A...403..857B}
}

@ARTICLE{Hales2014,
       author = {{Hales}, C.~A. and {Norris}, R.~P. and {Gaensler}, B.~M. and {Middelberg}, E. and {Chow}, K.~E. and {Hopkins}, A.~M. and {Huynh}, M.~T. and {Lenc}, E. and {Mao}, M.~Y.},
        title = "{ATLAS 1.4 GHz Data Release 2 - I. Observations of the CDF-S and ELAIS-S1 fields and methods for constructing differential number counts}",
      journal = {\mnras},
         year = 2014,
        month = jul,
       volume = {441},
       number = {3},
        pages = {2555-2592},
          doi = {10.1093/mnras/stu576},
archivePrefix = {arXiv},
       eprint = {1403.5307},
 primaryClass = {astro-ph.GA},
       adsurl = {https://ui.adsabs.harvard.edu/abs/2014MNRAS.441.2555H}
}

\appendix
\section*{APPENDIX A: CATALOGUE STRUCTURE}

This data release contains two FITS binary tables: (i) the \emph{source} catalogue
(\texttt{MeerKLASS\_UHF\_DR1\_SRL.fits}; hereafter SRL), and (ii) the \emph{Gaussian‐component}
catalogue (\texttt{MeerKLASS\_UHF\_DR1\_GAUL.fits}; hereafter GAUL). The imaging was organised into $3.2^{\circ}\!\times\!3.2^{\circ}$ tiles with
$\simeq0.1^{\circ}$ overlaps, each identified by an integer \texttt{Tile\_ID}. 

\subsection*{The SRL (source) catalogue}
The SRL table lists one row per \emph{source} after cross‐tile de‐duplication.
Columns appear in the following order:
\begin{itemize}\itemsep0.2em
  \item \textbf{Source\_Name} \,(\textit{string}) -- IAU identifier in the form
  \texttt{MeerKLASS-UHF\_DR1 JHHMMSS.S$\pm$DDMMSS.S}. Derived from the best astrometric
  position in ICRS (precision: RA to 0.1s, Dec to 0.1'').
  \item \textbf{N\_Gaus} \,(\textit{int}) -- number of Gaussian components associated with the source.
  \item \textbf{Source\_id} \,(\textit{string}) -- globally unique source identifier, constructed as \texttt{Tile\_\{Tile\_ID\}\_\{local Source\_id\}} (e.g. \texttt{Tile\_41\_1234}).
  This serves as the primary key used to join GAUL $\rightarrow$ SRL.
  \item \textbf{PyBDSF block} -- all native PyBDSF source columns from \textbf{RA} through
  \textbf{S\_Code} (inclusive), in PyBDSF’s original order. Typical fields include:
  \texttt{RA, E\_RA, DEC, E\_DEC} in degree; 
  \texttt{Total\_flux, E\_Total\_flux} in Jy;
  \texttt{Peak\_flux, E\_Peak\_flux} in Jy\,beam$^{-1}$, 
  deconvolved sizes/PA, local noise (\texttt{Isl\_rms}), and the PyBDSF source code
  \texttt{S\_Code} (e.g. single/multiple).
  \item \textbf{E\_Total\_flux\_combined} -- The total flux-density error obtained by adding in quadrature the PyBDSF fitting error and the flux-density terms from Section 6.2 of Chatterjee et al. 
  \item \textbf{Tile\_ID} \,(\textit{int}) -- integer tile identifier of the retained detection after de‐duplication.
  \item \textbf{Tile\_BMAJ}, \textbf{Tile\_BMIN} \,(\si{arcsec}), \textbf{Tile\_BPA} \,(\si{deg}) -- restoring‐beam FWHM major/minor axes and position angle for the corresponding tile,
  propagated from the tile image headers (FITS keywords \texttt{BMAJ/BMIN/BPA}).
  \item \textbf{center\_dist} \,(\si{arcmin}) -- great‐circle separation of the source from the
  tile centre (used only as a tiebreaker during de‐duplication).
\end{itemize}

\subsection*{The GAUL (Gaussian components) catalogue}
The GAUL table lists individual 2D Gaussian components fitted by PyBDSF and retained
for the SRL sources (components are taken from the same tile as the winning SRL entry; no
additional de‐duplication is applied at the component level). Key columns are:
\begin{itemize}\itemsep0.2em
  \item \textbf{Gaus\_id} \,(\textit{string}) -- globally unique component identifier
  \texttt{Tile\_\{Tile\_ID\}\_\{local Gaus\_id\}}.
  \item \textbf{Source\_id} \,(\textit{string}) -- parent SRL key, same format as in SRL
  (above). This is the foreign key for GAUL$\rightarrow$SRL joins.
  \item \textbf{Tile\_ID} \,(\textit{int}) -- tile identifier.
\end{itemize}
The GAUL table otherwise contains the standard PyBDSF component parameters:
component centre (RA, Dec), peak and integrated flux densities, deconvolved major/minor
axes and position angle, and shape parameters. The housekeeping fields \texttt{Isl\_id} and \texttt{Wave\_id} have been removed in the release version for clarity. Units follow PyBDSF conventions (positions in \si{deg}, peak in \si{Jy\,beam^{-1}}, integrated in \si{Jy}, sizes in \si{arcsec}, angles in \si{deg}).

\subsection*{Usage notes}

\begin{itemize}\itemsep0.2em
  \item \textbf{Astrometry/flux scale.} Accuracy is characterised in \autoref{subsec:astrophoto}; offsets are well below the $3''$ pixel scale, and fluxes agree with external surveys within expected spectral and resolution systematics (see main text).
  \item \textbf{Beam columns.} \texttt{Tile\_BMAJ/BMIN/BPA} describe the restoring beam of the tile supplying the retained detection; they are not per‐source beam fits.
  \item \textbf{De‐dup radius.} The $3''$ cross‐tile match radius corresponds to one pixel and was validated to balance completeness against false associations in overlap regions.
  \item \textbf{Source\_id/Gaus\_id strings.} Identifiers are case‐sensitive and include the tile prefix; treat them as strings in downstream analysis.
\end{itemize}

\section*{APPENDIX B: Map noise level and the point source flux error}
The smearing resulting from the delay centre being at a fixed azimuth and elevation has an unusual consequence on the relationship between the noise level on the map and the flux error for a point source. The impact of the fixed delay centre is that the complex visibilities that are averaged during the 2-second sampling period do not in general have the same phase, and so coherence is lost. The degree of coherence loss depends upon the fringe-rate of the visibilities, which is proportional to their $u$-coordinate. Therefore, in the $uv$-plane the visibility amplitude of a point source decreases according to a sinc function dependent on the $u$-coordinate. As a result, when the point source is mapped, its point-spread function is convolved with a top-hat in the RA direction, giving the smearing that is apparent in images. However, the impact of the fixed delay-centre on the visibility noise is very different. The visibility noises that are averaged during the 2-second sampling period all have independent, random phases, so a phase rotation during the integration has no effect on the expectation amplitude of the noise. Therefore, in the $uv$-plane the noise amplitude is unaffected by smearing, with the result that when the noise is mapped, its point-spread function is determined solely by the weighted-sampling of the $uv$-plane; it is not convolved with a top-hat in the RA direction. The result of this is that the solid angle of the point-spread function of a source is larger than the point-spread function of noise features on the map. This means that there are multiple independent noise contributions (independent noise “pixels”) to the flux determined for a point source, with the result that the error on the source flux is greater than the noise level on the map. This increase in error is a factor equal to the square root of the number of independent noise contributions, i.e. the square root of the ratio of the solid angle of the source point-spread function to the noise point-spread function. This effect has been accounted for in the catalogue tables, with the result that the flux errors are higher than the map noise levels.

\bsp	
\label{lastpage}
\end{document}